\documentclass[lettersize,journal]{IEEEtran}
\def\BibTeX{{\rm B\kern-.05em{\sc i\kern-.025em b}\kern-.08em
    T\kern-.1667em\lower.7ex\hbox{E}\kern-.125emX}}

\usepackage{xurl}
\usepackage{amsmath}
\usepackage{amssymb}
\usepackage{graphicx}
\usepackage{array}
\usepackage{booktabs}
\usepackage{multirow}
\usepackage{makecell}
\usepackage{tabularx}
\usepackage{longtable}
\usepackage{threeparttable}
\usepackage{ifthen}
\usepackage[font=footnotesize,labelfont=bf]{caption}
\usepackage{adjustbox}
\usepackage{pdflscape}
\usepackage{algorithm}
\usepackage{algpseudocode}
\usepackage[table]{xcolor}
\usepackage[resetlabels]{multibib}
\newcites{app}{Appendix References}
\usepackage{fvextra}
\usepackage{float}
\usepackage{tikz}
\usetikzlibrary{matrix}
\usepackage[hidelinks]{hyperref}

\definecolor{tgcmshade}{gray}{0.92}
\definecolor{kcInit}{HTML}{DDEBFF}
\definecolor{kcExec}{HTML}{FFE6D1}
\definecolor{kcPers}{HTML}{E2F3E6}
\definecolor{kcDef}{HTML}{FBE0E0}
\definecolor{kcCred}{HTML}{EADCF8}
\definecolor{kcDisc}{HTML}{FFF4C9}
\definecolor{kcLat}{HTML}{FCE4EC}
\definecolor{kcColl}{HTML}{D7F1F0}
\definecolor{kcCtwo}{HTML}{DCE7FF}
\definecolor{kcExf}{HTML}{E8E8E8}
\definecolor{kcInitH}{HTML}{C9DDFF}
\definecolor{kcExecH}{HTML}{FFD7B3}
\definecolor{kcPersH}{HTML}{CDEDD5}
\definecolor{kcDefH}{HTML}{F6CACA}
\definecolor{kcCredH}{HTML}{DCC6F4}
\definecolor{kcDiscH}{HTML}{FFEBA3}
\definecolor{kcLatH}{HTML}{F8BBD0}
\definecolor{kcCollH}{HTML}{BFE9E7}
\definecolor{kcCtwoH}{HTML}{C6D7FF}
\definecolor{kcExfH}{HTML}{D8D8D8}

\newcolumntype{Y}{>{\raggedright\arraybackslash}X}
\newcolumntype{N}[1]{>{\columncolor{#1}\centering\bfseries\arraybackslash}X}
\newcolumntype{B}{>{\bfseries\arraybackslash}l}

\newcolumntype{P}[2]{%
>{\columncolor{#1}%
\raggedright\arraybackslash%
\setlength{\parindent}{0pt}%
\setlength{\leftskip}{3pt}%
\setlength{\rightskip}{3pt plus 1fil}%
}p{#2}%
}

\newcolumntype{D}[1]{%
>{\raggedright\arraybackslash%
\setlength{\parindent}{0pt}%
\setlength{\leftskip}{3pt}%
\bfseries%
}p{#1}%
}

\begin{document}

\title{TGCM: Topic-Guided Consistency Modeling for One-Step Disentanglement of Interleaved APT Technique Sequences}


\author{Guo-Wei~Wong,
Ming-Chuan~Yang,
Shou-De~Lin,
Wang-Chien~Lee,
and Meng~Chang~Chen%
\IEEEcompsocitemizethanks{
\IEEEcompsocthanksitem G.-W. Wong and S.-D. Lin are with National Taiwan University, Taiwan.
\IEEEcompsocthanksitem M.-C. Yang is with National Taiwan Ocean University, Taiwan.
\IEEEcompsocthanksitem W.-C. Lee is with Pennsylvania State University, USA.
\IEEEcompsocthanksitem M. C. Chen is with Academia Sinica, Taiwan.
}
}

\markboth{SUBMITTED TO IEEE Transactions on Dependable and Secure Computing}%
{}

\maketitle
\begin{abstract}
In enterprise environments, multiple Advanced Persistent Threat (APT) campaigns may execute concurrently, producing audit logs in which events from different campaigns are interleaved without explicit campaign boundaries. We formulate this setting as the \emph{Unknown-$K$ Interleaved Sequence Demixing (UKISD)} problem, which seeks to recover coherent campaign episodes from interleaved observations when the number of concurrent campaigns is unknown. We represent each campaign episode as a sequence of MITRE ATT\&CK technique occurrences and formulate demixing as an occurrence-level assignment problem that reconstructs episode-consistent technique sequences. Existing statistical, provenance-based, and sequence-analysis approaches typically assume single-campaign observations or rely on local heuristics, limiting their effectiveness under severe interleaving, repeated techniques, post-abstraction noise, and unknown mixture cardinality.

We propose \textbf{Topic-Guided Consistency Modeling (TGCM)}, a consistency-inspired one-step demixing framework for UKISD. TGCM directly learns the inverse mapping from interleaved technique sequences to latent campaign episodes by jointly performing occurrence-level assignment and sequence reconstruction in a single forward pass. To improve semantic coherence, TGCM integrates lightweight topic guidance derived from MITRE ATT\&CK narratives with an embedding-space self-consistency objective, enabling efficient and robust one-step inference.

We evaluate TGCM on synthetic mixtures, mixed benchmark datasets, public DARPA engagement traces, and \textsc{CAPTure}, a controlled end-to-end emulation benchmark comprising 200 attack scenarios, 855 million audit events, and 25 ATT\&CK-aligned attack profiles across single-host and multi-host environments. Experimental results demonstrate that TGCM consistently improves occurrence-level assignment under heavy interleaving, repeated technique reuse, symbolic technique-extraction errors, and budgeted unknown-$K$ inference. TGCM further generalizes effectively to previously unseen benchmark datasets without retraining and maintains strong end-to-end performance when integrated with practical ATT\&CK extraction pipelines. These results establish TGCM as an effective post-abstraction decision-support framework for reasoning about concurrent APT campaigns in realistic operational environments. The implementation and experimental artifacts are publicly available at \url{https://irish-kw.github.io/TGCM_Website/}.
\end{abstract}


\section{Introduction}
Advanced Persistent Threats (APTs)~\cite{saha2024expertapts} pose a significant challenge to modern enterprise security due to their stealthy, multi-stage, and long-running nature. These campaigns often unfold over extended periods, interleaving malicious actions with benign system activities to evade detection. Prior work detects APT campaigns either directly from audit logs~\cite{ding2023airtag, xiong2020conan} or from intermediate representations derived from these logs~\cite{alsaheel2021atlas, lv2024trec, Li_2024, cheng2024kairos, goyal2023sometimes}. Many systems further abstract low-level system events into higher-level adversarial behaviors, commonly represented using MITRE ATT\&CK techniques, to facilitate downstream reasoning.
These approaches include provenance analysis, representation learning, graph-based matching, and related ATT\&CK-based reasoning techniques~\cite{milajerdi2019holmes,ding2023airtag,ProVG}.

This work considers an operational setting in which defenders analyze ATT\&CK-aligned technique sequences produced by upstream detection systems. Even after upstream detectors identify suspicious techniques, concurrent intrusions or activities by multiple operators may produce a mixed sequence of technique occurrences. We therefore study \emph{episode-level technique demixing}: determining which observed technique occurrences belong to the same underlying attack episode in order to support Security Operations Center (SOC) triage and incident scoping, rather than performing final threat-actor attribution.

Most existing approaches assume that each observed technique sequence corresponds to a single campaign or attack episode. In practice, however, multiple intrusion episodes may overlap within the same enterprise environment, causing their techniques to become temporally interleaved~\cite{Sophos2022Triple,unit42_exchange_timeline_2021,alperovitch2016bears,DART}. As illustrated in Table~\ref{tab:dual-interleaving-sequence}, a real incident may involve an Initial Access Broker (IAB), downstream ransomware operators, and other coordinated or loosely connected operators. Their activities may therefore produce a mixed technique sequence without explicit episode boundaries, while the underlying operations may be coordinated, loosely connected, or completely unrelated.
\begin{table}[ht]
\caption{Author-derived ATT\&CK mapping of the interleaved Karma--Conti incident reported by Gallagher~\cite{gallagher2022karma_conti}.}
\label{tab:dual-interleaving-sequence}
\centering
\scriptsize
\setlength{\tabcolsep}{1.5pt}
\renewcommand{\arraystretch}{0.82}

\begin{tabularx}{\columnwidth}{@{}r >{\raggedright\arraybackslash}X l l@{}}
\toprule
\# & Technique & Tactic & Actor \\
\midrule
1  & T1190 ProxyShell exploit       & Init. Access & Likely IAB \\
2  & T1505.003 Web Shell            & Persistence  & Likely IAB \\
3  & T1136 Create Account           & Persistence  & Likely IAB \\
4  & T1021.001 RDP                  & Lat. Move.   & Karma \\
5  & T1059.001 PowerShell           & Execution    & Likely Conti \\
6  & T1543.003 Windows Service      & Persistence  & Likely Conti \\
7  & T1560.001 Archive via Utility  & Collection   & Karma \\
8  & T1567.002 Cloud Exfiltration   & Exfil.       & Karma \\
9  & T1569.002 Service Execution    & Execution    & Karma \\
10 & T1562.001 Impair Defenses      & Def. Evasion & Conti \\
11 & T1047 WMI                      & Discovery    & Conti \\
12 & T1518.001 Security SW Discovery& Discovery    & Conti \\
13 & T1567.002 Cloud Exfiltration   & Exfil.       & Likely Conti \\
14 & T1486 Data Encryption          & Impact       & Conti \\
\bottomrule
\end{tabularx}
\end{table}

\begin{figure*}[ht]
    \centering
    \includegraphics[width=1.0\textwidth]{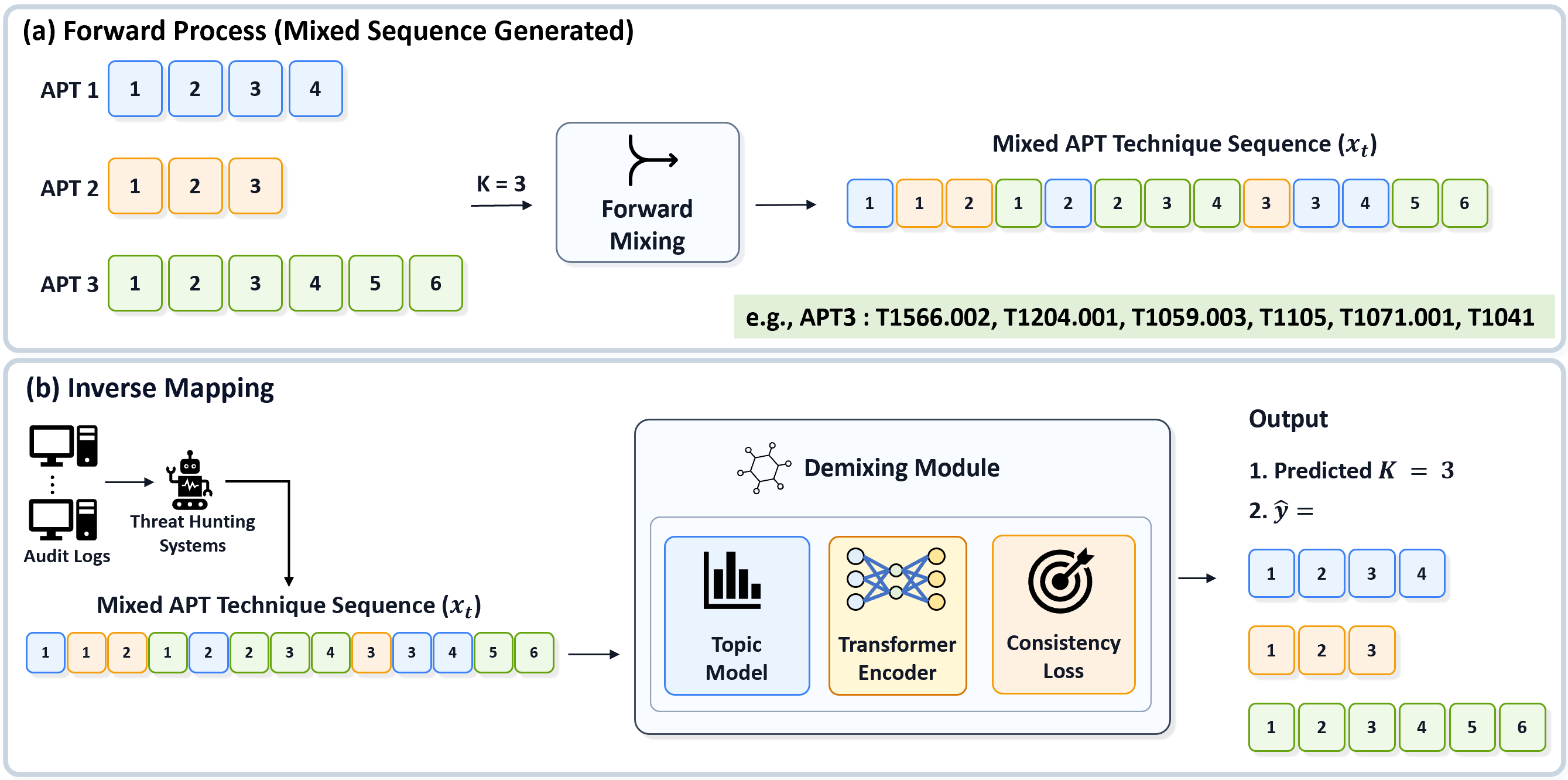}
     \caption{Overview of TGCM. (a) Forward mixing interleaves multiple campaign episodes into a mixed technique sequence $x_t$. (b) TGCM learns a one-step inverse mapping that estimates the active episode count $\hat{K}$ under a fixed decoding budget $K_{\max}$, predicts occurrence-level assignments $\hat{y}$, and reconstructs episode-level technique sequences.}
    \label{fig:model_overview}
\end{figure*}
These observations motivate treating an observed technique sequence as the superposition of multiple latent attack episodes. We formalize this problem as the \textbf{Unknown-$K$ Interleaved Sequence Demixing (UKISD)} problem. Given a mixed technique sequence, UKISD aims to recover the underlying campaign episodes, assign each technique occurrence to its corresponding episode, and determine the occupied episode slots under a fixed maximum decoding budget $K_{\max}$. UKISD is inherently difficult because its search space is combinatorial and closely related to the classical string-unshuffling problem~\cite{buss2014unshuffling,bulteau2020recognizing}.

Solving UKISD is challenging for several reasons. First, different APT campaigns frequently share common ATT\&CK techniques (e.g., T1059, Command and Scripting Interpreter), making frequency- or correlation-based heuristics unreliable when multiple campaigns overlap. Second, the abstraction from audit logs to ATT\&CK techniques is often noisy and incomplete, introducing additional ambiguity into occurrence-level episode assignment. Existing statistical, provenance-based, and sequence-analysis methods are not designed for this setting. Statistical methods such as Hidden Markov Models (HMMs)~\cite{ghosh2023blind} and Frequent Episode Mining (FEM)~\cite{fournier2022maxfem} primarily exploit local statistical patterns and struggle with long-range dependencies under heavy interleaving. Structural methods such as Object-Centric Petri Nets (OCPNs)~\cite{van2020discovering} enforce causal constraints but lack sufficient semantic information to distinguish shared techniques across different episodes. More generally, existing methods either assume a fixed number of underlying sources or analyze the mixed sequence holistically without explicitly recovering the latent campaign episodes.

Rather than viewing UKISD as a sequence classification or clustering, we formulate UKISD as an inverse demixing problem. As illustrated in Figure~\ref{fig:model_overview}(a), an observed interleaved technique sequence is generated by forward mixing multiple campaign episodes. The objective of UKISD is to reverse this process by recovering the underlying episode-level sequences. This perspective naturally accommodates an unknown number of episodes under a fixed decoding budget while preserving temporal dependencies among technique occurrences despite shared techniques and incomplete observations.
To solve UKISD, we propose \textbf{Topic-Guided Consistency Modeling (TGCM)}, a consistency-inspired framework for the one-step disentanglement of interleaved ATT\&CK technique sequences. TGCM formulates UKISD as a joint occurrence-level episode assignment and sequence restoration problem, as shown in Figure~\ref{fig:model_overview}(b). Rather than separating repeated techniques solely according to their identities, TGCM exploits their surrounding episode context to recover coherent episode-level sequences.

TGCM consists of three complementary components (Figure~\ref{fig:model_overview}(b)). First, a \emph{topic-guided semantic regularizer} derived from FASTopic~\cite{wu2024fastopic} pretrained on MITRE ATT\&CK narratives encourages semantically coherent episode reconstruction. Second, a \emph{Transformer encoder} captures contextual and long-range dependencies within the mixed technique sequence. Third, a \emph{self-consistency objective} inspired by consistency models~\cite{song2023consistency} regularizes the inverse mapping from mixed observations to episode-level sequences, enabling efficient single-step inference without the iterative sampling required by diffusion-based generative models~\cite{ho2020denoising,nichol2021improved,dhariwal2021diffusion}. Topic guidance and self-consistency serve as auxiliary regularizers, while the task-aligned formulation constitutes the primary modeling contribution. Our ablation study quantifies the contributions of the task-aligned formulation and the auxiliary regularizers separately.

Because real-world audit logs rarely provide ground-truth episode decompositions, TGCM is trained on controlled mixtures generated by forward mixing multiple single-episode technique sequences. We evaluate TGCM on synthetic datasets, benchmark scenarios, public DARPA traces, and the proposed \textsc{CAPTure} end-to-end emulation benchmark under an unknown number of episodes and a fixed maximum decoding budget. Experimental results demonstrate improved occurrence-level assignment across heavy interleaving, shared techniques, post-abstraction noise, and unseen attack scenarios. Ablation studies further show that the proposed task-aligned formulation contributes the largest performance gains, while topic guidance and self-consistency provide complementary improvements.

Our main contributions are summarized as follows:
\begin{itemize}
    \item \textbf{Problem formulation.} We formulate the Unknown-$K$ Interleaved Sequence Demixing (UKISD) problem for recovering latent campaign episodes from interleaved ATT\&CK technique sequences.

    \item \textbf{Task-aligned demixing framework.} We propose TGCM, which formulates UKISD as joint occurrence-level episode assignment and sequence restoration. TGCM supports an unknown number of episodes under a fixed maximum decoding budget and recovers episode-level sequences in a single inference step.

    \item \textbf{Semantic and self-consistency regularization.} TGCM incorporates topic guidance derived from MITRE ATT\&CK narratives and embedding-space self-consistency as auxiliary regularizers to improve semantic coherence and reconstruction stability.

    \item \textbf{Controlled end-to-end benchmark.} We introduce \textsc{CAPTure}, a controlled CALDERA--Procmon emulation benchmark containing single-host and multi-host attack scenarios. It enables end-to-end evaluation under unknown episode counts and noise introduced by upstream log-to-technique abstraction.

    \item \textbf{Comprehensive evaluation.} We evaluate TGCM on synthetic datasets, benchmark scenarios, public DARPA traces, and \textsc{CAPTure}. The results demonstrate consistent improvements in occurrence-level assignment across different levels of interleaving, shared techniques, post-abstraction noise, and unseen attack scenarios, while ablation studies distinguish the gains from the task-aligned formulation and auxiliary regularization.
\end{itemize}

\section{Preliminaries}
This section reviews the audit-log abstraction pipeline, introduces the technique-sequence representation used throughout the paper, and summarizes the assumptions underlying the UKISD formulation.

\subsection{Abstraction of Technique Sequences}
System-level audit logs record security-relevant operating system activities, including process execution, file access, registry modification, and network communication. Such logs are collected by widely deployed platforms such as Procmon, ETW, Sysmon, Linux audit, and CamFlow. A log event is commonly represented as a triple $\langle$subject, operation, object$\rangle$. In practice, audit logs are high-volume and heterogeneous, making direct campaign-level reasoning over raw events both computationally expensive and semantically fragile.

To support campaign-level analysis, audit logs are commonly abstracted into ordered MITRE ATT\&CK technique sequences~\cite{mitre_attack}. Throughout this paper, we refer to these ordered representations simply as \emph{technique sequences}. For example, an APT campaign may be represented as
\textit{Spearphishing Attachment} $\rightarrow$ \textit{Command and Scripting Interpreter} $\rightarrow$ \textit{Credential Dumping} $\rightarrow$ \textit{Exfiltration Over C2 Channel}. This abstraction enables campaign-level analysis while remaining consistent with analyst workflows and existing SOC practices. 
Representative upstream systems perform this abstraction by mapping audit logs to ATT\&CK-aligned technique sequences:

\begin{itemize}
\item \textbf{Rule-based provenance analysis.}
Early systems~\cite{milajerdi2019holmes,milajerdi2019poirot,zeng2021watson} construct dependency graphs and apply expert-defined rules to infer high-level attack scenarios.
\item \textbf{Graph-based learning.}
Subsequent approaches incorporate causal reasoning~\cite{kwon2018mci,hassan2020omegalog} or graph representation learning~\cite{han2020unicorn,cheng2024kairos} for anomaly detection and attack inference.
\item \textbf{Sequence-based modeling.}
ATLAS~\cite{alsaheel2021atlas} models attack evidence as ordered sequences, enabling chronological reasoning for multi-stage APT analysis.
\item \textbf{Fine-grained technique recognition.}
Recent methods, such as TREC~\cite{lv2024trec} and NODLINK~\cite{Li_2024}, focus on identifying individual techniques via few-shot learning or online sequence summarization.
\end{itemize}

These upstream systems are complementary to TGCM. They transform audit logs into ATT\&CK technique sequences, whereas TGCM operates on the resulting sequences to recover interleaved campaign episodes.

\subsection{Problem Characteristics}

Compared with isolated single-campaign sequences, mixed technique sequences encountered in practice exhibit several characteristics relevant to UKISD:

\begin{itemize}
\item \textbf{Scale and low signal-to-noise ratio.} Enterprise audit logs are large and dominated by benign activities, making the extraction of technique sequences inherently error-prone.
\item \textbf{Interleaving and technique reuse.} Multiple campaign episodes may overlap temporally, and commonly used techniques may appear across different episodes, introducing inherent assignment ambiguity.
\item \textbf{Incomplete observability.} Campaign traces may be partially observed due to limited logging coverage, sensor failures, or collection constraints.
\end{itemize}
These characteristics jointly motivate the UKISD formulation introduced in Section~\ref{sec:task_formulation}.

\subsection{Scope and Assumptions}
We explicitly distinguish operational reality from the scope of this work.
\begin{itemize}

\item \textbf{Post-abstraction setting.}
TGCM operates on ATT\&CK technique sequences extracted by upstream systems rather than raw audit logs. Unless otherwise stated, all inputs refer to this post-abstraction representation.
\item \textbf{Limits of identifiability.}
When extracted sequences are severely corrupted, latent campaigns may become \emph{difficult to identify}, as the available observations may not provide sufficient signal for reliable inference.
\item \textbf{Partial observability.}
Observed sequences may be incomplete due to logging gaps or collection limitations.
\item \textbf{Technique-sequence input.}
TGCM uses extracted ATT\&CK technique identifiers, with benign events filtered upstream. Unless otherwise stated, TGCM denotes this post-abstraction, technique-sequence setting.
\item \textbf{Unknown number of campaign episodes.}
Unlike many sequence separation methods that assume a fixed number of sources, UKISD does not assume prior knowledge of the number of underlying campaign episodes. TGCM instead performs demixing under a fixed decoding budget $K_{\max}$ and estimates the active episode count during inference.
\item \textbf{Episode-level decomposition.}
TGCM separates mixed technique sequences into behaviorally coherent campaign episodes rather than attributing activities to specific threat actors. Recovered episodes may encompass activities conducted by a single operator, multiple independent operators, or multiple coordinated operators active during the same time period. Consequently, UKISD should be interpreted as an episode decomposition problem rather than a threat-actor attribution problem.
\item \textbf{Observation quality.}
The formulation assumes noise-free mixed sequences generated by interleaving complete campaign traces. This assumption is relaxed in Section~\ref{Sec:CAPTure}, where upstream extraction errors and partial observability are considered through end-to-end evaluation.
\end{itemize}

These preliminaries establish the foundation for the UKISD formulation and the TGCM framework presented in the next sections.

\section{Threat Model and Problem Formulation}
\label{sec:threat_model}
This section specifies the adversary model and defender capabilities, formalizes the UKISD problem, and analyzes its inherent challenges.

\subsection{Adversary Model}
\label{sec:adversary_model}

We consider one or more intrusion operators executing multi-stage activities within a shared enterprise environment. These activities may belong to independent campaigns, coordinated operations, or downstream stages.
We assume:
\begin{itemize}
    \item \textbf{Overlapping campaign episodes.} Multiple campaign episodes may overlap temporally and interleave in the extracted technique sequence. Episodes are separable for operational analysis, but need not correspond to independent threat actors.
    \item \textbf{Technique reuse.} Techniques may be reused across campaign episodes (e.g., \textit{T1105} in Figure~\ref{fig:sequence_interleave}), creating intrinsic assignment ambiguity.
    \item \textbf{Post-abstraction noise.} Observed technique sequences may contain missing, substituted, or spurious techniques due to imperfect upstream log-to-technique abstraction.
\end{itemize}
Accordingly, the threat model focuses on episode-level demixing under temporal interleaving, technique reuse, and post-abstraction sequence noise. Upstream telemetry collection and technique extraction are assumed to have been completed.

\subsection{Defender Capabilities}
\label{sec:defender_capabilities}
At inference time, the defender observes only an ATT\&CK technique sequence abstracted by upstream systems, without access to raw audit logs, provenance graphs, or ground-truth episode boundaries.
To isolate the core UKISD problem, we first evaluate TGCM using controlled mixtures generated from complete single-campaign technique sequences and later assess robustness under missing, substituted, and spurious technique perturbations.

\newlength{\tokW}
\setlength{\tokW}{14mm}

\newcommand{\TokBox}[2]{\colorbox{#1}{\makebox[\tokW][c]{\rule{0pt}{2.0ex}#2}}}
\newcommand{\IdleTok}{\TokBox{green!20}{Idle}}
\newcommand{\BlueTok}[1]{\TokBox{cyan!25}{#1}}
\newcommand{\OrangeTok}[1]{\TokBox{orange!25}{#1}}
\newcommand{\Legend}[2]{\colorbox{#1}{\strut #2}}

\begin{figure*}[t]
\centering
\setlength{\tabcolsep}{2pt}
\renewcommand{\arraystretch}{1.25}
\begin{adjustbox}{max width=\textwidth}
\begin{tabular}{@{}p{0.20\textwidth}*{11}{c}@{}}
\toprule
\textbf{Case} & \multicolumn{11}{c}{\textbf{Technique sequence extracted by the upstream system}} \\
\midrule

(a) Single episode ($i$) &
\IdleTok & \IdleTok & \OrangeTok{T1566.001} & \IdleTok &
\OrangeTok{T1069.003} & \OrangeTok{T1497.001} & \OrangeTok{T1003.001} &
\OrangeTok{T1224} & \IdleTok & \OrangeTok{T1201} & \OrangeTok{T1105} \\

(b) Single episode ($i{+}1$) &
\BlueTok{T1566.001} & \BlueTok{T1059.001} & \IdleTok & \BlueTok{T1105} &
\IdleTok & \IdleTok & \IdleTok & \IdleTok & \BlueTok{T1112} & \IdleTok & \IdleTok \\

(c) Mixture of (a) and (b) &
\BlueTok{T1566.001} & \BlueTok{T1059.001} & \OrangeTok{T1566.001} & \BlueTok{T1105} &
\OrangeTok{T1069.003} & \OrangeTok{T1497.001} & \OrangeTok{T1003.001} &
\OrangeTok{T1224} & \BlueTok{T1112} & \OrangeTok{T1201} & \OrangeTok{T1105} \\

\bottomrule
\end{tabular}
\end{adjustbox}
\caption{Technique occurrences are ordered from left to right. Case (c) is formed by interleaving two distinct campaign episodes (\Legend{cyan!25}{blue} and \Legend{orange!25}{orange}). Blank cells represent visualization-only temporal gaps and are omitted when constructing the observed technique sequence.}
\label{fig:sequence_interleave}
\end{figure*}

\subsection{UKISD Problem Formulation}
\label{sec:task_formulation}
Let $\mathcal{V}$ denote the vocabulary of MITRE ATT\&CK techniques. Given an observed mixed technique sequence
\[
x_{\text{obs}} = [w_1, w_2, \dots, w_n], \quad w_i \in \mathcal{V}
\]
the goal is to jointly infer:
\begin{itemize}
    \item The active episode slots within a maximum budget $K_{\max}$, and
    \item A technique-level assignment vector $\hat{y} \in \{0,1,\dots,K_{\max}\}^n$, where $\hat{y}_i$ indicates the inferred episode associated with technique occurrence $w_i$ (with $0$ optionally denoting residual or unassigned elements).
\end{itemize}
Equivalently, the output may be represented as a set of order-preserving episode-level subsequences induced from $x_{\text{obs}}$ by the assignment vector $\hat y$.
Episode-slot identifiers are assigned deterministically: constructed mixtures follow the source-campaign order used to form $x_0$, whereas naturally interleaved traces follow the first observed occurrence of each annotated episode. Once assigned, the same identifiers are retained throughout data construction, training, and evaluation.

Figure~\ref{fig:sequence_interleave} illustrates UKISD. Cases (a) and (b) represent two latent campaign episodes, while Case (c) shows their observed interleaving. The objective is to recover the latent episodes and assign each technique occurrence accordingly.
For example, a correct solution yields $K{=}2$ and an assignment vector $\hat{y}=[2,2,1,2,1,1,1,1,2,1,1]$, where $1$ denotes the orange campaign and $2$ denotes the blue campaign, that separates the mixed sequence into the two underlying episodes.

\subsection{Problem Analysis}
\label{sec:problem_analysis}
Given a mixed technique sequence, the defender must jointly infer the occupied episode slots and assign each technique occurrence to its latent episode. This problem exhibits structural non-identifiability: multiple interleavings,  shared techniques, and unknown episode cardinality jointly create ambiguity that cannot be resolved by local heuristics or purely discriminative models.

A practical solution should satisfy the following requirements:
\begin{itemize}
    \item (R1) \textbf{Order awareness}: preserve campaign-internal temporal structure.
    \item (R2) \textbf{Budgeted $K$ inference}: operate without prior knowledge of the episode count under a fixed maximum decoding budget.
    \item (R3) \textbf{Robustness}: handle technique reuse, partial observations, and post-abstraction perturbations.
    \item (R4) \textbf{Efficiency}: scale to long technique sequences.
    \item (R5) \textbf{Post-abstraction demixing}: operate on upstream ATT\&CK technique sequences across heterogeneous logging environments.
\end{itemize}

Existing approaches address subsets of these requirements but do not simultaneously support unknown episode cardinality, order-preserving demixing, technique-sequence inputs, and robustness to post-abstraction perturbations.
Bag-of-techniques methods discard temporal structure~\cite{lee2023camp2vec}. HMM-based models assume a fixed model order and require specifying the number of latent states in advance~\cite{baum1970maximization}; even extensions such as factorial HMMs still rely on a predefined number of latent chains~\cite{ghahramani1995factorial}. Provenance-based approaches depend on system-level attributes that are unavailable in this setting~\cite{milajerdi2019holmes, han2020unicorn}. Discriminative sequence models typically encode mixed observations holistically rather than explicitly disentangling latent campaign episodes~\cite{du2017deeplog, guo2021logbert}.

These observations motivate viewing UKISD as a \emph{generative disentanglement} problem, where the goal is to recover latent campaign episodes by inverting their observed superposition under structural and semantic inductive biases.
This formulation motivates the TGCM framework presented in the next section.
\section{TGCM: A Consistency-Inspired Demixing Framework}
\label{sec:method}

\subsection{Overview}
\label{sec:overview}

To satisfy the UKISD requirements defined in Section~\ref{sec:problem_analysis}, we propose Topic-Guided Consistency Modeling (TGCM), a consistency-inspired one-step demixing framework that operates on post-abstraction ATT\&CK technique sequences
(satisfying \textit{R5: Post-abstraction demixing}). We summarize the symbols and their definitions in Appendix Table~\ref{tab:notation}.

TGCM treats the observed technique sequence as the superposition of multiple latent campaign episodes and learns a direct inverse mapping that simultaneously reconstructs canonical episode-level technique sequences and predicts occurrence-level episode assignments.
The recovered subsequences support incident scoping and analyst triage, not definitive threat-actor attribution. The main architectural commitment is task-aligned occurrence-level episode assignment with sequence restoration; the topic and consistency terms below are auxiliary regularizers rather than the sole explanation for TGCM's performance advantage. The framework consists of three components, explicitly designed to satisfy the remaining requirements (R1--R4):

\begin{itemize}
    \item \textbf{Sequence restoration and assignment} (Section~\ref{sec:objectives}).
    TGCM jointly optimizes reconstruction fidelity and semantic consistency to reconstruct the clean sequence and predict technique-level assignments $\hat{y}$, allowing budgeted estimation of the active episode count under $K_{\max}$ (\textit{R2: Budgeted \(K\) inference}).

    \item \textbf{One-step consistency-inspired demixing}(Section~\ref{sec:consistency_model}).
    A self-consistency objective inspired by consistency models~\cite{song2023consistency} learns a direct inverse mapping from a mixed sequence $x_t$ to a clean sequence $\hat{x}_0$, enabling efficient single-step inference (\textit{R4: Efficiency}).

    \item \textbf{Semantic regularization} (Section~\ref{sec:FASTopic}).
    A neural topic model (FASTopic~\cite{wu2024fastopic}) captures latent semantic structure from MITRE ATT\&CK narratives, serving as a \emph{semantic regularizer} that encourages coherent kill-chain progression and helps reduce ambiguity under technique reuse (\textit{R1: Order-awareness} and \textit{R3: Robustness}).
\end{itemize}

Appendix~\ref{sec:infotheory} provides a conceptual information-theoretic reading of these losses.

\subsection{Topic Modeling with FASTopic}
\label{sec:FASTopic}

Although occurrence-level assignment provides the primary supervision for UKISD, technique reuse and partial observations often introduce ambiguity because multiple episode decompositions may explain the same observed technique sequence. To reduce this ambiguity, TGCM incorporates a semantic regularizer derived from ATT\&CK technique descriptions. Rather than predicting campaign identities, the regularizer biases reconstruction toward behaviorally coherent technique combinations that are semantically consistent with complete attack episodes, thereby improving kill-chain preservation (R1) and robustness to technique reuse (R3).

We instantiate this semantic regularizer using FASTopic~\cite{wu2024fastopic}, a neural topic model, trained over ATT\&CK technique descriptions encoded by a cybersecurity-oriented language model. For each single-APT campaign in the training corpus, we concatenate the textual descriptions of all ATT\&CK techniques into a campaign-level document (Appendix~\ref{app:tech_desc_mapping}). FASTopic then learns a latent topic representation that captures high-level semantic structure across complete campaigns rather than individual techniques.

Let $K_{\mathrm{topic}}$ denote the number of latent topics, $\Phi\in\mathbb{R}^{K_{\mathrm{topic}}\times |\mathcal{V}_{\mathrm{word}}|}$ denote the learned topic--word logit matrix, and $\theta(d)$ denote the topic distribution inferred for campaign document $d$. Because TGCM operates on ATT\&CK techniques instead of words, we project the learned word-level topic distribution onto the ATT\&CK technique vocabulary through a sparse mapping matrix $M$, which aggregates normalized description tokens associated with each technique.
The resulting technique-level topic prior is
\begin{equation}
\label{eq:TETM}
\omega(d)
=
\mathrm{Norm}\!\left(
\theta(d)^{\top}
\mathrm{softmax}_{\mathrm{word}}(\Phi)
M
\right)
\end{equation}
\noindent where $\omega(d)$ is a probability distribution over ATT\&CK techniques in the vocabulary $\mathcal{V}$ and $\mathrm{Norm}(\cdot)$ normalizes the resulting vector into a valid probability distribution. Padding and residual symbols receive zero prior probability.

Importantly, $\theta(d)$ represents only a semantic topic allocation and is never used as an episode label. During supervised TGCM training, source-campaign identifiers are used solely to retrieve frozen topic distributions that serve as semantic priors. They neither reveal the latent episode assignments nor supervise the occurrence-level assignment objective.

Let $\ell_{\mathrm{rec},i}$ denote the reconstruction logits produced by TGCM before semantic fusion. The topic prior is incorporated through log-linear fusion,
\begin{equation}
\label{eq:fused}
\ell_{\mathrm{fused},i}
=
\ell_{\mathrm{rec},i}
+
\lambda
\log(\omega(d_i)+\epsilon)
\end{equation}
where $\lambda$ controls the strength of semantic regularization and $\epsilon$ ensures numerical stability. This log-linear fusion injects ATT\&CK-derived semantic priors into the reconstruction logits. It regularizes technique prediction but does not influence episode assignment, which is learned solely by the occurrence-level assignment head.

The reconstruction cross-entropy is computed once from the topic-fused logits and the canonical clean sequence:
\begin{equation}
\label{eq:ce_topic}
\ell_{\mathrm{CE}}
=
\mathrm{CE}
(\ell_{\mathrm{fused}},x_0)
\end{equation}

In addition to logit fusion, the implementation uses two auxiliary topic regularizers. Let $\mathcal{R}$ denote the set of valid episode instances in a mini-batch and let $\mathcal{V}_{\mathrm{topic}}\subseteq\mathcal{V}$ denote the technique vocabulary supported by the topic prior. For episode instance $(b,r)\in\mathcal{R}$, let $q^{\mathrm{bow}}_{b,r,v}$ be the sequence-aggregated reconstructed probability of technique $v$ and let $\omega_{b,r,v}$ be the corresponding FASTopic-induced technique prior. The bag-of-techniques matching loss is
\begin{equation}
\label{eq:topic_bow}
\mathcal{L}_{\mathrm{bow}}
=
-\frac{1}{|\mathcal{R}|}
\sum_{(b,r)\in\mathcal{R}}
\sum_{v\in\mathcal{V}_{\mathrm{topic}}}
\omega_{b,r,v}
\log\!\left(q^{\mathrm{bow}}_{b,r,v}+\epsilon\right)
\end{equation}

Let $\theta_{b,r,k}$ be the frozen FASTopic allocation of episode instance $(b,r)$ for topic $k$, and let $\hat{\theta}_{b,r,k}$ be the topic distribution inferred from the reconstructed technique probabilities. The topic-space alignment loss is
\begin{equation}
\label{eq:topic_align}
\mathcal{L}_{\mathrm{align}}
=
-\frac{1}{|\mathcal{R}|}
\sum_{(b,r)\in\mathcal{R}}
\sum_{k=1}^{K_{\mathrm{topic}}}
\theta_{b,r,k}
\log\!\left(\hat{\theta}_{b,r,k}+\epsilon\right)
\end{equation}

The complete auxiliary topic regularizer is
\begin{equation}
\label{eq:topic_objective}
\mathcal{L}_{\mathrm{topic\_reg}}
=
w_{\mathrm{topic\_bow}}\mathcal{L}_{\mathrm{bow}}
+
w_{\mathrm{topic\_topic}}\mathcal{L}_{\mathrm{align}}
\end{equation}
The topic prior therefore affects reconstruction in two ways: logit fusion before the single reconstruction cross-entropy and the two auxiliary topic regularizers above. The reconstruction cross-entropy is included exactly once in the total objective; the fusion coefficient $\lambda$ modifies reconstruction logits and is not an additional loss weight.

During deployment, TGCM replaces training-time topic priors with neutral topic conditioning. Consequently, the semantic regularizer improves reconstruction using only learned ATT\&CK semantics, without relying on campaign identities or other ground-truth information unavailable in operational settings.

\subsection{Consistency-Inspired One-Step Inverse Demixing}
\label{sec:consistency_model}

To enable efficient single-step disentanglement (R4), TGCM learns a direct one-step inverse mapping from a mixed technique sequence to its canonical clean representation. Rather than iteratively refining the reconstruction, the proposed consistency-inspired model estimates the clean sequence in a single inference pass while preserving the semantic regularization introduced in the previous subsection.
We employ a consistency-inspired model $f_\theta(x_t,t)$ that directly estimates the clean sequence $\hat{x}_0$ from a trajectory state $x_t$. Each technique in $x_t$ is mapped to a $d$-dimensional embedding via a learnable input matrix $W_{\mathrm{in}} \in \mathbb{R}^{|\mathcal{V}|\times d}$, yielding $E_t=\mathrm{Embed}(x_t; W_{\mathrm{in}})$. The embeddings are augmented with positional encoding and processed by a Transformer encoder backbone to capture contextual dependencies.

The backbone is conditioned on the discrete trajectory index through an AdaLN-style modulation applied once at the Transformer encoder input. Let $e_t=\mathrm{Embed}_{\mathrm{time}}(t)$ denote the timestep embedding and let $P$ denote the positional encoding. The modulated encoder input is
\begin{equation}
\label{eq:adanorm}
\widetilde{E}_t=\mathrm{LN}(E_t+P),\qquad
H_t=\bigl(1+\gamma(e_t)\bigr)\odot\widetilde{E}_t+b(e_t)
\end{equation}
where $\gamma(\cdot)$ and $b(\cdot)$ are learned scale and shift projections derived from the timestep embedding, and $\mathrm{LN}(\cdot)$ denotes Layer Normalization~\cite{ba2016layer}. The Transformer encoder processes $H_t$. This AdaLN-style modulation is timestep-conditioned rather than topic-conditioned. FASTopic guidance enters reconstruction through the logit fusion in Equation~\ref{eq:fused} and the auxiliary losses $\mathcal{L}_{\mathrm{bow}}$ and $\mathcal{L}_{\mathrm{align}}$.
Rather than predicting the clean sequence directly, TGCM estimates an embedding-space residual that removes the effect of forward mixing. The reconstructed embedding is obtained as
\begin{equation}
\label{eq:logits}
\hat{E}_0
=
E_t
-
a(t)\,
\hat{r}_{\theta}(E_t,t),
\qquad
\mathrm{logits}
=
\hat{E}_0W_{\mathrm{out}}^{\top}
\end{equation}
where $\hat{r}_{\theta}$ denotes the predicted residual, $W_{\mathrm{out}}$ is the output vocabulary projection matrix, and $a(t)$ controls the correction strength along the forward-mixing trajectory. In our implementation,
\[
a(t)=\frac{t-1}{T-1}
\]
so that the implementation uses $x_1=x_0$ as the clean anchor: $a(1)=0$, while $t\in\{2,\ldots,T\}$ indexes progressively stronger interleaving and larger residual corrections. Input and output embeddings may optionally be tied by setting $W_{\mathrm{out}}=W_{\mathrm{in}}$.

The implementation samples $t\in\{1,\ldots,T\}$. The first indexed state $x_1$ is identical to the canonical clean target $x_0$, and nonzero mixing begins at $t=2$. The index $t$ therefore denotes a controlled forward-trajectory level rather than physical time or an observable corruption level. During training, consistency learning encourages representations sampled from different trajectory levels to reconstruct a common clean sequence. Consequently, inference does not require estimating the true mixing index of an observed sequence.

TGCM follows the consistency-learning principle of mapping multiple forward-mixing states to a shared canonical representation, but adopts a lightweight parameter-sharing implementation rather than a full continuous-time consistency model with an EMA teacher or PF-ODE trajectory.

\subsection{Self-Consistency Regularization}
While the reconstruction model estimates the canonical clean sequence from a single mixed input, multiple forward-mixing states may originate from the same latent campaign episodes. To ensure that the learned inverse mapping is independent of the particular mixing state, TGCM introduces a self-consistency objective that encourages these forward-mixing states to converge to a common canonical representation.
This improves the stability and efficiency of one-step inverse demixing (R4).

Specifically, two trajectory states $(x_{t_1},x_{t_2})$ are sampled from the same forward-mixing trajectory, where $t_1,t_2\in\{1,\ldots,T\}$ are independently selected indices and need not be adjacent. Their reconstructed embeddings are encouraged to agree through
\begin{equation}
\label{eq:consistency_loss}
\mathcal{L}_{\mathrm{consist}}
=
\mathbb{E}
\left[
\left\|
\hat{E}_{0,\theta}(x_{t_1},t_1)
-
\mathrm{sg}
\!\left(
\hat{E}_{0,\theta}(x_{t_2},t_2)
\right)
\right\|_2^2
\right]
\end{equation}
where $\mathrm{sg}(\cdot)$ denotes the stop-gradient operator applied to the target branch.

The consistency objective is imposed entirely on the continuous embedding representation before discrete decoding. Consequently, it regularizes the latent inverse mapping without requiring back-propagation through discrete technique predictions or hard episode labels.
Rather than supervising individual technique predictions, this objective encourages different forward-mixing states of the same campaign to reconstruct an identical canonical embedding. As a result, the learned inverse mapping becomes invariant to the particular mixing stage, allowing TGCM to perform stable one-step reconstruction without estimating the true forward-mixing index during inference.

\subsection{Reconstruction and Assignment Objectives}
\label{sec:aux_object}

Recovering latent campaign episodes requires both assigning each technique occurrence to its source episode (R3) and reconstructing the corresponding canonical technique sequence (R1). Accordingly, TGCM jointly optimizes an occurrence-level assignment objective and a reconstruction objective.

The primary supervision is provided by the occurrence-level assignment objective,
\begin{equation}
\mathcal{L}_{\mathrm{assign}}
=
\mathbb{E}
\left[
\mathrm{CE}
(
\mathrm{logits}_{\mathrm{assign}},
y_z
)
\right]
\end{equation}
where $y_z$ denotes the reference episode label for each technique occurrence. This objective directly supervises the occurrence-to-episode mapping required by UKISD and serves as the principal task-aligned learning signal.

During training, reconstruction cross-entropy is computed separately for each valid reference episode slot and then averaged across the active slots. Let $\mathcal{A}_{B}$ denote the non-residual episode-slot labels present in a mini-batch. Using the topic-fused logits from Equation~\ref{eq:fused}, the implemented reconstruction objective is
\begin{equation}
\label{eq:reconstruction_loss}
\mathcal{L}_{\mathrm{rec}}
=
\frac{1}{|\mathcal{A}_{B}|}
\sum_{a\in\mathcal{A}_{B}}
\mathrm{CE}\!\left(
\ell_{\mathrm{fused}}[y_z=a],
x_0[y_z=a]
\right)
\end{equation}

The reconstruction objective restores the canonical technique sequence within each reference episode slot, while the assignment objective determines the episode membership of every technique occurrence. Together, these two objectives directly optimize the two primary goals of UKISD. The topic-aware and self-consistency objectives introduced in the previous subsections provide complementary semantic and structural regularization.

\subsection{Topic-Aware Regularization and Overall Objective}
\label{sec:objectives}

The preceding subsections describe five loss terms that jointly address the requirements of UKISD. The occurrence-level assignment objective provides the primary task supervision, the reconstruction objective restores canonical technique sequences, the consistency objective stabilizes one-step inverse demixing, and the two topic losses provide bag-of-techniques and topic-space regularization.

The overall training objective used by the implementation is
\begin{equation}
\label{eq:total_loss}
\begin{aligned}
\mathcal{L} ={}&
w_{\mathrm{consist}}\mathcal{L}_{\mathrm{consist}}
+
w_{\mathrm{ce}}\mathcal{L}_{\mathrm{rec}}
+
w_{\mathrm{aptid}}\mathcal{L}_{\mathrm{assign}}
\\
&+
w_{\mathrm{topic\_bow}}\mathcal{L}_{\mathrm{bow}}
+
w_{\mathrm{topic\_topic}}\mathcal{L}_{\mathrm{align}}
\end{aligned}
\end{equation}

Thus, $\mathcal{L}_{\mathrm{rec}}$ is the only reconstruction cross-entropy term in Equation~\ref{eq:total_loss}. Topic logit fusion changes the logits used by this term, while $\mathcal{L}_{\mathrm{bow}}$ and $\mathcal{L}_{\mathrm{align}}$ are separate auxiliary losses. In ablations, the corresponding topic weights are set to zero and the consistency term is disabled when required. Equations~\ref{eq:TETM}--\ref{eq:total_loss} therefore match the loss computation used by the formal training code path.

\subsection{Inference Procedure}
\label{sec:inference_procedure}
At inference time, the defender observes only a mixed technique sequence
$x_{\mathrm{obs}}$, while the true number of latent campaign episodes $K$,
their composition, and the corresponding forward diffusion trajectory are all
unknown. Unlike training, inference does not rely on source-campaign topic
information. Instead, TGCM performs a single one-step demixing pass to
simultaneously predict the reconstructed clean sequence $\hat{x}_0$ and the
assignment probabilities
$p_\theta(y_i=k \mid x_{\mathrm{obs}})$ for each technique occurrence and
candidate assignment slot $k\in\{1,\ldots,K_{\max}\}$.

The latent assignment of each technique occurrence is first determined by
maximum-probability decoding:
\begin{equation}
\hat{y}_i=\arg\max_{k\in\{1,\ldots,K_{\max}\}}
p_\theta(y_i=k \mid x_{\mathrm{obs}})
\end{equation}

The estimated number of active campaign episodes is then obtained by counting
the occupied assignment slots:
\begin{equation}
\hat{K}
=
\left|
\left\{
k\in\{1,\ldots,K_{\max}\}
:
\exists\, i \text{ such that } \hat{y}_i=k
\right\}
\right|
\label{eq:k_estimation}
\end{equation}
where $\hat{K}$ represents the number of occupied latent slots within the
predefined capacity $K_{\max}$. This formulation enables TGCM to infer the
effective number of concurrent campaign episodes without requiring the true
value of $K$ during deployment. The reported model is trained with
$K_{\max}=6$; increasing this capacity requires resizing the assignment head
and retraining, and performance beyond six concurrent episodes remains
unevaluated.
\section{Empirical Studies}
\label{sec:empirical_studies}
This section evaluates whether TGCM satisfies the five requirements of the Unknown-$K$ Interleaved Sequence Demixing (UKISD) problem introduced in Section~\ref{sec:task_formulation}. Specifically, we evaluate (R1) reconstruction of canonical campaign episodes, (R2) estimation of the active episode count under a fixed decoding budget, (R3) occurrence-level episode assignment, (R4) efficient one-step inverse demixing, and (R5) generalization to realistic deployment settings without campaign-specific prior knowledge.

To this end, we conduct three complementary evaluations. First, we use SAGA-generated synthetic mixtures to provide fully controlled supervision for sequence reconstruction, technique-occurrence assignment, and budgeted unknown-$K$ estimation. Second, we evaluate the zero-shot generalization capability of TGCM on public benchmark datasets, including ATLAS, NODLINK, ProvCon, DARPA TC-E3, and DARPA TC-E5. Single-campaign datasets are transformed into controlled multi-campaign test cases using the proposed structure-aware forward-mixing process, whereas DARPA TC-E5 is evaluated directly because it naturally exhibits interleaved multi-campaign behavior. Third, we evaluate TGCM on \textsc{CAPTure}, an end-to-end evaluation dataset constructed from CALDERA--Procmon audit logs by an upstream ATT\&CK technique extraction pipeline. Unlike symbolic benchmark datasets, \textsc{CAPTure} preserves the practical challenges of real deployments, including single-host and multi-host intrusion scenarios, unknown mixture cardinality, and realistic symbolic extraction noise propagated from the upstream extraction process.

Unless otherwise stated, TGCM is trained and evaluated with \(K_{\max}=6\). Extending the decoding capacity beyond six concurrent episodes requires resizing the assignment head and retraining the model; performance for \(K_{\max}>6\) remains unevaluated.
During supervised training on SAGA, FASTopic uses source-campaign topic lookup to learn semantic representations. During inference, however, all zero-shot, DARPA TC-E5, and \textsc{CAPTure} evaluations use the neutral topic-conditioning vector described in Section~\ref{sec:FASTopic}, ensuring that inference relies only on information available in operational deployments without requiring campaign identities, episode labels, or mixture metadata.
Together, these evaluations examine TGCM under progressively more realistic conditions: (i) controlled supervised learning with occurrence-level ground truth, (ii) transfer to unseen campaigns and benchmark datasets, and (iii) end-to-end deployment with unknown mixture cardinality and noisy upstream technique extraction.

\subsection{Structure-Aware Forward Process}
\label{sec:datasets_mixing}
\begin{figure*}[t]
    \centering
    \includegraphics[width=1.0\textwidth]{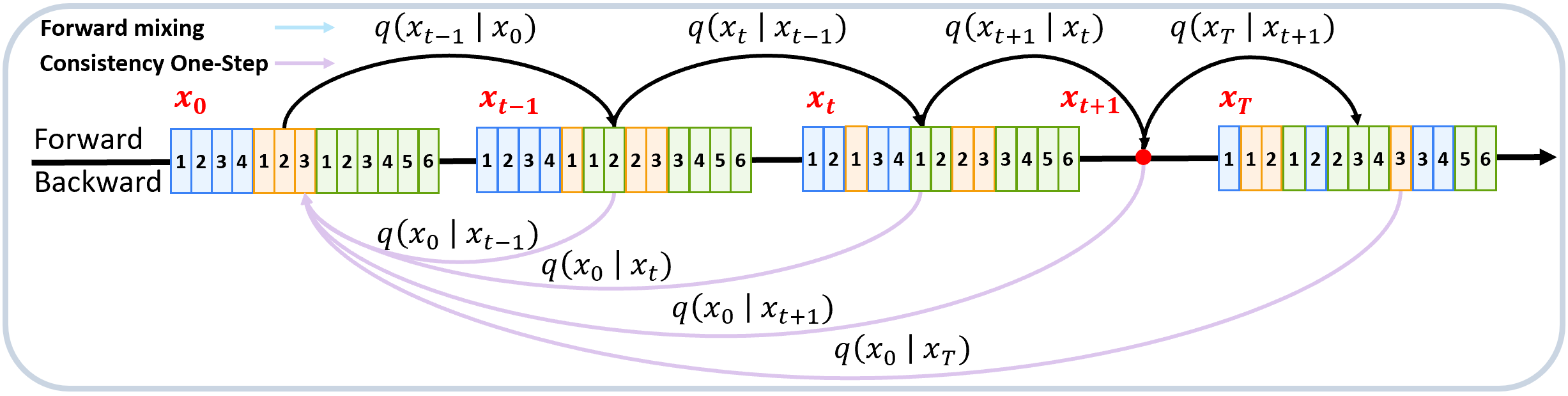}
    \caption{
    Forward mixing and single-step demixing in TGCM. A canonical sequence $x_0$, formed by concatenating $K$ single-APT campaigns, is progressively transformed into mixed states ${x_t}$ via a structure-aware forward process. Given an observed mixed sequence with unknown mixing level, TGCM applies a consistency-based inverse mapping to recover the canonical clean sequence $\hat{x}_0$ and predict technique-level assignments $\hat{y}$ in a single step.
    }
    \label{fig:forback}
\end{figure*}

Naturally interleaved APT campaigns with occurrence-level ground-truth episode assignments are rarely available. To construct supervised UKISD training and evaluation data while preserving a realistic campaign structure, we employ the structure-aware forward mixing process.
Rather than randomly shuffling techniques across campaigns, the forward process progressively transforms multiple canonical campaign episodes into increasingly interleaved observations while preserving the temporal structure of every latent episode.

The forward process operates at the granularity of \emph{ordered technique occurrences}. Each campaign episode is represented as an ordered ATT\&CK technique sequence, where every occurrence is associated with both its technique identity and its source-episode label. During forward mixing, contiguous technique blocks are relocated across campaign boundaries to increase interleaving while preserving three invariants:
\begin{enumerate}
\item the identity of every technique occurrence;
\item the occurrence-level source-episode assignment; and
\item the within-episode ordering of every latent campaign.
\end{enumerate}
These invariants ensure that the forward process modifies only the observable interleaving pattern while preserving the latent campaign episodes. Consequently, the inverse demixing task is to recover the original canonical episodes rather than to reconstruct new sequences.

\begin{algorithm}[t]
\caption{High-level structure-aware forward mixing}
\label{alg:structure_aware_mixing}
\begin{algorithmic}[1]
\Require Single-campaign technique sequences $\{s_1,\ldots,s_K\}$, maximum trajectory index $T$
\Ensure Trajectory states $\{(x_t,y_t)\}_{t=1}^{T}$ with $(x_1,y_1)=(x_0,y_0)$
\State Construct the canonical target
\[
x_0 = s_1 \Vert s_2 \Vert \cdots \Vert s_K,
\]
and initialize $y_{0,i}=k$ if the $i$-th occurrence originates from $s_k$.
\State Set $(x_1,y_1)\gets(x_0,y_0)$ as the indexed clean anchor.
\For{$t=2,\ldots,T$}
    \State Select an interaction budget $m_t$ and a block size $b_t$.
    \State Starting from $(x_{t-1},y_{t-1})$, choose contiguous blocks from different campaigns.
    \State Move or insert selected blocks across campaign boundaries to increase interleaving.
    \State Reject any move that changes the relative order of occurrences within the same campaign.
    \State Output the resulting mixed sequence and source labels as $(x_t,y_t)$.
\EndFor
\end{algorithmic}
\end{algorithm}

Algorithm~\ref{alg:structure_aware_mixing} summarizes the forward process. The canonical target $x_0$ is formed by concatenating the selected campaign episodes, and the implementation stores the identical sequence as the first indexed trajectory state $x_1=x_0$. For the three campaigns illustrated in Figure~\ref{fig:forback}, the canonical sequence is

\[
x_0=[1,2,3,4]\Vert[1,2,3]\Vert[1,2,3,4,5,6]
\]

with corresponding occurrence-level episode labels

\[
y_0=[1,1,1,1]\Vert[2,2,2]\Vert[3,3,3,3,3,3]
\]

For $t\in\{2,\ldots,T\}$, each forward-mixing step increases cross-episode interaction by relocating contiguous blocks. Early nonzero mixing stages produce coarse interleaving using relatively large blocks (e.g., $x_{t-1}$ in Figure~\ref{fig:forback}) and limited interactions, whereas later stages generate progressively finer interleaving through smaller blocks and more frequent interactions. Consequently, the forward process defines a controlled trajectory whose first indexed state is clean and whose later states become increasingly interleaved.

The key property of the forward process is that it preserves the internal ordering of every latent campaign episode. Let $\Pi_k(x_t,y_t)$ denote the ordered subsequence obtained by selecting all technique occurrences whose source label equals $k$. Then
\begin{equation}
\begin{aligned}
\Pi_k(x_0,y_0)=\Pi_k(x_t,y_t)=s_k
\\
\forall k\in\{1,\ldots,K\},
\forall t\in\{1,\ldots,T\}
\end{aligned}
\end{equation}

Therefore, although the observed sequence becomes increasingly interleaved, every latent campaign episode remains recoverable as an order-preserving subsequence. This property ensures that occurrence-level source labels remain valid throughout the forward process, providing direct supervision for episode assignment. At the same time, by preserving the temporal progression of each campaign, the generated mixtures remain structurally consistent with realistic APT behaviors instead of arbitrary shuffled sequences. Detailed schedules and operator definitions are provided in Appendix~\ref{app:forward_mixed}.

During supervised training, we first sample a mixture cardinality $K$, select $K$ canonical campaign episodes, and generate trajectory states indexed by $t\in\{1,\ldots,T\}$, where $t=1$ is the clean anchor and $t>1$ denotes nonzero interleaving. Pairs of mixed observations originating from the same episodes supervise the consistency objective, while the occurrence-level reference labels provide direct supervision for episode assignment. These labels define the fixed training targets $y_z$ used by the assignment objective. Episode slots are deterministically assigned according to the source-campaign order used to construct each mixture, and the same identifiers are retained throughout the forward trajectory, training, and evaluation.

Synthetic forward mixing is applied only when naturally interleaved data are unavailable. Datasets such as DARPA TC-E5 and \textsc{CAPTure} already contain naturally interleaved campaign behaviors and are therefore evaluated directly without synthetic mixing.

\subsection{Metrics}
\label{sec:metrics}

The objective of the Unknown-$K$ Interleaved Sequence Demixing (UKISD) problem is to recover the latent campaign episode associated with each observed ATT\&CK technique occurrence. We therefore use two complementary metric categories:

\begin{itemize}
\item \textbf{Occurrence-level assignment metrics}: Accuracy (Acc), Macro Precision (P), Macro Recall (R), and Macro-F1, which directly evaluate episode assignment for individual technique occurrences.

\item \textbf{Partition-quality diagnostics}: the Fowlkes--Mallows Index (FMI) and Normalized Mutual Information (NMI), which measure global agreement between the predicted and ground-truth episode partitions.
\end{itemize}

Let $N$ denote the number of mixed sequences. For sequence $i$ of length $n_i$, let $y_{i,j}$ and $\hat{y}_{i,j}$ denote the ground-truth and predicted episode labels of occurrence $j$, respectively. Let $\mathcal{A}$ denote the set of non-residual episode labels in the evaluation set, excluding label $0$ when a residual channel is used.

\noindent
\textbf{Statistical reporting.}
Unless otherwise specified, results are reported as the mean and standard deviation over repeated runs. Confidence intervals and paired statistical tests are reported when paired repeated measurements are available. Time denotes wall-clock inference time measured using two NVIDIA A100 80GB GPUs, an AMD EPYC 7282 16-Core Processor, and 1TB of system memory.

\noindent
\textbf{Occurrence-level assignment metrics.}

\emph{Occurrence-level assignment accuracy (Acc)} measures the fraction of technique occurrences assigned to their correct latent episodes under the fixed episode-slot labels established during mixture construction or benchmark annotation. To avoid domination by long sequences, accuracy is computed per sequence and then averaged:
\begin{equation}
\mathrm{Acc}
=
\frac{1}{N}
\sum_{i=1}^{N}
\left(
\frac{1}{n_i}
\sum_{j=1}^{n_i}
\mathbb{I}\!\left[
\hat{y}_{i,j}=y_{i,j}
\right]
\right)
\end{equation}

We further report macro-averaged precision, recall, and Macro-F1. For each episode $a\in\mathcal{A}$,
\begin{equation}
P_a=\frac{\mathrm{TP}_a}{\mathrm{TP}_a+\mathrm{FP}_a},
\quad
R_a=\frac{\mathrm{TP}_a}{\mathrm{TP}_a+\mathrm{FN}_a},
\quad
F_{1,a}=\frac{2P_aR_a}{P_a+R_a}
\end{equation}
where $\mathrm{TP}_a$, $\mathrm{FP}_a$, and $\mathrm{FN}_a$ denote the true positives, false positives, and false negatives for episode $a$, respectively. The macro-averaged scores are
\begin{equation}
M_{\mathrm{macro}}
=\frac{1}{|\mathcal{A}|}
\sum_{a\in\mathcal{A}} M_a,
\qquad
M\in\{P,R,F_1\}
\end{equation}

\medskip
\noindent
\textbf{Partition-quality diagnostics.}

\emph{Fowlkes--Mallows Index (FMI)} evaluates pairwise agreement between predicted and ground-truth partitions:
\begin{equation}
\mathrm{FMI}
=\sqrt{
\frac{\mathrm{TP}}
{\mathrm{TP}+\mathrm{FP}}
\cdot
\frac{\mathrm{TP}}
{\mathrm{TP}+\mathrm{FN}}
}
\end{equation}
where $\mathrm{TP}$, $\mathrm{FP}$, and $\mathrm{FN}$ count occurrence pairs that are correctly co-assigned, incorrectly co-assigned, or incorrectly separated. FMI ranges from 0 to 1, with higher values indicating stronger agreement.

\emph{Normalized Mutual Information (NMI)} measures the shared information between predicted partition $U$ and ground-truth partition $V$:
\begin{equation}
\mathrm{NMI}(U,V)
=\frac{2I(U;V)}
{H(U)+H(V)}
\end{equation}
where $I(U;V)$ is their mutual information and $H(\cdot)$ denotes entropy. NMI ranges from 0 to 1, with higher values indicating stronger partition agreement.

Accuracy, Macro Precision, Macro Recall, and Macro-F1 are treated as the primary metrics because they directly evaluate occurrence-level episode assignment. FMI and NMI serve as complementary diagnostics of the recovered partition structure, since clustering agreement may remain high even when individual occurrences are assigned to incorrect episode slots.

\begin{table*}[ht!]
\centering
\caption{\textbf{Representative comparison of demixing approaches.}
Methods are compared according to five architectural and algorithmic characteristics relevant to the UKISD problem: (1) \textbf{budgeted $K$}: inference without prior knowledge of the number of mixed campaign episodes; (2) \textbf{Seq.}: explicit modeling of sequential dependencies; (3) \textbf{GPU}: GPU-enabled parallel processing; (4) \textbf{Efficiency}: computational efficiency and scalability; and (5) \textbf{Effectiveness}: capability to model complex, non-linear, and long-range dependencies. The ratings summarize the intrinsic properties and expected suitability of each approach for UKISD rather than empirical benchmark performance, which is reported separately in the experimental results. Symbols denote the level of support: $\bullet$ (high/full), $\circ$ (partial/medium), and $-$ (low/none).}
\label{tab:comprehensive_comparison}
\resizebox{\textwidth}{!}{%
\begin{tabular}{l l l c c c l l}
\toprule
\textbf{Category} & \textbf{Group} & \textbf{Method} & \textbf{Budgeted $K$} & \textbf{Seq.} & \textbf{GPU} & \textbf{Efficiency} & \textbf{Effectiveness} \\
\midrule
\multirow{8}{*}{\textbf{Cat. I: Unsupervised}} & \multirow{8}{*}{\textit{Pattern Mining}} 
 & Smith-Waterman (SW)~\cite{smith1981identification, schmidt2024cudasw++} & $-$ & $\bullet$ & $\bullet$ & Low ($O(L^2)$) & Medium (Local only) \\
 & & OCPN / iLPM~\cite{van2023object, tax2016mining} & $\bullet$ & $\bullet$ & $-$ & Low (CPU-bound) & Medium (Rule-structured) \\
 & & EC-SA (ECC family)~\cite{bayomie2019probabilistic} & $\bullet$ & $\circ$ & $-$ & Medium (Probabilistic) & Medium (Case-level) \\
 & & Frequent Episode (FEM)~\cite{fournier2022maxfem} & $\bullet$ & $\circ$ & $-$ & Low (Combinatorial) & Medium (Window-limited) \\
 & & TOA Histograms~\cite{xie2023novel, chen2025radar} & $-$ & $\circ$ & $-$ & High ($O(N)$) & Low (Periodicity-based) \\
 & & HMM~\cite{baum1970maximization} & $\bullet$ & $\circ$ & $\circ$ & Medium (Iterative) & Medium (Markovian) \\
 & & Factorial HMM (FHMM)~\cite{ghahramani1995factorial} & $\bullet$ & $\circ$ & $\circ$ & Low (State Exp.) & Medium (Concurrent) \\
 & & Uniform Random (UR) & $-$ & $-$ & $\bullet$ & High ($O(N)$) & Low (Chance-level) \\
\midrule
\multirow{6}{*}{\textbf{Cat. II: Deep \& Statistical}} & \multirow{3}{*}{\textit{Group A: Deep}} 
 & SepFormer~\cite{subakan2021attention} & $-$ & $\bullet$ & $\bullet$ & High (GPU) & High (Global Seq.) \\
 & & DANet~\cite{chen2017deep} & $\bullet$ & $\bullet$ & $\bullet$ & High (GPU) & High (Attractor) \\
 & & MossFormer2~\cite{zhao2024mossformer2} & $-$ & $\bullet$ & $\bullet$ & High (Hybrid) & High \\
 \cmidrule{2-8}
 & \multirow{2}{*}{\textit{Group B: Statistical}} 
 & DECOMPOSE~\cite{bottcher2018trace} & $\bullet$ & $-$ & $\bullet$ & Medium (Matrix Fac.) & Medium (No order) \\
 & & TD-DMD~\cite{nedzhibov2025blind} & $-$ & $\circ$ & $\bullet$ & Medium (SVD) & Medium (Linear) \\
 \cmidrule{2-8}
 & \textit{Group C: GenAI} & LLMs (ChatGPT/Gemini) & $\circ$ & $\bullet$ & $\bullet$ & Low (Latency) & Medium (Prompt-only) \\
\midrule
\textbf{Proposed} & \textbf{Ours} & \textbf{TGCM} & $\bullet$ & $\bullet$ & $\bullet$ & \textbf{High (GPU)} & \textbf{High} \\
\bottomrule
\end{tabular}%
}
\end{table*}

\begin{figure*}[ht!]
    \centering
    \includegraphics[width=1.0\textwidth]{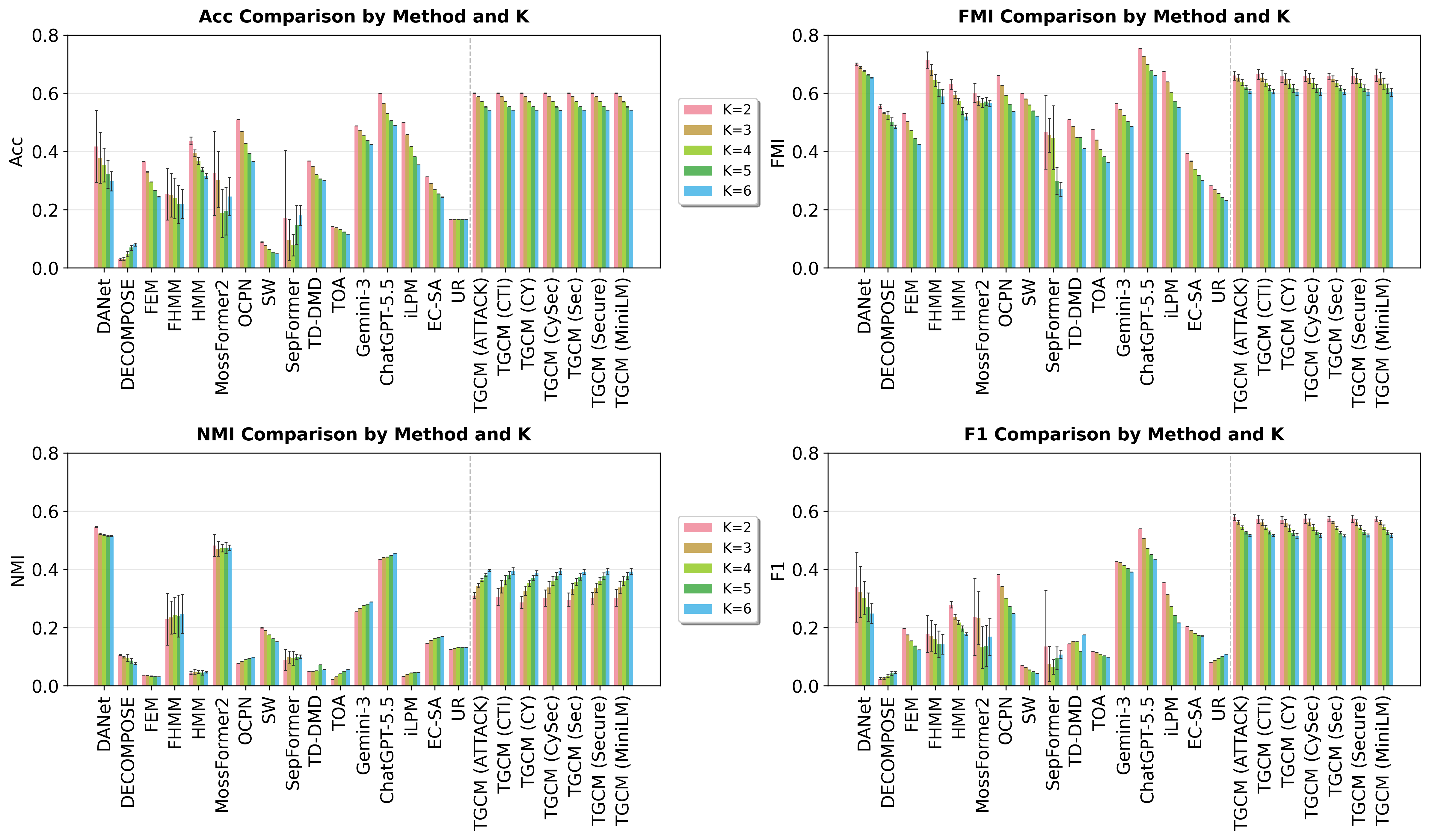}
    \caption{\textbf{Performance comparison of representative baselines and TGCM under blind unknown-$K$ demixing.}
The four subplots report (top-left) occurrence-level assignment accuracy (Acc), (top-right) Fowlkes--Mallows Index (FMI), (bottom-left) Normalized Mutual Information (NMI), and (bottom-right) Macro-F1. Results are grouped by the ground-truth number of campaign episodes ($K\in\{2,\ldots,6\}$), which is used only for stratified reporting and is never provided during inference. Classical probabilistic baselines search candidate mixture sizes over $K\in[2,6]$, neural baselines use the same decoding budget ($K_{\max}=6$) as TGCM, and prompt-based LLMs infer the number of episodes directly from the input prompt. Black error bars indicate the standard deviation over repeated runs where available; ChatGPT-5.5 and Gemini-3 were each evaluated once and therefore have no repeated-run error bars. Methods to the right of the dashed line denote TGCM variants using different semantic embedding backbones (ATTACK, CTI, CY, CYSec, Sec, Secure, and MiniLM). \textit{UR} denotes the Uniform Random baseline. Overall, TGCM consistently achieves the strongest occurrence-level assignment performance across different mixture cardinalities, demonstrating the effectiveness of directly modeling the UKISD problem.}
    \label{fig:performance_comparison}
\end{figure*}

\subsection{Baseline Comparisons}

Since UKISD is newly formulated, no existing method directly satisfies all of its requirements, including unknown-cardinality estimation, occurrence-level assignment, order-preserving demixing, and one-step inference. We therefore compare TGCM with representative baselines from six methodological families. These include alignment and process-mining methods, such as Smith--Waterman~\cite{smith1981identification,schmidt2024cudasw++}, OCPN~\cite{van2023object}, EC-SA~\cite{bayomie2019probabilistic}, iLPM~\cite{tax2016mining}, and FEM~\cite{fournier2022maxfem}; statistical and probabilistic methods, including TOA histograms~\cite{xie2023novel,chen2025radar}, HMM, and FHMM~\cite{baum1970maximization,ghahramani1995factorial}; neural sequence-separation models, including SepFormer~\cite{subakan2021attention}, MossFormer2~\cite{zhao2024mossformer2}, and DANet~\cite{chen2017deep}; algebraic decomposition methods, including DECOMPOSE~\cite{bottcher2018trace} and Time-Delayed DMD~\cite{nedzhibov2025blind}; prompt-based LLM reasoning; and Uniform Random as the chance-level reference. Table~\ref{tab:comprehensive_comparison} summarizes their methodological characteristics, while Figure~\ref{fig:performance_comparison} reports their quantitative performance.

All methods are evaluated under the same \emph{budgeted unknown-$K$} protocol. The ground-truth mixture cardinality is used only to stratify results and is never provided during inference. Classical probabilistic methods search over candidate values $K\in[2,6]$, while TGCM and the adapted neural baselines use the same Up-to-6 decoding budget ($K_{\max}=6$). Neural baselines operate on learnable ATT\&CK technique embeddings, whereas prompt-based LLMs infer the episode count and assignments directly from the input prompt.

For the LLM baselines, ChatGPT-5.5 and Gemini-3 are evaluated using a fixed few-shot JSON prompt. Each model is executed once over the evaluation set, and malformed or length-mismatched outputs are counted as invalid. The complete prompt is provided in Appendix~\ref{app:llm_prompt_template}; these results are treated as prompt-only references rather than fully optimized LLM systems.

\begin{table*}[h!]
\centering
\caption{Kill-chain coverage of the zero-shot benchmarks and \textsc{CAPTure}, measured by the number of unique techniques mapped to each ATT\&CK tactic. The complete zero-shot dataset mappings are reported in Appendix Table~\ref{tab:kill_chain_final_v4}, while the \textsc{CAPTure} profile-level mapping is reported in Appendix Table~\ref{tab:kill_chain_capture_apt}. Zero entries indicate unavailable tactic coverage in the evaluated data rather than absence of the tactic in real-world campaigns.}
\footnotesize
\setlength{\tabcolsep}{3pt}
\renewcommand{\arraystretch}{1.25}
\arrayrulecolor{black!20}
\setlength{\arrayrulewidth}{0.35pt}

\begin{tabularx}{\textwidth}{|B|N{kcInit}|N{kcExec}|N{kcPers}|N{kcDef}|N{kcCred}|N{kcDisc}|N{kcLat}|N{kcColl}|N{kcCtwo}|N{kcExf}|}
\hline
\multicolumn{1}{|c|}{\normalfont\bfseries Dataset} &
\cellcolor{kcInitH}\normalfont\bfseries Init.\ Access &
\cellcolor{kcExecH}\normalfont\bfseries Execution &
\cellcolor{kcPersH}\normalfont\bfseries Persistence &
\cellcolor{kcDefH}\normalfont\bfseries Def.\ Evasion &
\cellcolor{kcCredH}\normalfont\bfseries Cred.\ Access &
\cellcolor{kcDiscH}\normalfont\bfseries Discovery &
\cellcolor{kcLatH}\normalfont\bfseries Lat.\ Move. &
\cellcolor{kcCollH}\normalfont\bfseries Collection &
\cellcolor{kcCtwoH}\normalfont\bfseries C2 &
\cellcolor{kcExfH}\normalfont\bfseries Exfiltration
\tabularnewline\hline
\multicolumn{11}{|c|}{\textit{Zero-shot datasets}}
\tabularnewline\hline
ATLAS       & 0 & 3 & 1 & 3 & 0 & 1 & 0 & 0 & 2 & 0 \tabularnewline\hline
NODLINK     & 0 & 1 & 0 & 0 & 0 & 9 & 0 & 1 & 1 & 0 \tabularnewline\hline
ProvCon     & 1 & 0 & 1 & 0 & 1 & 1 & 0 & 0 & 1 & 0 \tabularnewline\hline
DARPA TC-E3 & 2 & 6 & 3 & 1 & 2 & 9 & 0 & 1 & 3 & 1 \tabularnewline\hline
DARPA TC-E5 & 1 & 4 & 1 & 3 & 2 & 9 & 0 & 2 & 4 & 2 \tabularnewline\hline

\multicolumn{11}{|c|}{\textit{End-to-end controlled emulation benchmark}}
\tabularnewline\hline
\textsc{CAPTure} & 8 & 28 & 24 & 14 & 0 & 58 & 11 & 17 & 13 & 1
\tabularnewline\hline

\end{tabularx}

\label{tab:kill_chain_mapping}
\end{table*}

\begin{table*}[ht!]
\centering
\caption{\textbf{Zero-shot performance across five benchmark datasets under the fixed-slot evaluation.} Each cell reports the mean with standard deviation shown as a subscript over five repeated runs.}
\label{tab:zero_shot_master}
\resizebox{\textwidth}{!}{
\begin{tabular}{l|c|cc|cc|cc|cc|cc}
\toprule
\multirow{2}{*}{\textbf{Dataset}} & \multirow{2}{*}{\textbf{Model}} & \multicolumn{2}{c|}{\textbf{K=2}} & \multicolumn{2}{c|}{\textbf{K=3}} & \multicolumn{2}{c|}{\textbf{K=4}} & \multicolumn{2}{c|}{\textbf{K=5}} & \multicolumn{2}{c}{\textbf{K=6}} \\
 & & \textbf{Acc} $\uparrow$ & \textbf{Macro-F1} $\uparrow$ & \textbf{Acc} $\uparrow$ & \textbf{Macro-F1} $\uparrow$ & \textbf{Acc} $\uparrow$ & \textbf{Macro-F1} $\uparrow$ & \textbf{Acc} $\uparrow$ & \textbf{Macro-F1} $\uparrow$ & \textbf{Acc} $\uparrow$ & \textbf{Macro-F1} $\uparrow$ \\
\midrule
\multicolumn{12}{c}{\textbf{Forward-mixing zero-shot}} \\
\midrule
\multirow{2}{*}{\textbf{ATLAS}} & TGCM & $\mathbf{0.535}_{\mathbf{0.014}}^{\dagger}$ & $\mathbf{0.448}_{\mathbf{0.065}}$ & $\mathbf{0.371}_{\mathbf{0.011}}^{\dagger}$ & $\mathbf{0.251}_{\mathbf{0.024}}$ & $\mathbf{0.268}_{\mathbf{0.011}}^{\dagger}$ & $0.150_{0.024}$ & N/A & N/A & N/A & N/A \\
 & DANet & $0.416_{0.044}$ & $0.352_{0.051}$ & $0.294_{0.031}$ & $0.228_{0.029}$ & $0.230_{0.012}$ & $\mathbf{0.162}_{\mathbf{0.012}}$ & N/A & N/A & N/A & N/A \\
\midrule
\multirow{2}{*}{\textbf{NODLINK}} & TGCM & $0.493_{0.010}$ & $0.345_{0.038}$ & $\mathbf{0.328}_{\mathbf{0.007}}^{\dagger}$ & $0.190_{0.029}$ & N/A & N/A & N/A & N/A & N/A & N/A \\
 & DANet & $\mathbf{0.505}_{\mathbf{0.087}}$ & $\mathbf{0.449}_{\mathbf{0.082}}$ & $0.296_{0.012}$ & $\mathbf{0.221}_{\mathbf{0.024}}$ & N/A & N/A & N/A & N/A & N/A & N/A \\
\midrule
\multirow{2}{*}{\textbf{ProvCon}} & TGCM & $\mathbf{0.449}_{\mathbf{0.051}}$ & $0.298_{0.052}$ & $\mathbf{0.310}_{\mathbf{0.021}}$ & $0.172_{0.020}$ & $\mathbf{0.241}_{\mathbf{0.017}}$ & $0.110_{0.014}$ & $\mathbf{0.195}_{\mathbf{0.013}}$ & $0.086_{0.011}$ & $\mathbf{0.174}_{\mathbf{0.008}}$ & $0.077_{0.011}$ \\
 & DANet & $0.381_{0.099}$ & $\mathbf{0.347}_{\mathbf{0.102}}$ & $0.268_{0.063}$ & $\mathbf{0.202}_{\mathbf{0.048}}$ & $0.216_{0.019}$ & $\mathbf{0.138}_{\mathbf{0.015}}$ & $0.187_{0.061}$ & $\mathbf{0.107}_{\mathbf{0.037}}$ & $0.161_{0.017}$ & $\mathbf{0.079}_{\mathbf{0.010}}$ \\
\midrule
\multirow{2}{*}{\textbf{DARPA TC-E3}} & TGCM & $\mathbf{0.504}_{\mathbf{0.021}}$ & $0.357_{0.015}$ & $\mathbf{0.347}_{\mathbf{0.019}}$ & $0.199_{0.018}$ & $\mathbf{0.256}_{\mathbf{0.022}}$ & $0.125_{0.013}$ & $\mathbf{0.201}_{\mathbf{0.017}}$ & $0.089_{0.008}$ & $\mathbf{0.165}_{\mathbf{0.009}}$ & $0.071_{0.007}$ \\
 & DANet & $0.425_{0.072}$ & $\mathbf{0.379}_{\mathbf{0.076}}$ & $0.306_{0.029}$ & $\mathbf{0.239}_{\mathbf{0.036}}$ & $0.228_{0.027}$ & $\mathbf{0.150}_{\mathbf{0.016}}$ & $0.199_{0.032}$ & $\mathbf{0.122}_{\mathbf{0.024}}$ & $0.163_{0.014}$ & $\mathbf{0.098}_{\mathbf{0.006}}^{\dagger}$ \\
\midrule
\midrule
\multicolumn{12}{c}{\textbf{In-the-wild interleaved zero-shot}} \\
\midrule
\multirow{2}{*}{\textbf{DARPA TC-E5}} & TGCM & $\mathbf{0.536}_{\mathbf{0.095}}$ & $\mathbf{0.458}_{\mathbf{0.122}}$ & N/A & N/A & $\mathbf{0.506}_{\mathbf{0.090}}$ & $\mathbf{0.434}_{\mathbf{0.115}}$ & N/A & N/A & N/A & N/A \\
 & DANet & $0.450_{0.147}$ & $0.365_{0.157}$ & N/A & N/A & $0.440_{0.133}$ & $0.362_{0.143}$ & N/A & N/A & N/A & N/A \\
\bottomrule
\end{tabular}}
\vspace{0.25em}
\begin{minipage}{0.98\textwidth}
\footnotesize $\uparrow$ indicates that a higher value is better. Bold indicates the better mean between TGCM and DANet for the same dataset, $K$, and metric. Superscript $^{\dagger}$ marks cases where the reported 95\% confidence intervals are separated in the favorable direction.
\end{minipage}
\end{table*}

\subsection{Impact of Topic Guidance and Consistency}
\label{sec:ablation}
\begin{table}[h]
\centering
\caption{Normalized gain across mixture complexity $K$.}
\label{tab:early_gain_by_k}
\footnotesize

\begin{tabular*}{\columnwidth}{@{}c@{\extracolsep{\fill}}c c c c@{}}
\toprule
$K$ & Topic+Consistency & Topic & Consistency & Base \\
\midrule
2 & \textbf{18.9\%} & \textbf{18.9\%} & 16.5\% & 16.4\% \\
4 & \textbf{18.1\%} & 17.7\% & 17.0\% & 16.8\% \\
6 & \textbf{15.0\%} & 14.4\% & 13.5\%  & 13.4\% \\
\bottomrule
\end{tabular*}
\vspace{0.25em}
\begin{minipage}{0.98\columnwidth}
\footnotesize Values are normalized gains over the uniform-random baseline.
\end{minipage}
\end{table}

This experiment evaluates whether the topic-guided semantic regularizer and the consistency objective improve one-step inverse demixing beyond the task-aligned base model. Using the same training and evaluation protocol, we conduct a factorial ablation with four variants: (i) \textit{Base}, which disables both components; (ii) \textit{+Topic} (Equations~\ref{eq:fused}--\ref{eq:topic_objective}), which enables only topic guidance; (iii) \textit{+Consistency} (Equation~\ref{eq:consistency_loss}), which enables only the consistency objective; and (iv) \textit{+Topic+Consistency}, which corresponds to the full TGCM model. Unless otherwise stated, all remaining hyperparameters are held fixed.

Table~\ref{tab:early_gain_by_k} reports the normalized gain of each variant over the Uniform Random baseline at representative mixture cardinalities. For an occurrence-level assignment score \(M\), the gain is defined as
\[
100\times
\frac{M_{\mathrm{variant}}-M_{\mathrm{UR}}}
     {1-M_{\mathrm{UR}}}
\]
where \(M_{\mathrm{variant}}\) is the score of the evaluated variant and \(M_{\mathrm{UR}}\) is the Uniform Random score under the same \(K\).

The full \textit{+Topic+Consistency} configuration is numerically the best or tied for the best across all reported mixture sizes. Topic guidance provides a larger individual improvement by encouraging semantically coherent episode reconstruction, whereas consistency regularization yields a smaller but complementary gain by stabilizing occurrence-level assignments across different interleaving states. Their combination, therefore, offers the most reliable improvement over the base formulation. Nevertheless, the normalized gain decreases as \(K\) increases, indicating that high-cardinality mixtures remain more difficult even when both regularizers are enabled. Embedding-type sensitivity and feature-importance diagnostics are reported in Appendix~\ref{app:embedding_type_effect}.

\subsection{Zero-shot Evaluation on Benchmarks}

We evaluate whether TGCM, trained only on SAGA-generated synthetic campaigns, generalizes to unseen campaign sources and benchmark datasets without fine-tuning. All campaigns are mapped to MITRE ATT\&CK technique sequences, with coverage summarized in Table~\ref{tab:kill_chain_mapping} and complete mappings provided in Appendix Table~\ref{tab:kill_chain_final_v4}. TGCM is compared with \textbf{DANet}~\cite{chen2017deep}, the strongest neural baseline identified in the preceding experiments.

ATLAS~\cite{alsaheel2021atlas}, NODLINK~\cite{Li_2024}, ProvCon~\cite{provcon25}, and DARPA TC-E3~\cite{darpae3} contain single-campaign traces; we therefore apply the proposed forward-mixing process only to construct controlled multi-campaign test cases. For each supported cardinality ($K\le6$), results are averaged over five independent runs. Table~\ref{tab:zero_shot_master} shows that TGCM achieves higher occurrence-level assignment accuracy than DANet across most settings, although Macro-F1 varies more across datasets because of differences in campaign complexity and technique distributions. Complete results are reported in Appendix Table~\ref{tab:landscape_full_metrics}.

DARPA TC-E5~\cite{darpae5} provides naturally interleaved campaigns spanning multiple hosts and days, so no synthetic mixing is applied. We manually annotate occurrence-level episode assignments from the official documentation, assign episode slots according to the first observed occurrence of each annotated episode, and evaluate cases containing two or four concurrent campaigns. TGCM continues to outperform DANet, indicating that its inverse demixing capability generalizes beyond SAGA-generated mixtures. The results are consistent with the kill-chain coverage in Table~\ref{tab:kill_chain_mapping}: topic guidance helps disambiguate repeated techniques, while consistency regularization supports coherent episode reconstruction across diverse interleaving patterns.

\subsection{Robustness to Technique Perturbations}
\label{subsec:technique_perturbation_robustness}
\begingroup
\newcommand{\dc}[2]{\makebox[3.00em][r]{$#1$}\makebox[0.30em][l]{$#2$}}
\newcommand{\cimark}{^{\dagger}}
\begin{table*}[t]
\centering
\caption{DARPA TC-E5 robustness under technique perturbations. Each entry reports $\Delta M = M_{\mathrm{TGCM}} - M_{\mathrm{DANet}}$; positive values favor TGCM and negative values favor DANet. Superscript $\dagger$ marks a two-sided 95\% paired Student-$t$ confidence interval that excludes zero.}
\label{tab:noise_robustness}
\footnotesize
\setlength{\tabcolsep}{1.2pt}
\renewcommand{\arraystretch}{0.82}
\begin{tabular*}{\textwidth}{@{\extracolsep{\fill}}llccccccccc@{}}
\toprule
\textbf{Noise} & \textbf{$K$} & $\boldsymbol{\rho=.1}$ & $\boldsymbol{\rho=.2}$ & $\boldsymbol{\rho=.3}$ & $\boldsymbol{\rho=.4}$ & $\boldsymbol{\rho=.5}$ & $\boldsymbol{\rho=.6}$ & $\boldsymbol{\rho=.7}$ & $\boldsymbol{\rho=.8}$ & $\boldsymbol{\rho=.9}$ \\
\midrule
\multicolumn{11}{@{}l}{\textbf{(a) Accuracy difference $\Delta\mathrm{Acc}=\mathrm{Acc}_{\mathrm{TGCM}}-\mathrm{Acc}_{\mathrm{DANet}}$}} \\
Missing & 2 & \dc{\mathbf{+0.078}}{} & \dc{\mathbf{+0.088}}{} & \dc{\mathbf{+0.061}}{} & \dc{\mathbf{+0.016}}{} & \dc{\mathbf{+0.098}}{} & \dc{\mathbf{+0.090}}{\cimark} & \dc{\mathbf{+0.159}}{} & \dc{-0.022}{} & \dc{\mathbf{+0.040}}{} \\
Missing & 4 & \dc{\mathbf{+0.060}}{} & \dc{\mathbf{+0.084}}{} & \dc{\mathbf{+0.080}}{} & \dc{\mathbf{+0.024}}{} & \dc{\mathbf{+0.085}}{} & \dc{\mathbf{+0.112}}{} & \dc{\mathbf{+0.127}}{} & \dc{-0.043}{} & \dc{\mathbf{+0.017}}{} \\
Confusion & 2 & \dc{\mathbf{+0.061}}{} & \dc{\mathbf{+0.109}}{} & \dc{\mathbf{+0.162}}{\cimark} & \dc{\mathbf{+0.124}}{} & \dc{\mathbf{+0.111}}{} & \dc{\mathbf{+0.175}}{\cimark} & \dc{\mathbf{+0.156}}{} & \dc{\mathbf{+0.110}}{} & \dc{\mathbf{+0.152}}{\cimark} \\
Confusion & 4 & \dc{\mathbf{+0.038}}{} & \dc{\mathbf{+0.065}}{} & \dc{\mathbf{+0.147}}{\cimark} & \dc{\mathbf{+0.090}}{} & \dc{\mathbf{+0.071}}{} & \dc{\mathbf{+0.158}}{\cimark} & \dc{\mathbf{+0.137}}{} & \dc{\mathbf{+0.092}}{} & \dc{\mathbf{+0.149}}{\cimark} \\
Insertion & 2 & \dc{\mathbf{+0.049}}{} & \dc{\mathbf{+0.083}}{} & \dc{-0.092}{} & \dc{\mathbf{+0.018}}{} & \dc{-0.126}{} & \dc{-0.028}{} & \dc{-0.023}{} & \dc{-0.035}{} & \dc{\mathbf{+0.033}}{} \\
Insertion & 4 & \dc{-0.006}{} & \dc{\mathbf{+0.057}}{} & \dc{-0.118}{} & \dc{-0.015}{} & \dc{-0.129}{} & \dc{-0.056}{} & \dc{-0.049}{} & \dc{-0.045}{} & \dc{\mathbf{+0.008}}{} \\
\midrule
\multicolumn{11}{@{}l}{\textbf{(b) Macro-F1 difference $\Delta\mathrm{MacroF1}=\mathrm{MacroF1}_{\mathrm{TGCM}}-\mathrm{MacroF1}_{\mathrm{DANet}}$}} \\
Missing & 2 & \dc{\mathbf{+0.082}}{} & \dc{\mathbf{+0.102}}{} & \dc{\mathbf{+0.075}}{} & \dc{\mathbf{+0.036}}{} & \dc{\mathbf{+0.093}}{} & \dc{\mathbf{+0.093}}{\cimark} & \dc{\mathbf{+0.090}}{} & \dc{-0.098}{} & \dc{\mathbf{+0.006}}{} \\
Missing & 4 & \dc{\mathbf{+0.077}}{} & \dc{\mathbf{+0.093}}{} & \dc{\mathbf{+0.081}}{} & \dc{\mathbf{+0.031}}{} & \dc{\mathbf{+0.065}}{} & \dc{\mathbf{+0.099}}{} & \dc{\mathbf{+0.065}}{} & \dc{-0.116}{} & \dc{-0.029}{} \\
Confusion & 2 & \dc{\mathbf{+0.080}}{} & \dc{\mathbf{+0.124}}{} & \dc{\mathbf{+0.187}}{\cimark} & \dc{\mathbf{+0.184}}{\cimark} & \dc{\mathbf{+0.132}}{} & \dc{\mathbf{+0.192}}{\cimark} & \dc{\mathbf{+0.198}}{\cimark} & \dc{\mathbf{+0.131}}{} & \dc{\mathbf{+0.150}}{\cimark} \\
Confusion & 4 & \dc{\mathbf{+0.048}}{} & \dc{\mathbf{+0.078}}{} & \dc{\mathbf{+0.162}}{\cimark} & \dc{\mathbf{+0.141}}{\cimark} & \dc{\mathbf{+0.088}}{} & \dc{\mathbf{+0.163}}{\cimark} & \dc{\mathbf{+0.166}}{\cimark} & \dc{\mathbf{+0.111}}{} & \dc{\mathbf{+0.132}}{\cimark} \\
Insertion & 2 & \dc{\mathbf{+0.071}}{} & \dc{\mathbf{+0.063}}{} & \dc{-0.089}{} & \dc{\mathbf{+0.020}}{} & \dc{-0.134}{} & \dc{-0.058}{} & \dc{-0.070}{} & \dc{-0.049}{} & \dc{-0.004}{} \\
Insertion & 4 & \dc{\mathbf{+0.022}}{} & \dc{\mathbf{+0.041}}{} & \dc{-0.104}{} & \dc{-0.008}{} & \dc{-0.125}{} & \dc{-0.069}{} & \dc{-0.077}{} & \dc{-0.053}{} & \dc{-0.018}{} \\
\bottomrule
\end{tabular*}
\vspace{0.10em}
\begin{minipage}{0.98\textwidth}
\scriptsize Boldface marks positive deltas. Accuracy is the primary occurrence-level assignment metric; Macro-F1 highlights balance across recovered episodes under noise.
\end{minipage}
\end{table*}
\endgroup

The previous zero-shot experiments assume that the ATT\&CK technique sequence has already been extracted correctly. In practice, however, upstream ATT\&CK abstraction may omit techniques, misclassify one technique as another, or introduce spurious techniques originating from benign or unrelated activities. This experiment evaluates whether TGCM remains effective under such symbolic extraction errors while keeping the ground-truth episode assignments unchanged. The objective is to assess robustness to imperfect ATT\&CK abstraction rather than robustness to raw telemetry evasion or adversarial log manipulation.

We consider three representative perturbation families. \textit{Missing} randomly removes observed technique occurrences to simulate incomplete evidence or extractor false negatives. \textit{Confusion} replaces a technique with another ATT\&CK technique to model technique-level misclassification. \textit{Insertion} injects additional technique occurrences to represent extractor false positives or unrelated background activities that survive the abstraction process. The perturbation rate $\rho$ controls the corruption level, and results are reported separately for different mixture cardinalities because larger mixtures contain more reused techniques and more ambiguous episode boundaries.

Table~\ref{tab:noise_robustness} reports the absolute performance difference,
\[
\Delta M = M_{\mathrm{TGCM}} - M_{\mathrm{DANet}}
\]
where positive values indicate that TGCM achieves higher performance than DANet under the same perturbed input. TGCM consistently maintains the largest advantage under \textit{Confusion}, demonstrating that semantic topic guidance effectively disambiguates substituted techniques. The performance gain remains largely positive under \textit{Missing}, indicating resilience to incomplete technique observations. In contrast, \textit{Insertion} is the most challenging perturbation because additional techniques introduce ambiguous evidence that may alter episode boundaries. Overall, these results suggest that TGCM is robust to the symbolic extraction errors commonly encountered during ATT\&CK abstraction, while insertion-heavy scenarios remain the principal failure mode and motivate future improvements to the abstraction pipeline.

\subsection{End-to-End Evaluation on \textsc{CAPTure} }
\label{Sec:CAPTure}
\begin{table*}[t]
\centering
\caption{\textbf{End-to-end evaluation on \textsc{CAPTure} after pooling single-host and multi-host results with Up-to-6 decoding ($K_{\max}=6$).} Each cell reports mean$_{\text{std}}$ over five seeds; superscript $\dagger$ marks cells whose 95\% confidence interval is separated from the counterpart in the favorable direction.}
\label{tab:capture_upstream}
\footnotesize
\setlength{\tabcolsep}{4pt}
\renewcommand{\arraystretch}{1.05}
\begin{tabular*}{\textwidth}{@{\extracolsep{\fill}}llccccc@{}}
\toprule
\multicolumn{7}{l}{\textbf{(a) Accuracy} $\uparrow$} \\
\midrule
Extractor & Model & Mix=2 & Mix=3 & Mix=4 & Mix=5 & Mix=6 \\
\midrule
SFM & TGCM & $\mathbf{0.617}_{\mathbf{0.028}}^{\dagger}$ & $\mathbf{0.431}_{\mathbf{0.056}}$ & $\mathbf{0.394}_{\mathbf{0.029}}$ & $\mathbf{0.281}_{\mathbf{0.050}}$ & $\mathbf{0.244}_{\mathbf{0.057}}$ \\
SFM & DANet & $0.484_{0.036}$ & $0.357_{0.025}$ & $0.323_{0.038}$ & $0.262_{0.015}$ & $0.214_{0.021}$ \\
Zoomer & TGCM & $\mathbf{0.572}_{\mathbf{0.049}}$ & $0.392_{0.047}$ & $\mathbf{0.322}_{\mathbf{0.043}}$ & $0.234_{0.017}$ & $0.226_{0.022}$ \\
Zoomer & DANet & $0.550_{0.064}$ & $\mathbf{0.449}_{\mathbf{0.035}}$ & $0.306_{0.038}$ & $\mathbf{0.237}_{\mathbf{0.021}}$ & $\mathbf{0.253}_{\mathbf{0.061}}$ \\
TREC & TGCM & $\mathbf{0.605}_{\mathbf{0.051}}$ & $\mathbf{0.447}_{\mathbf{0.056}}$ & $\mathbf{0.340}_{\mathbf{0.031}}$ & $\mathbf{0.265}_{\mathbf{0.015}}$ & $\mathbf{0.234}_{\mathbf{0.013}}$ \\
TREC & DANet & $0.570_{0.090}$ & $0.375_{0.040}$ & $0.299_{0.069}$ & $0.241_{0.038}$ & $0.217_{0.040}$ \\
\midrule
\multicolumn{7}{l}{\textbf{(b) Macro-F1} $\uparrow$} \\
\midrule
Extractor & Model & Mix=2 & Mix=3 & Mix=4 & Mix=5 & Mix=6 \\
\midrule
SFM & TGCM & $\mathbf{0.419}_{\mathbf{0.039}}$ & $\mathbf{0.250}_{\mathbf{0.012}}$ & $\mathbf{0.202}_{\mathbf{0.010}}$ & $\mathbf{0.144}_{\mathbf{0.009}}$ & $\mathbf{0.114}_{\mathbf{0.008}}$ \\
SFM & DANet & $0.360_{0.038}$ & $0.225_{0.009}$ & $0.181_{0.017}$ & $0.140_{0.011}$ & $0.106_{0.007}$ \\
Zoomer & TGCM & $0.435_{0.039}$ & $0.246_{0.021}$ & $\mathbf{0.190}_{\mathbf{0.019}}$ & $\mathbf{0.144}_{\mathbf{0.010}}$ & $\mathbf{0.117}_{\mathbf{0.008}}$ \\
Zoomer & DANet & $\mathbf{0.460}_{\mathbf{0.079}}$ & $\mathbf{0.287}_{\mathbf{0.022}}$ & $0.171_{0.013}$ & $0.141_{0.013}$ & $0.114_{0.021}$ \\
TREC & TGCM & $0.412_{0.018}$ & $0.239_{0.021}$ & $0.173_{0.011}$ & $0.141_{0.009}$ & $\mathbf{0.118}_{\mathbf{0.014}}$ \\
TREC & DANet & $\mathbf{0.484}_{\mathbf{0.106}}$ & $\mathbf{0.266}_{\mathbf{0.036}}$ & $\mathbf{0.181}_{\mathbf{0.041}}$ & $\mathbf{0.141}_{\mathbf{0.019}}$ & $0.114_{0.017}$ \\
\midrule
\multicolumn{7}{l}{\textbf{(c) Cardinality error ($K$-MAE)} $\downarrow$} \\
\midrule
Extractor & Model & Mix=2 & Mix=3 & Mix=4 & Mix=5 & Mix=6 \\
\midrule
SFM & TGCM & $0.590_{0.286}$ & $1.155_{0.299}$ & $1.845_{0.089}$ & $2.195_{0.135}$ & $2.970_{0.156}$ \\
SFM & DANet & \textsc{N.S.} & \textsc{N.S.} & \textsc{N.S.} & \textsc{N.S.} & \textsc{N.S.} \\
Zoomer & TGCM & $0.265_{0.146}$ & $0.875_{0.071}$ & $1.480_{0.125}$ & $2.055_{0.255}$ & $2.500_{0.225}$ \\
Zoomer & DANet & \textsc{N.S.} & \textsc{N.S.} & \textsc{N.S.} & \textsc{N.S.} & \textsc{N.S.} \\
TREC & TGCM & $0.525_{0.242}$ & $1.185_{0.160}$ & $1.875_{0.102}$ & $2.265_{0.177}$ & $3.085_{0.135}$ \\
TREC & DANet & \textsc{N.S.} & \textsc{N.S.} & \textsc{N.S.} & \textsc{N.S.} & \textsc{N.S.} \\
\bottomrule
\end{tabular*}
\vspace{0.25em}
\begin{minipage}{0.98\textwidth}
\footnotesize $\uparrow$ indicates that a higher value is better, while $\downarrow$ indicates that a lower value is better. Bold indicates the better value between TGCM and DANet for each upstream extractor and mixture size. DANet does not provide a native cardinality estimate. Its external silhouette-based KMeans search is used only to obtain clustering assignments under unknown-$K$ inference and is not considered an intrinsic episode-count prediction; therefore, DANet $K$-MAE is reported as N.S.
\end{minipage}
\end{table*}

We evaluate TGCM in an end-to-end setting with raw audit logs, imperfect upstream log-to-technique extraction, and unknown episode cardinality. We construct \textsc{CAPTure}, a CALDERA-collected benchmark covering single-host and multi-host environments with $K=2$ to $6$. It contains 200 scenarios and approximately 855 million events, including 852.4 million benign and 2.65 million malicious events generated from 25 ATT\&CK-aligned profiles. To our knowledge, \textsc{CAPTure} is the first public benchmark combining end-to-end audit logs, upstream ATT\&CK extraction, interleaved intrusions, and occurrence-level demixing labels under unknown cardinality. 
Its aggregate kill-chain coverage is summarized in Table~\ref{tab:kill_chain_mapping}, with complete mappings, collection details, and statistics provided in Appendices~\ref{Appendix:CAPTure} and~\ref{app:capture_dataset_statistics}. Episode slots are assigned by the first observed occurrence of each annotated episode.

We independently apply SFM~\cite{huang2025cascade}, Zoomer~\cite{qiu2025zoomer}, and TREC~\cite{lv2024trec} to extract ATT\&CK technique sequences, which are processed by TGCM and DANet without fine-tuning or access to the true $K$. TGCM uses Up-to-6 decoding ($K_{\max}=6$) and estimates occupied slots, whereas DANet does not intrinsically estimate cardinality.

Table~\ref{tab:capture_upstream} reports pooled single-host and multi-host results, with complete per-extractor results in Appendix Table~\ref{tab:capture_upstream_detail}. TGCM achieves higher occurrence-level assignment accuracy than DANet across most settings, with the clearest gains on SFM and TREC. Macro-F1 varies more across extractors, indicating greater sensitivity to upstream extraction characteristics.

Accuracy remains the primary unknown-$K$ metric because it directly measures occurrence-level assignment, while Macro-F1, NMI, and FMI serve as complementary partition diagnostics. Table~\ref{tab:capture_upstream} also reports the MAE between inferred $\hat{K}$ and ground-truth $K$. TGCM estimates $\hat{K}$ from occupied slots, whereas DANet requires the cluster count externally. TGCM is therefore positioned primarily as an occurrence-level demixing module, with cardinality estimation as a complementary capability under Up-to-6 decoding.
\section{Related Work}
\label{sec:related_work}

Our problem lies at the intersection of attack investigation, process mining, blind source separation, and generative modeling. We summarize the closest research threads and distinguish them from TGCM.

\noindent
\textbf{Attack investigation and process mining.}
Provenance-based systems~\cite{hossain2017sleuth,alsaheel2021atlas,hassan2019nodoze} and ATT\&CK-oriented pipelines~\cite{lv2024trec,jia2024magic,li2022attackg,yang2022flexible} detect suspicious behaviors from raw telemetry and abstract them into attack evidence. Event-case correlation methods similarly associate events with latent cases, but often depend on process models or attribute constraints~\cite{pourmirza2017correlation,bayomie2019probabilistic,bayomie2023event,lichtenstein2021attribute,pegoraro2022resolving}. TGCM instead addresses a downstream problem: disentangling temporally interleaved ATT\&CK technique sequences after symbolic abstraction, under unknown mixture cardinality and ambiguous episode assignments. It is therefore complementary to upstream detection and provenance analysis.

\noindent
\textbf{Blind source separation.}
Classical BSS methods, including ICA and NMF, assume linearity or statistical independence~\cite{jutten1991blind,hyvarinen2000independent,lee1999learning}, which are unsuitable for discrete and order-sensitive technique sequences. Radar de-interleaving separates signals primarily through timing information~\cite{chen2025radar}, whereas TGCM combines sequence structure with semantic regularization for non-additive APT technique interleaving.

\noindent
\textbf{Diffusion, consistency models, and semantic priors.}
Diffusion and consistency models provide general denoising frameworks~\cite{austin2021structured,li2022diffusion,song2023consistency}. TGCM treats interleaving as a forward corruption process and adopts a consistency-inspired objective for single-step inverse demixing, rather than iterative generation. FASTopic-derived priors~\cite{blei2003latent,grootendorst2022bertopic,wu2024fastopic} are further used as training-time semantic regularizers instead of static features.

\noindent
\textbf{Conversation disentanglement.}
Conversation disentanglement also separates interleaved sources~\cite{shen2006thread,kummerfeld2019large}, but typically predicts clusters or reply graphs. TGCM additionally targets canonical, order-preserving campaign episode recovery.
\section{Limitations and Conclusion}

We formulate the UKISD problem, which recovers latent campaign episodes from interleaved ATT\&CK technique sequences under overlapping execution, technique reuse, post-abstraction noise, and unknown mixture cardinality. We propose TGCM, a consistency-inspired one-step framework that jointly performs occurrence-level assignment and sequence reconstruction through topic-guided semantic regularization and embedding-space self-consistency.

Experiments on synthetic mixtures, public benchmarks, DARPA engagement traces, and the end-to-end \textsc{CAPTure} benchmark show that TGCM improves occurrence-level assignment under heavy interleaving, repeated techniques, symbolic extraction errors, and budgeted unknown-$K$ inference. It also generalizes to unseen campaigns without fine-tuning and integrates with practical ATT\&CK extraction pipelines. TGCM therefore serves as a post-abstraction decision-support module that separates mixed technique streams into episode-consistent sequences for downstream SOC investigation.

TGCM is primarily designed for occurrence-level demixing, while cardinality estimation remains complementary under the current Up-to-6 decoding budget. The present framework assumes reasonably accurate ATT\&CK abstraction and does not explicitly address adaptive mimicry or chaff generation. Future work includes unseen techniques, larger decoding budgets, and validation in long-running enterprise deployments.

\clearpage
\makeatletter
\if@twocolumn\else\twocolumn\fi
\makeatother

\bibliographystyle{IEEEtran}
\bibliography{ref,appendix_ref}

\clearpage
\section*{Generative AI Usage}
Generative AI tools were used for language polishing, grammar checking, clarity improvements, and supervised drafting or checking of auxiliary manuscript artifacts such as formatting scripts and table-generation utilities. All technical claims, experimental results, citations, artifact descriptions, and conclusions were verified, curated, and approved by the authors. No generative AI tool was used to fabricate results, create unsupported references, or replace author judgment in scientific analysis.

\clearpage
\appendices
\onecolumn
\section{Training and Validation Data Statistics}
\label{app:length_stats}

Table~\ref{tab:length_stats} reports statistics for $K\in\{2,\dots,6\}$. The per-source lengths remain stable (mean $\approx 10.8$), indicating comparable single-campaign trace granularity, while the interleaved-mixed sequence lengths increase with $K$ ($\approx 21.6 \rightarrow 34.6$) because the observation contains more sources (approximately additive in length). Note that \emph{Count} in the source block aggregates total sampled sources, whereas in the mixed block it denotes the number of mixed instances ($N$).
\begin{table}[h]
\centering
\caption{Length statistics of (i) single-APT source technique sequences before mixing and (ii) mixed technique sequences after interleaving ($K\in\{2,\dots,6\}$). In the source block, \textit{Count} is the total number of sampled source campaigns aggregated over all mixed instances, i.e., $\sum_{i=1}^{n} K_i$ (not necessarily $K\times n$ since $K_i$ may be smaller than $K$). In the mixed block, \textit{Count} denotes the number of mixed instances ($n$).}
\setlength{\tabcolsep}{4pt}
\renewcommand{\arraystretch}{1.05}

\begin{tabular*}{\columnwidth}{@{}@{\extracolsep{\fill}}c c r c c c@{}}
\toprule
$K$ & Split & Count & Min & Mean$\pm$Std & Max \\
\midrule
\multicolumn{6}{@{}l@{}}{\textbf{Source single-APT length}} \\
\midrule
\multirow{2}{*}{2} & Train & 204,800 & 5.00 & 10.82 $\pm$ 1.42 & 13.00 \\
                   & Valid & 204,800 & 5.00 & 10.84 $\pm$ 1.40 & 13.00 \\
\multirow{2}{*}{3} & Train & 230,551 & 5.00 & 10.82 $\pm$ 1.41 & 13.00 \\
                   & Valid & 230,450 & 5.00 & 10.81 $\pm$ 1.42 & 13.00 \\
\multirow{2}{*}{4} & Train & 261,898 & 5.00 & 10.82 $\pm$ 1.41 & 13.00 \\
                   & Valid & 261,622 & 5.00 & 10.82 $\pm$ 1.43 & 13.00 \\
\multirow{2}{*}{5} & Train & 293,847 & 5.00 & 10.82 $\pm$ 1.42 & 13.00 \\
                   & Valid & 294,387 & 5.00 & 10.82 $\pm$ 1.41 & 13.00 \\
\multirow{2}{*}{6} & Train & 326,992 & 5.00 & 10.81 $\pm$ 1.42 & 13.00 \\
                   & Valid & 327,597 & 5.00 & 10.82 $\pm$ 1.42 & 13.00 \\
\midrule
\multicolumn{6}{@{}l@{}}{\textbf{Mixed APT length}} \\
\midrule
\multirow{2}{*}{2} & Train & 102,400 & 11.00 & 21.64 $\pm$ 2.00 & 26.00 \\
                   & Valid & 102,400 & 12.00 & 21.68 $\pm$ 1.98 & 26.00 \\
\multirow{2}{*}{3} & Train & 102,400 & 12.00 & 24.37 $\pm$ 5.14 & 39.00 \\
                   & Valid & 102,400 & 11.00 & 24.33 $\pm$ 5.14 & 39.00 \\
\multirow{2}{*}{4} & Train & 102,400 & 11.00 & 27.68 $\pm$ 7.76 & 52.00 \\
                   & Valid & 102,400 & 12.00 & 27.63 $\pm$ 7.74 & 52.00 \\
\multirow{2}{*}{5} & Train & 102,400 & 10.00 & 31.05 $\pm$ 10.31 & 64.00 \\
                   & Valid & 102,400 & 10.00 & 31.12 $\pm$ 10.33 & 64.00 \\
\multirow{2}{*}{6} & Train & 102,400 & 11.00 & 34.53 $\pm$ 12.86 & 75.00 \\
                   & Valid & 102,400 & 11.00 & 34.61 $\pm$ 12.89 & 76.00 \\
\bottomrule
\end{tabular*}
\label{tab:length_stats}
\end{table}

\clearpage
\section{Notation Table}
\label{app:notation_table_appendix}
\begin{table}[h!]
\centering
\caption{Summary of Notation}
\label{tab:notation}
\scriptsize
\setlength{\tabcolsep}{4pt}
\renewcommand{\arraystretch}{1.12}
\begin{tabular}{p{0.30\linewidth}|p{0.64\linewidth}}
\toprule
\textbf{Symbol} & \textbf{Description} \\
\midrule

$x_{\mathrm{obs}}$ & Observed mixed technique sequence at deployment. \\
$\mathcal{V}$ & ATT\&CK technique vocabulary; $x\in\mathcal{V}^n$. \\
$w_i$ & The $i$-th observed technique occurrence. \\
$n$ & Sequence length. \\
$x_0$ & Canonical clean sequence formed by concatenating $K$ episode-level technique sequences. \\
$\hat{x}_0$ & Restored clean sequence predicted by TGCM. \\
$x_t$ & Forward-trajectory state at index $t\in\{1,\ldots,T\}$; $x_1=x_0$ is the clean anchor and nonzero mixing begins at $t=2$. \\
$T$ & Maximum forward-trajectory index. \\
$y_z$ & Fixed occurrence-level reference episode labels used for supervised training. \\
$\hat{y}$ & Predicted technique-level episode labels. \\
$K,\hat{K}$ & True and estimated numbers of latent campaign episodes. \\
$K_{\max}$ & Maximum episode-slot budget used in unknown-$K$ inference. \\
$b,k$ & Batch index and episode-slot index. \\

\midrule
$q(x_t\mid x_{t-1})$ & Forward Markov kernel for $t\in\{2,\ldots,T\}$. \\
$q_{1\rightarrow t}(x_t\mid x_1)$ & Transition from the indexed clean anchor $x_1=x_0$ to $x_t$. \\
$t_1,t_2$ & Two indices from the same forward trajectory. \\
$\alpha(t)$ & Mixing intensity schedule. \\
$N_{\mathrm{pairs}}(t)$ & Number of activated interacting episode pairs. \\
$S(t)$ & Number of blocks per activated pair. \\
$L(t)$ & Expected block length. \\
$\mathcal{K}$ & Set of active episodes in a mixture. \\
$\mathcal{P}_t$ & Set of interacting episode pairs at timestep $t$. \\
$\mathrm{Seg}(\cdot)$ & Segmentation operator. \\
$\mathrm{Interleave}(\cdot,\cdot)$ & Blockwise interleaving operator. \\
$\Vert$ & Concatenation operator. \\

\midrule
$K_{\mathrm{topic}}$ & Number of latent topics. \\
$E_{\mathrm{topic}}$ & Text encoder used by FASTopic. \\
$\Phi$ & FASTopic topic--word logit matrix. \\
$\theta(d)$ & Soft topic allocation for document $d$. \\
$M$ & Word-to-technique mapping matrix used to project FASTopic topics into the ATT\&CK technique vocabulary. \\
$\omega(d)$ & Topic-induced technique prior for document $d$. \\
$\ell_{\mathrm{rec}},\ell_{\mathrm{fused}}$ & Reconstruction and topic-fused logits. \\
$\lambda$ & Topic-prior logit-fusion strength. \\
$e_t$ & Learned timestep embedding used by the AdaLN-style encoder-input modulation. \\

\midrule
$f_\theta$ & Consistency-inspired model parameterized by $\theta$. \\
$d$ & Embedding dimension. \\
$W_{\mathrm{in}}$ & Input technique embedding table. \\
$E_t$ & Technique embeddings of $x_t$, i.e., $E_t=\mathrm{Embed}(x_t;W_{\mathrm{in}})$. \\
$H_t$ & Timestep-modulated Transformer encoder input. \\
$\mathrm{LN}(\cdot)$ & Layer Normalization. \\
$\gamma,b$ & Scale and shift projections derived from the timestep embedding $e_t$. \\
$\hat{r}_\theta$ & Predicted embedding-space residual. \\
$a(t)$ & Time-dependent residual gate. \\
$\hat{E}_0$ & Estimated clean embedding sequence. \\
$W_{\mathrm{out}}$ & Output vocabulary projection matrix. \\

\midrule
$\mathcal{L}_{\mathrm{consist}}$ & Embedding-space consistency loss. \\
$\mathcal{L}_{\mathrm{rec}}$ & Technique reconstruction cross-entropy computed once from topic-fused logits. \\
$\mathcal{L}_{\mathrm{assign}}$ & Occurrence-level episode-assignment loss. \\
$\mathcal{L}_{\mathrm{bow}}$ & Bag-of-techniques matching loss. \\
$\mathcal{L}_{\mathrm{align}}$ & Topic-space alignment loss. \\
$\mathcal{L}_{\mathrm{topic\_reg}}$ & $w_{\mathrm{topic\_bow}}\mathcal{L}_{\mathrm{bow}}+w_{\mathrm{topic\_topic}}\mathcal{L}_{\mathrm{align}}$. \\
$\mathcal{L}$ & Overall training objective; reconstruction cross-entropy appears only through $\mathcal{L}_{\mathrm{rec}}$. \\
$w_{\mathrm{consist}},w_{\mathrm{ce}},w_{\mathrm{aptid}},w_{\mathrm{topic\_bow}},w_{\mathrm{topic\_topic}}$ & Scalar weights of the corresponding loss terms. \\

\midrule
$\mathcal{A}$ & Set of non-residual episode IDs for evaluation. \\
$\mathrm{Acc}$ & Mean per-sequence occurrence-level episode-assignment accuracy under fixed episode-slot labels. \\
$\mathrm{FMI},\mathrm{NMI}$ & Partition-quality metrics. \\
$P,R,F_1$ & Macro-averaged precision, recall, and Macro-F1. \\

\bottomrule
\end{tabular}
\end{table}

\clearpage
\section{Implementation and Hyperparameter Details}
\label{app:implementation_details}

To make the experimental configuration auditable, we report the main numeric settings used in the released TGCM training and FASTopic scripts. TGCM and all neural baselines are trained on the same SAGA-generated forward-mixed training split unless otherwise stated, and evaluation datasets are never used for fine-tuning in the unseen-dataset transfer experiments. Table~\ref{tab:implementation_details} reports the principal hyperparameters used for the main TGCM results; Table~\ref{tab:fastopic_details} reports the corresponding FASTopic topic-model settings.

\begin{table}[h]
\centering
\caption{Principal TGCM hyperparameters used for the main reported runs.}
\label{tab:implementation_details}
\scriptsize
\setlength{\tabcolsep}{3pt}
\renewcommand{\arraystretch}{1.08}
\begin{tabular}{p{0.34\linewidth}|p{0.58\linewidth}}
\toprule
\textbf{Component} & \textbf{Setting} \\
\midrule
Transformer backbone & 8 layers, 8 attention heads, $d_{\mathrm{model}}=64$, feed-forward dimension 128, dropout 0.15. \\
Conditioning & AdaLN-style encoder-input modulation derived from the timestep embedding; topic guidance is applied through reconstruction-logit fusion and auxiliary topic losses. \\
Optimization & AdamW (fused), learning rate $2\times10^{-3}$, weight decay 0.015. \\
Scheduler & CosineAnnealingWarmRestarts with $T_0=10$, $T_{\mathrm{mult}}=1$, and $\eta_{\min}=10^{-4}$. \\
Batch/training & Batch size 1024, gradient accumulation 4, 1000 epochs, gradient clipping 1.0, label smoothing 0.05, bfloat16 AMP. \\
Forward process & $T=10$ indexed trajectory states; $t=1$ is the clean anchor and nonzero mixing begins at $t=2$. \\
Decoding budget & $K_{\max}=6$ for the main unknown-$K$ setting; training samples $K\in\{2,\ldots,6\}$. \\
Loss weights & $w_{\mathrm{consist}}=1.0$, $w_{\mathrm{ce}}=1.0$, $w_{\mathrm{aptid}}=1.0$, $w_{\mathrm{topic\_bow}}=1.0$, $w_{\mathrm{topic\_topic}}=1.0$. \\
Loss interpretation & $w_{\mathrm{ce}}$ weights the single reconstruction cross-entropy; $w_{\mathrm{topic\_bow}}$ and $w_{\mathrm{topic\_topic}}$ weight separate auxiliary topic losses. The fusion coefficient $\lambda$ modifies reconstruction logits and is not an additional loss term. \\
Topic fusion & Logit-fusion weight $\lambda=0.2$. \\
Main topic setting & $K_{\mathrm{topic}}=3$ with SecBERT as the default topic encoder; other encoders are used only in ablations. \\
\bottomrule
\end{tabular}
\end{table}

\begin{table}[h]
\centering
\caption{FASTopic settings used to construct the topic prior.}
\label{tab:fastopic_details}
\scriptsize
\setlength{\tabcolsep}{3pt}
\renewcommand{\arraystretch}{1.08}
\begin{tabular}{p{0.34\linewidth}|p{0.58\linewidth}}
\toprule
\textbf{Component} & \textbf{Setting} \\
\midrule
Encoder & Frozen sentence-transformer encoder; BERT-family encoders use 768-d embeddings, while all-MiniLM-L6-v2 uses 384-d embeddings. \\
Training & 2000 epochs, Adam optimizer, learning rate $2\times10^{-3}$, full-batch training. \\
Topics & Main $K_{\mathrm{topic}}=3$. \\
Topic words & 20 top words per topic, vocabulary size 10000, minimum term frequency 1. \\
Regularization & DT$_\alpha=3.0$, TW$_\alpha=2.0$, topic temperature 1.0. \\
Projection & Document projection MLP with hidden sizes $(256,128,256)$, GELU, LayerNorm, dropout 0.1. \\
Alignment/distillation & $\lambda_{\mathrm{align}}=0.05$, $\lambda_{\mathrm{kd\_topic}}=0.5$, $\lambda_{\mathrm{kd\_word}}=0.5$, $\tau_{\mathrm{kd}}=1.5$. \\
Split & 80/20 train/test split with seed 42. \\
\bottomrule
\end{tabular}
\end{table}

This appendix is intended to prevent ambiguity about which parameters affect the reported numbers. The default script value of 300 epochs is overridden by the formal training script to 1000 epochs. Zero-shot, DARPA TC-E5, and \textsc{CAPTure} evaluations should be reproduced with source-campaign topic lookup disabled and neutral topic conditioning enabled.

\clearpage
\section{Structure-Aware Forward Mixing: Full Definitions}
\label{app:forward_mixed}

This appendix provides the exact schedules and operator definitions for the structure-aware forward mixing kernel $q(x_t\mid x_{t-1})$ used in Section~\ref{sec:datasets_mixing}.
The implementation indexes trajectory states by $t\in\{1,\dots,T\}$, sets $x_1=x_0$ as the clean anchor, uses $t_0=1$ as the onset index, and denotes the saturation point by $t^\star$. Nonzero mixing begins at $t=2$.

\subsection{Interaction Schedule}

We define a normalized progress variable $u(t)\in[0,1]$ and a cosine-ramp mixing intensity $\alpha(t)\in[0,1]$:
\begin{align}
u(t) &= \min\!\Bigl(1,\max\!\bigl(0,\tfrac{t-t_0}{t^\star-t_0}\bigr)\Bigr), \\
\alpha(t) &= \tfrac{1}{2}\bigl(1-\cos(\pi u(t))\bigr).
\end{align}
The mixing intensity controls: the number of activated disjoint pairs $N_{\text{pairs}}(t)$, the number of blocks per pair $S(t)$, and the expected block length $L(t)$:
\begin{align}
N_{\text{pairs}}(t) &= \mathrm{round}\!\bigl(N_{\min}+\alpha(t)(N_{\max}-N_{\min})\bigr), \\
S(t) &= \mathrm{round}\!\bigl(S_{\min}+\alpha(t)(S_{\max}-S_{\min})\bigr), \\
L(t) &= L_0+\alpha(t)(L_1-L_0).
\end{align}
Here $N_{\text{pairs}}(t)\in[N_{\min},N_{\max}]$ is the number of activated interacting pairs at step $t$,
$S(t)\in[S_{\min},S_{\max}]$ is the number of blocks per activated pair,
and $L(t)$ is the expected block length interpolated from $L_0$ to $L_1$ (larger $t$ yields shorter blocks when $L_1<L_0$).

\subsection{Granularity Schedule (Segmentation Operator)}

For any sequence $A$, the segmentation operator
\begin{equation}
(A_1,\dots,A_{S(t)})=\mathrm{Seg}\!\bigl(A;S(t),L(t)\bigr)
\end{equation}
partitions $A$ into $S(t)$ \emph{consecutive} blocks that preserve order:
\begin{equation}
A = A_1 \Vert A_2 \Vert \cdots \Vert A_{S(t)}.
\end{equation}
We require $\mathrm{Seg}(\cdot)$ to satisfy:
(i) blocks are contiguous and order-preserving,
(ii) the concatenation recovers the original sequence,
and (iii) the expected block length is controlled by $L(t)$ (i.e., $\mathbb{E}[|A_i|]\approx L(t)$),
so increasing $t$ induces finer segmentation (shorter blocks).

The same segmentation is applied to the paired sequence $B$:
\begin{equation}
(B_1,\dots,B_{S(t)})=\mathrm{Seg}\!\bigl(B;S(t),L(t)\bigr).
\end{equation}

\subsection{Mixed Operator (Pair Sampling and Interleaving)}

Let $\mathcal{K}$ be the set of active campaigns at step $t$. We sample a set of \emph{disjoint} interacting pairs
\begin{equation}
\mathcal{P}_t \subseteq \mathcal{K}\times \mathcal{K},\qquad |\mathcal{P}_t|=N_{\text{pairs}}(t),
\end{equation}
where disjoint means each campaign participates in at most one pair in $\mathcal{P}_t$ at a given $t$.

Given segmented blocks of a selected pair $(A,B)\in\mathcal{P}_t$, we define the interleaving operator:
\begin{equation}
\mathrm{Interleave}(A,B)=A_1\Vert B_1\Vert \cdots \Vert A_{S(t)}\Vert B_{S(t)}.
\end{equation}
This operator preserves within-block order while altering the global temporal arrangement across campaigns.

For each nonzero mixing step $t\in\{2,\ldots,T\}$, we apply $\mathrm{Seg}$ and $\mathrm{Interleave}$ to every $(A,B)\in\mathcal{P}_t$ to produce updated sequences,
while campaigns not in $\mathcal{P}_t$ remain unchanged. The resulting global sequence after processing all pairs defines $x_t$ and thus the Markov kernel $q(x_t\mid x_{t-1})$.

\subsection{Label Propagation (Technique-to-Campaign)}

Recall the technique-level campaign labels $y_0\in\{0,\dots,K\}^n$ defined on the clean concatenation $x_0$, with $(x_1,y_1)=(x_0,y_0)$ in the indexed trajectory.
Since the forward operator reorders techniques via block interleaving, we propagate labels by applying the \emph{same block-level permutation} to $y$ as is applied to $x$ at each step:
if $x_t$ is obtained from $x_{t-1}$ by concatenating a sequence of blocks in some order, then $y_t$ is obtained by concatenating the corresponding label blocks in the identical order.
This ensures each technique in $x_t$ retains its originating campaign ID, including the explicit handling of technique reuse by allowing identical techniques to carry different IDs under different contexts.

\subsection{Trajectory Sampling for Training}

Starting from the indexed clean anchor $x_1=x_0$, we compose the one-step kernel to obtain $q_{1\rightarrow t}(x_t\mid x_1)$ and sample training pairs along the same trajectory:
\begin{equation}
(x_{t_1}, x_{t_2}, y, K, t_1, t_2), \quad x_t \sim q_{1\rightarrow t}(x_t \mid x_1),
\quad t,t_1,t_2\in\{1,\ldots,T\}.
\end{equation}
In practice, one may equivalently carry a single label sequence aligned to each sampled $x_t$ (i.e., $y_t$), since the forward process deterministically induces the same reindexing on technique and labels given the sampled schedules and pairings.

\clearpage
\section{Conceptual Information-Theoretic Interpretation}
\label{sec:infotheory}
This appendix gives a conceptual lens for the TGCM objective rather than a formal information-theoretic proof. We do not derive a variational lower bound, an identifiability guarantee, or a convergence result. Instead, the chain-rule view below explains why occurrence-level assignment, sequence reconstruction, topic regularization, and self-consistency are complementary training signals for UKISD.

A useful decomposition is to view the observation as carrying information about both the episode assignment $y$ and the clean episode-ordered sequence $x_0$:
\begin{equation}
I(x_{\mathrm{obs}}; x_0, y) = I(x_{\mathrm{obs}}; y) + I(x_{\mathrm{obs}}; x_0 \mid y).
\end{equation}
Under this interpretation, the supervised assignment loss targets the occurrence-to-episode component, while reconstruction and topic regularization encourage the predicted sequence to remain compatible with ATT\&CK-level technique semantics. The consistency-inspired loss does not optimize mutual information directly; it regularizes estimates sampled from different forward-mixing states to share a common embedding-space clean-state representation.

\begin{enumerate}
    \item \textbf{Episode assignment.} $\mathcal{L}_{\mathrm{assign}}$ provides the direct task-aligned supervision signal for occurrence-level episode labels.
    \item \textbf{Sequence restoration.} $\mathcal{L}_{\mathrm{rec}}$ applies the single reconstruction cross-entropy to topic-fused logits and encourages recovery of the canonical clean sequence.
    \item \textbf{Topic regularization.} $\mathcal{L}_{\mathrm{bow}}$ and $\mathcal{L}_{\mathrm{align}}$ encourage reconstructed technique distributions to remain compatible with the FASTopic-derived technique prior and topic allocation.
    \item \textbf{Embedding-space invariance.} $\mathcal{L}_{\mathrm{consist}}$ encourages estimates from different mixing levels to agree in continuous embedding space before discrete decoding.
\end{enumerate}
Thus, the information-theoretic view should be read as an explanatory analogy for the objective design, not as evidence that the losses optimize mutual information exactly.

\clearpage
\section{Benchmark Dataset Coverage}
\label{app:benchmark_dataset_coverage}
\begin{table*}[h!]
\centering
\caption{Mapping of techniques to kill-chain tactics across zero-shot evaluation datasets.}
\setlength{\tabcolsep}{0pt}
\renewcommand{\arraystretch}{1.4}
\arrayrulecolor{black!20}
\setlength{\arrayrulewidth}{0.5pt}

\newlength{\wDataset}
\newlength{\wPhase}
\setlength{\wDataset}{1.9cm}
\setlength{\wPhase}{\dimexpr(\linewidth-\wDataset-12\arrayrulewidth)/10\relax}

\begin{tabular}{|D{\wDataset}|
P{kcInit}{\wPhase}|P{kcExec}{\wPhase}|P{kcPers}{\wPhase}|P{kcDef}{\wPhase}|P{kcCred}{\wPhase}|
P{kcDisc}{\wPhase}|P{kcLat}{\wPhase}|P{kcColl}{\wPhase}|P{kcCtwo}{\wPhase}|P{kcExf}{\wPhase}|}
\hline

Dataset &
\multicolumn{1}{c|}{\cellcolor{kcInitH}\bfseries Init.\ Acc.} &
\multicolumn{1}{c|}{\cellcolor{kcExecH}\bfseries Execution} &
\multicolumn{1}{c|}{\cellcolor{kcPersH}\bfseries Persist.} &
\multicolumn{1}{c|}{\cellcolor{kcDefH}\bfseries Def.\ Eva.} &
\multicolumn{1}{c|}{\cellcolor{kcCredH}\bfseries Cred.\ Acc.} &
\multicolumn{1}{c|}{\cellcolor{kcDiscH}\bfseries Discovery} &
\multicolumn{1}{c|}{\cellcolor{kcLatH}\bfseries Lat.\ Move.} &
\multicolumn{1}{c|}{\cellcolor{kcCollH}\bfseries Collect.} &
\multicolumn{1}{c|}{\cellcolor{kcCtwoH}\bfseries C2} &
\multicolumn{1}{c|}{\cellcolor{kcExfH}\bfseries Exfil.}
\tabularnewline\hline

ATLAS &
-- &
T1204.002\newline T1059.003\newline T1059.001 &
T1053.005 &
T1497.001\newline T1055.001\newline T1562.001 &
-- &
T1016 &
-- &
-- &
T1071.001\newline T1090.001 &
--
\tabularnewline\hline

NODLINK &
-- &
T1047 &
-- &
-- &
-- &
T1482\newline T1083\newline T1057\newline T1018\newline T1082\newline T1124\newline T1033\newline T1016\newline T1049 &
-- &
T1119 &
T1105 &
--
\tabularnewline\hline

ProvCon &
T1566.001 &
-- &
T1574.001 &
-- &
T1003 &
T1087.001 &
-- &
-- &
T1105 &
--
\tabularnewline\hline

DARPA TC-E3 &
T1566.001\newline T1566.002 &
T1059.004\newline T1204.002\newline T1059.001\newline T1059.003\newline T1047\newline T1203 &
T1543.003\newline T1053.005\newline T1176 &
T1089 &
T1003.008\newline T1555.003 &
T1057\newline T1082\newline T1016\newline T1033\newline T1049\newline T1007\newline T1135\newline T1018\newline T1046 &
-- &
T1005 &
T1105\newline T1071\newline T1071.001 &
T1041
\tabularnewline\hline

DARPA TC-E5 &
T1189 &
T1204.001\newline T1047\newline T1059\newline T1059.003 &
T1197 &
T1055.001\newline T1070.004\newline T1055 &
T1003.001\newline T1003.008 &
T1083\newline T1033\newline T1057\newline T1082\newline T1049\newline T1046\newline T1016\newline T1007\newline T1087 &
-- &
T1074\newline T1005 &
T1071.001\newline T1071\newline T1071.004\newline T1105 &
T1048\newline T1041
\tabularnewline\hline

\end{tabular}
\label{tab:kill_chain_final_v4}
\end{table*}

\clearpage
\section{Complete Zero-shot Evaluation Results}
\label{app:complete_zero_shot_results}
\begin{table*}[h]
\centering
\caption{Complete zero-shot evaluation results across all reported metrics under the fixed-slot evaluation. This appendix table expands the compact main-paper summary in Table~\ref{tab:zero_shot_master}; each cell reports TGCM / DANet as mean$_{\mathrm{std}}$. Superscript $\dagger$ marks the statistically favored cell when the corresponding 95\% confidence interval is separated from the counterpart.}
\label{tab:landscape_full_metrics}
\scriptsize
\renewcommand{\arraystretch}{1.1}
\setlength{\tabcolsep}{3.5pt}

\begin{threeparttable}
\begin{adjustbox}{width=\textwidth,center}
\begin{tabular}{l|c|c|ccccc}
\toprule
\multirow{2}{*}{\textbf{Dataset}} & \multirow{2}{*}{\textbf{Metric}} & \multirow{2}{*}{\textbf{Model}} & \multicolumn{5}{c}{\textbf{Number of Campaigns ($K$)}} \\
\cmidrule(lr){4-8}
 & & & \textbf{2} & \textbf{3} & \textbf{4} & \textbf{5} & \textbf{6} \\
\midrule
\multirow{7}{*}{\textbf{ATLAS}}
 & Acc $\uparrow$  & TGCM / DANet & $\mathbf{0.535}_{\mathbf{0.014}}^{\dagger}$ / $0.416_{0.044}$ & $\mathbf{0.371}_{\mathbf{0.011}}^{\dagger}$ / $0.294_{0.031}$ & $\mathbf{0.268}_{\mathbf{0.011}}^{\dagger}$ / $0.230_{0.012}$ & -- & -- \\
 & FMI $\uparrow$  & TGCM / DANet & $\mathbf{0.612}_{\mathbf{0.060}}$ / $0.537_{0.029}$ & $\mathbf{0.549}_{\mathbf{0.040}}^{\dagger}$ / $0.447_{0.022}$ & $\mathbf{0.476}_{\mathbf{0.026}}^{\dagger}$ / $0.376_{0.027}$ & -- & -- \\
 & NMI $\uparrow$  & TGCM / DANet & $0.089_{0.053}$ / $\mathbf{0.163}_{\mathbf{0.091}}$ & $0.183_{0.023}$ / $\mathbf{0.214}_{\mathbf{0.029}}$ & $0.170_{0.043}$ / $\mathbf{0.171}_{\mathbf{0.044}}$ & -- & -- \\
 & P $\uparrow$  & TGCM / DANet & $\mathbf{0.462}_{\mathbf{0.100}}$ / $0.394_{0.043}$ & $0.215_{0.033}$ / $\mathbf{0.228}_{\mathbf{0.022}}$ & $0.118_{0.022}$ / $\mathbf{0.148}_{\mathbf{0.011}}$ & -- & -- \\
 & R $\uparrow$  & TGCM / DANet & $\mathbf{0.534}_{\mathbf{0.014}}^{\dagger}$ / $0.373_{0.059}$ & $\mathbf{0.372}_{\mathbf{0.010}}^{\dagger}$ / $0.274_{0.036}$ & $\mathbf{0.268}_{\mathbf{0.009}}^{\dagger}$ / $0.223_{0.014}$ & -- & -- \\
 & Macro-F1 $\uparrow$  & TGCM / DANet & $\mathbf{0.448}_{\mathbf{0.065}}$ / $0.352_{0.051}$ & $\mathbf{0.251}_{\mathbf{0.024}}$ / $0.228_{0.029}$ & $0.150_{0.024}$ / $\mathbf{0.162}_{\mathbf{0.012}}$ & -- & -- \\
 & Time $\downarrow$  & TGCM / DANet & $0.010_{0.000}$ / $\mathbf{0.002}_{\mathbf{0.002}}^{\dagger}$ & $0.009_{0.001}$ / $\mathbf{0.001}_{\mathbf{0.001}}^{\dagger}$ & $0.010_{0.001}$ / $\mathbf{0.001}_{\mathbf{0.001}}^{\dagger}$ & -- & -- \\
\midrule
\multirow{7}{*}{\textbf{NODLINK}}
 & Acc $\uparrow$  & TGCM / DANet & $0.493_{0.010}$ / $\mathbf{0.505}_{\mathbf{0.087}}$ & $\mathbf{0.328}_{\mathbf{0.007}}^{\dagger}$ / $0.296_{0.012}$ & -- & -- & -- \\
 & FMI $\uparrow$  & TGCM / DANet & $\mathbf{0.674}_{\mathbf{0.029}}^{\dagger}$ / $0.584_{0.040}$ & $\mathbf{0.527}_{\mathbf{0.037}}$ / $0.518_{0.023}$ & -- & -- & -- \\
 & NMI $\uparrow$  & TGCM / DANet & $0.027_{0.060}$ / $\mathbf{0.224}_{\mathbf{0.024}}^{\dagger}$ & $0.113_{0.075}$ / $\mathbf{0.273}_{\mathbf{0.037}}^{\dagger}$ & -- & -- & -- \\
 & P $\uparrow$  & TGCM / DANet & $0.278_{0.069}$ / $\mathbf{0.489}_{\mathbf{0.090}}^{\dagger}$ & $0.144_{0.037}$ / $\mathbf{0.219}_{\mathbf{0.035}}$ & -- & -- & -- \\
 & R $\uparrow$  & TGCM / DANet & $\mathbf{0.500}_{\mathbf{0.000}}$ / $0.479_{0.083}$ & $\mathbf{0.335}_{\mathbf{0.006}}^{\dagger}$ / $0.290_{0.021}$ & -- & -- & -- \\
 & Macro-F1 $\uparrow$  & TGCM / DANet & $0.345_{0.038}$ / $\mathbf{0.449}_{\mathbf{0.082}}$ & $0.190_{0.029}$ / $\mathbf{0.221}_{\mathbf{0.024}}$ & -- & -- & -- \\
 & Time $\downarrow$  & TGCM / DANet & $0.010_{0.001}$ / $\mathbf{0.001}_{\mathbf{0.001}}^{\dagger}$ & $0.010_{0.001}$ / $\mathbf{0.001}_{\mathbf{0.001}}^{\dagger}$ & -- & -- & -- \\
\midrule
\multirow{7}{*}{\textbf{ProvCon}}
 & Acc $\uparrow$  & TGCM / DANet & $\mathbf{0.449}_{\mathbf{0.051}}$ / $0.381_{0.099}$ & $\mathbf{0.310}_{\mathbf{0.021}}$ / $0.268_{0.063}$ & $\mathbf{0.241}_{\mathbf{0.017}}$ / $0.216_{0.019}$ & $\mathbf{0.195}_{\mathbf{0.013}}$ / $0.187_{0.061}$ & $\mathbf{0.174}_{\mathbf{0.008}}$ / $0.161_{0.017}$ \\
 & FMI $\uparrow$  & TGCM / DANet & $0.596_{0.046}$ / $\mathbf{0.697}_{\mathbf{0.042}}$ & $0.505_{0.019}$ / $\mathbf{0.525}_{\mathbf{0.044}}$ & $0.436_{0.018}$ / $\mathbf{0.454}_{\mathbf{0.024}}$ & $0.390_{0.014}$ / $\mathbf{0.413}_{\mathbf{0.030}}$ & $0.364_{0.033}$ / $\mathbf{0.382}_{\mathbf{0.027}}$ \\
 & NMI $\uparrow$  & TGCM / DANet & $0.044_{0.021}$ / $\mathbf{0.657}_{\mathbf{0.045}}^{\dagger}$ & $0.136_{0.090}$ / $\mathbf{0.514}_{\mathbf{0.051}}^{\dagger}$ & $0.163_{0.097}$ / $\mathbf{0.466}_{\mathbf{0.030}}^{\dagger}$ & $0.216_{0.107}$ / $\mathbf{0.473}_{\mathbf{0.034}}^{\dagger}$ & $0.313_{0.075}$ / $\mathbf{0.450}_{\mathbf{0.040}}$ \\
 & P $\uparrow$  & TGCM / DANet & $0.242_{0.055}$ / $\mathbf{0.366}_{\mathbf{0.112}}$ & $0.129_{0.023}$ / $\mathbf{0.180}_{\mathbf{0.039}}$ & $0.078_{0.012}$ / $\mathbf{0.113}_{\mathbf{0.020}}$ & $0.062_{0.012}$ / $\mathbf{0.081}_{\mathbf{0.033}}$ & $0.054_{0.009}$ / $\mathbf{0.058}_{\mathbf{0.008}}$ \\
 & R $\uparrow$  & TGCM / DANet & $\mathbf{0.448}_{\mathbf{0.064}}$ / $0.359_{0.100}$ & $\mathbf{0.325}_{\mathbf{0.024}}$ / $0.278_{0.064}$ & $\mathbf{0.242}_{\mathbf{0.024}}$ / $0.224_{0.023}$ & $\mathbf{0.203}_{\mathbf{0.014}}$ / $0.199_{0.048}$ & $\mathbf{0.177}_{\mathbf{0.013}}$ / $0.161_{0.017}$ \\
 & Macro-F1 $\uparrow$  & TGCM / DANet & $0.298_{0.052}$ / $\mathbf{0.347}_{\mathbf{0.102}}$ & $0.172_{0.020}$ / $\mathbf{0.202}_{\mathbf{0.048}}$ & $0.110_{0.014}$ / $\mathbf{0.138}_{\mathbf{0.015}}$ & $0.086_{0.011}$ / $\mathbf{0.107}_{\mathbf{0.037}}$ & $0.077_{0.011}$ / $\mathbf{0.079}_{\mathbf{0.010}}$ \\
 & Time $\downarrow$  & TGCM / DANet & $0.009_{0.001}$ / $\mathbf{0.001}_{\mathbf{0.001}}^{\dagger}$ & $0.010_{0.001}$ / $\mathbf{0.001}_{\mathbf{0.001}}^{\dagger}$ & $0.011_{0.001}$ / $\mathbf{0.001}_{\mathbf{0.001}}^{\dagger}$ & $0.010_{0.001}$ / $\mathbf{0.001}_{\mathbf{0.001}}^{\dagger}$ & $0.010_{0.001}$ / $\mathbf{0.001}_{\mathbf{0.001}}^{\dagger}$ \\
\midrule
\multirow{7}{*}{\textbf{DARPA TC-E3}}
 & Acc $\uparrow$  & TGCM / DANet & $\mathbf{0.504}_{\mathbf{0.021}}$ / $0.425_{0.072}$ & $\mathbf{0.347}_{\mathbf{0.019}}$ / $0.306_{0.029}$ & $\mathbf{0.256}_{\mathbf{0.022}}$ / $0.228_{0.027}$ & $\mathbf{0.201}_{\mathbf{0.017}}$ / $0.199_{0.032}$ & $\mathbf{0.165}_{\mathbf{0.009}}$ / $0.163_{0.014}$ \\
 & FMI $\uparrow$  & TGCM / DANet & $\mathbf{0.654}_{\mathbf{0.031}}$ / $0.523_{0.081}$ & $\mathbf{0.557}_{\mathbf{0.033}}$ / $0.482_{0.079}$ & $\mathbf{0.500}_{\mathbf{0.022}}$ / $0.486_{0.056}$ & $\mathbf{0.470}_{\mathbf{0.017}}$ / $0.469_{0.036}$ & $\mathbf{0.482}_{\mathbf{0.031}}$ / $0.464_{0.060}$ \\
 & NMI $\uparrow$  & TGCM / DANet & $0.046_{0.021}$ / $\mathbf{0.353}_{\mathbf{0.076}}^{\dagger}$ & $0.102_{0.038}$ / $\mathbf{0.343}_{\mathbf{0.080}}^{\dagger}$ & $0.127_{0.025}$ / $\mathbf{0.367}_{\mathbf{0.057}}^{\dagger}$ & $0.161_{0.030}$ / $\mathbf{0.373}_{\mathbf{0.039}}^{\dagger}$ & $0.252_{0.052}$ / $\mathbf{0.381}_{\mathbf{0.046}}^{\dagger}$ \\
 & P $\uparrow$  & TGCM / DANet & $0.312_{0.015}$ / $\mathbf{0.403}_{\mathbf{0.086}}$ & $0.161_{0.023}$ / $\mathbf{0.235}_{\mathbf{0.039}}$ & $0.102_{0.016}$ / $\mathbf{0.137}_{\mathbf{0.019}}$ & $0.070_{0.007}$ / $\mathbf{0.110}_{\mathbf{0.025}}^{\dagger}$ & $0.055_{0.005}$ / $\mathbf{0.087}_{\mathbf{0.007}}^{\dagger}$ \\
 & R $\uparrow$  & TGCM / DANet & $\mathbf{0.511}_{\mathbf{0.016}}^{\dagger}$ / $0.402_{0.067}$ & $\mathbf{0.347}_{\mathbf{0.008}}$ / $0.299_{0.039}$ & $\mathbf{0.257}_{\mathbf{0.016}}^{\dagger}$ / $0.214_{0.017}$ & $\mathbf{0.202}_{\mathbf{0.011}}$ / $0.187_{0.034}$ & $\mathbf{0.169}_{\mathbf{0.008}}$ / $0.161_{0.009}$ \\
 & Macro-F1 $\uparrow$  & TGCM / DANet & $0.357_{0.015}$ / $\mathbf{0.379}_{\mathbf{0.076}}$ & $0.199_{0.018}$ / $\mathbf{0.239}_{\mathbf{0.036}}$ & $0.125_{0.013}$ / $\mathbf{0.150}_{\mathbf{0.016}}$ & $0.089_{0.008}$ / $\mathbf{0.122}_{\mathbf{0.024}}$ & $0.071_{0.007}$ / $\mathbf{0.098}_{\mathbf{0.006}}^{\dagger}$ \\
 & Time $\downarrow$  & TGCM / DANet & $0.009_{0.001}$ / $\mathbf{0.001}_{\mathbf{0.001}}^{\dagger}$ & $0.010_{0.001}$ / $\mathbf{0.001}_{\mathbf{0.001}}^{\dagger}$ & $0.010_{0.000}$ / $\mathbf{0.001}_{\mathbf{0.001}}^{\dagger}$ & $0.010_{0.001}$ / $\mathbf{0.001}_{\mathbf{0.001}}^{\dagger}$ & $0.010_{0.000}$ / $\mathbf{0.001}_{\mathbf{0.001}}^{\dagger}$ \\
\midrule
\multirow{7}{*}{\textbf{DARPA TC-E5}}
 & Acc $\uparrow$  & TGCM / DANet & $\mathbf{0.536}_{\mathbf{0.095}}$ / $0.450_{0.147}$ & -- & $\mathbf{0.506}_{\mathbf{0.090}}$ / $0.440_{0.133}$ & -- & -- \\
 & FMI $\uparrow$  & TGCM / DANet & $0.586_{0.042}$ / $\mathbf{0.604}_{\mathbf{0.090}}$ & -- & $0.561_{0.038}$ / $\mathbf{0.614}_{\mathbf{0.073}}$ & -- & -- \\
 & NMI $\uparrow$  & TGCM / DANet & $0.145_{0.014}$ / $\mathbf{0.285}_{\mathbf{0.097}}^{\dagger}$ & -- & $0.168_{0.022}$ / $\mathbf{0.332}_{\mathbf{0.093}}^{\dagger}$ & -- & -- \\
 & P $\uparrow$  & TGCM / DANet & $\mathbf{0.556}_{\mathbf{0.129}}$ / $0.402_{0.157}$ & -- & $\mathbf{0.521}_{\mathbf{0.121}}$ / $0.385_{0.154}$ & -- & -- \\
 & R $\uparrow$  & TGCM / DANet & $\mathbf{0.508}_{\mathbf{0.146}}$ / $0.369_{0.171}$ & -- & $\mathbf{0.487}_{\mathbf{0.139}}$ / $0.378_{0.146}$ & -- & -- \\
 & Macro-F1 $\uparrow$  & TGCM / DANet & $\mathbf{0.458}_{\mathbf{0.122}}$ / $0.365_{0.157}$ & -- & $\mathbf{0.434}_{\mathbf{0.115}}$ / $0.362_{0.143}$ & -- & -- \\
 & Time $\downarrow$  & TGCM / DANet & $0.012_{0.004}$ / $\mathbf{0.002}_{\mathbf{0.002}}^{\dagger}$ & -- & $0.010_{0.001}$ / $\mathbf{0.001}_{\mathbf{0.001}}^{\dagger}$ & -- & -- \\
\bottomrule
\end{tabular}
\end{adjustbox}
\begin{tablenotes}[flushleft]
\footnotesize
\item Bold indicates the better mean between TGCM and DANet for the same dataset, $K$, and metric. Time is measured as wall-clock inference time; lower is \\ better. FMI and NMI are retained as secondary partition-structure diagnostics.
\end{tablenotes}
\end{threeparttable}
\end{table*}

\clearpage
\section{\textsc{CAPTure} Dataset Construction and Labeling}
\label{Appendix:CAPTure}
We construct \textsc{CAPTure}, a host-level audit dataset for studying interleaved mixed-intrusion behaviors under controlled end-to-end emulation noise. The dataset is built with MITRE CALDERA v5.3.0~\citeapp{caldera} for attack orchestration and Process Monitor (Procmon)~\citeapp{procmon} for host-side event collection. \textsc{CAPTure} includes both single-host and multi-host emulation settings in VMware, with Sandcat agents deployed on the participating Windows hosts. It should therefore be interpreted as a controlled emulation benchmark rather than a full enterprise SOC deployment trace.

Our attack pool is curated from locally available CALDERA adversaries and abilities, using MITRE ATT\&CK group pages as the primary reference and retaining only techniques that can be executed by the available CALDERA abilities. The resulting pool contains 25 ATT\&CK-aligned CALDERA profiles: 19 APT profiles and 6 APT-like ransomware-style profiles that exhibit multi-stage intrusion TTPs. Table~\ref{tab:kill_chain_capture_apt} summarizes the executable attack-profile-to-tactic mapping, showing that different profiles contribute distinct technique subsets across tactic categories and thus provide the behavioral diversity needed to construct interleaved mixed-intrusion traces.

To generate mixed traces, multiple CALDERA operations are allowed to run concurrently, so techniques from different profiles may overlap within a single execution window. Procmon is started before attack execution and stopped after all abilities have completed, producing complete host traces from the participating hosts for each run. Each trace is then exported to CSV, copied out of the virtual machine environment, and preserved before restoring the snapshot for the next run.

Benign background events are produced by the Windows hosts and supporting emulation environment during each run, including routine OS, service, registry, file-system, and process-management activity observed by Procmon. We do not inject benign labels by down-sampling or post-hoc balancing; instead, benign events are the non-attack events that remain after conservative CALDERA-context matching. This design yields the strong benign/malicious imbalance reported in the corpus statistics and reflects the post-abstraction challenge faced by upstream technique extractors.

For multi-host runs, traces from participating VMs are exported separately and merged using their recorded timestamps after collection. Because the environment is controlled but not intended to model sub-millisecond distributed causality, we interpret the resulting order as coarse technique-level temporal order rather than precise event-level causality across hosts. TGCM consumes extracted ATT\&CK technique sequences, so the evaluation targets episode-level ordering and assignment rather than fine-grained cross-host happens-before reconstruction.

We organize the collected data by mixing level, with separate per-run statistics reported for Mix-2 through Mix-6 subsets. These tables summarize the constituent profile combinations, total event volume, trace size, benign and malicious event counts, and malicious ratios. Together, they quantify the scale and sparsity of the collected traces and provide empirical context for the labeling procedure described next.

For reproducibility, the public artifact releases the implementation, experiment configurations, normalized technique sequences, episode labels, labeling-rule templates, and evaluation scripts. The full raw Procmon traces are substantially larger and may require additional storage and sanitization checks; we therefore treat their release separately from the lightweight technique-level benchmark used for downstream demixing replication.

\subsection{Technique-Level Event Labeling}

To convert raw Procmon traces into auditable technique-level evidence, we align two sources for each run: CALDERA execution metadata and exported Procmon CSV events. CALDERA records the execution context of each ability instance, including the attributed attack profile, ability identifier, mapped ATT\&CK technique, process context, and execution time window. Procmon records the corresponding host-side event stream, including timestamps, process names, operations, accessed paths, and event details. Event labeling is therefore performed by matching Procmon events to plausible CALDERA executions under a shared execution context, rather than by relying on isolated keywords alone.

\begin{figure}[h]
    \centering
    \includegraphics[width=1.0\textwidth]{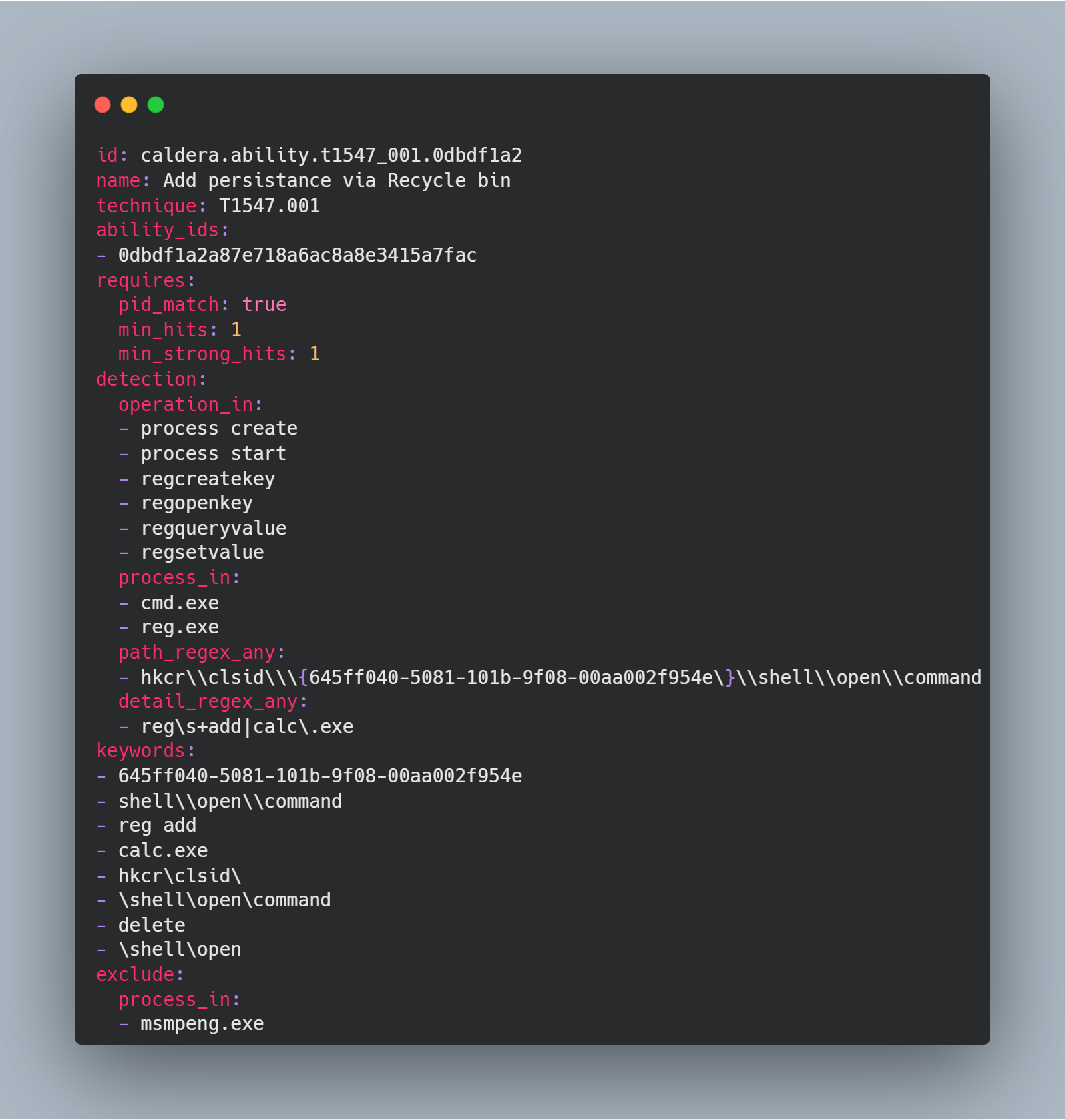}
   \caption{Example of a manually crafted labeling rule used in \textsc{CAPTure}. We write such rules for CALDERA abilities to detect their execution traces in raw audit logs and assign the corresponding ATT\&CK technique labels. Each rule specifies required conditions, event-operation types, process constraints, regex-based matching fields, keywords, and exclusions to support ability-level dataset annotation.}
    \label{fig:t1547001_yaml}
\end{figure}

The matching logic is implemented as a human-curated YAML rule base~(e.g. Figure~\ref{fig:t1547001_yaml}). In its current version, the rule base contains 198 rule files, covering 83 unique ATT\&CK techniques and 198 unique CALDERA abilities, with all rule files successfully parsed without errors, as summarized in Table~\ref{tab:procmon_rule_summary}. Coverage is intentionally uneven across techniques, reflecting the fact that some techniques are represented by only a few executable CALDERA abilities, whereas others recur across multiple abilities in our attack pool. At the technique level, the median coverage is 2 abilities per technique: 39 techniques are supported by a single ability, 15 by two abilities, 21 by three to four abilities, and 8 by at least five abilities. Representative high-coverage techniques include T1057 (10 abilities), T1016/T1018 (9 each), T1059.001/T1005 (8 each), and T1082/T1548.002/T1033 (5 each). When both ability-specific and technique-level rules are available, we prefer the former because they better capture the concrete execution semantics of a particular CALDERA ability.

Operationally, labeling proceeds in three stages. First, for each CALDERA execution instance, we generate candidate Procmon events that fall within the relevant execution context, primarily using PID consistency when available; for techniques that naturally induce cross-process effects, we allow broader process-chain conditions. Second, we apply rule-based filtering to retain only events whose operations, process names, accessed artifacts, or textual details are consistent with the expected behavior of the candidate technique, while also enforcing exclusion conditions for common non-target patterns. Third, we emit technique-level labels deterministically: events that satisfy a retained rule are assigned to the corresponding ATT\&CK technique and attack profile, whereas events that do not satisfy any retained rule are treated as non-attack background for the purpose of constructing confident technique-level labels.

This rule-based labeling is designed to be conservative and auditable, but it should not be interpreted as an independently adjudicated incident-response ground truth. The current pipeline does not generate a separate ambiguous label class: only deterministically matched rule outputs are used as confident technique-level malicious labels. Events without a retained rule match remain part of the collected raw trace statistics but are not used as ground-truth technique tokens for downstream demixing evaluation.

For each labeled event, we retain traceable evidence including the attributed attack profile, ability, technique, and rule-matching rationale, so that annotations remain auditable after labeling. Importantly, these event-level labels are not used as direct inputs to TGCM. Instead, they provide an intermediate supervision layer for constructing and evaluating upstream technique-recognition models, whose predicted technique sequences are then passed to TGCM. This design is consistent with the scope of our framework: TGCM operates after the abstraction/extraction stage, rather than directly on raw host telemetry.

\begin{table}[!h]
\centering
\begin{minipage}{0.62\textwidth}
\centering
\caption{Summary statistics of the Procmon rule base used for technique-level event labeling.}
\label{tab:procmon_rule_summary}
\begin{tabular}{lc}
\toprule
Metric & Count \\
\midrule
YAML rule files & 198 \\
Unique ATT\&CK techniques & 83 \\
Unique CALDERA abilities & 198 \\
Median abilities per technique & 2 \\
\bottomrule
\end{tabular}
\end{minipage}
\end{table}

\subsection{Handling High-Frequency Normal-Looking Behaviors}

A practical challenge in host-level tracing is that some ATT\&CK techniques naturally resemble routine system activity. Discovery-related behaviors, for example, may generate large numbers of file, directory, registry, or process-access events that are also common in benign workloads. Labeling such events solely by operation type would therefore introduce substantial false positives.

We address this issue with a conservative, context-aware policy. Common operations are accepted only when they are tied to campaign-linked execution context, such as a matched process context together with supporting artifact or command clues. High-frequency techniques are required to satisfy stronger contextual evidence than rare or highly distinctive behaviors. When the available evidence is insufficient for a rule match, the event is not emitted as a confident technique-level malicious label. This conservative design improves annotation reliability in controlled mixed traces, particularly for common system activities that may appear malicious when viewed in isolation.

\clearpage
\section{\textsc{CAPTure} Dataset Statistics}
\label{app:capture_dataset_statistics}
This appendix reports the executable attack-profile pool and per-run statistics for the \textsc{CAPTure} Mix-2 through Mix-6 subsets. The current corpus contains 200 scenarios totaling 855,086,945 events, including 852,441,108 benign events and 2,645,837 malicious events. The executable pool consists of 25 ATT\&CK-aligned profiles: 19 APT profiles and 6 APT-like ransomware-style profiles. Table~\ref{tab:kill_chain_capture_apt} reports the executable attack-profile-to-tactic mapping, and Tables~\ref{tab:capture_mix2_stats}--\ref{tab:capture_mix6_stats} report the per-run corpus statistics with subset-level totals. These tables describe the collected audit-log corpus itself and are therefore separated from the downstream demixing evaluation tables.

\begin{table*}[!htbp]
\centering
\scriptsize
\caption{Attack-profile-to-tactic mapping in \textsc{CAPTure}. Rows denote the 25 ATT\&CK-aligned profiles included in our CALDERA-based attack pool, consisting of 19 APT profiles and 6 APT-like ransomware-style profiles. Columns correspond to the ATT\&CK tactic categories used in this study. Each cell lists executable ATT\&CK technique IDs supported by the corresponding profile under that tactic. Empty cells indicate that no supported technique from that profile is included in the given tactic category.}
\setlength{\tabcolsep}{1pt}
\renewcommand{\arraystretch}{1.05}
\arrayrulecolor{black!20}
\setlength{\arrayrulewidth}{0.4pt}

\setlength{\wDataset}{1.95cm}
\setlength{\wPhase}{\dimexpr(\linewidth-\wDataset-11\arrayrulewidth-20\tabcolsep)/9\relax}

\begin{tabular}{|D{\wDataset}|
P{kcInit}{\wPhase}|P{kcExec}{\wPhase}|P{kcPers}{\wPhase}|P{kcDef}{\wPhase}|
P{kcDisc}{\wPhase}|P{kcLat}{\wPhase}|P{kcColl}{\wPhase}|P{kcCtwo}{\wPhase}|P{kcExf}{\wPhase}|}
\hline

Profile &
\multicolumn{1}{c|}{\cellcolor{kcInitH}\bfseries Init.\ Acc.} &
\multicolumn{1}{c|}{\cellcolor{kcExecH}\bfseries Execution} &
\multicolumn{1}{c|}{\cellcolor{kcPersH}\bfseries Persist.} &
\multicolumn{1}{c|}{\cellcolor{kcDefH}\bfseries Def.\ Eva.} &
\multicolumn{1}{c|}{\cellcolor{kcDiscH}\bfseries Discovery} &
\multicolumn{1}{c|}{\cellcolor{kcLatH}\bfseries Lat.\ Move.} &
\multicolumn{1}{c|}{\cellcolor{kcCollH}\bfseries Collect.} &
\multicolumn{1}{c|}{\cellcolor{kcCtwoH}\bfseries C2} &
\multicolumn{1}{c|}{\cellcolor{kcExfH}\bfseries Exfil.}
\tabularnewline\hline

Admin338 & T1566.001 & T1204.002 & -- & -- & T1087.001, T1083, T1069.001, T1016, T1049, T1007 & -- & -- & -- & --
\tabularnewline\hline
CobaltGroup & T1566.001 & T1204.002 & -- & -- & T1046 & -- & -- & T1219 & --
\tabularnewline\hline
FIN7 & T1566.001 & T1204.002 & T1547.001, T1053.005 & -- & -- & -- & -- & T1105 & --
\tabularnewline\hline
Gamaredon & T1566.001 & T1204.002, T1047 & T1547.001, T1053.005 & T1112 & T1082 & -- & -- & T1071.001 & --
\tabularnewline\hline
GorgonGroup & T1566.001 & T1204.002, T1059.001 & T1547.001, T1547.009 & T1055.002, T1562.001, T1564.003 & -- & -- & -- & -- & --
\tabularnewline\hline
Higaisa & T1566.001 & T1204.002 & T1547.001, T1053.005 & T1036.004 & T1082, T1016 & -- & -- & -- & --
\tabularnewline\hline
Patchwork & T1566.001 & T1204.002, T1059.001 & T1547.001 & T1548.002 & T1033, T1518.001 & -- & T1005 & -- & --
\tabularnewline\hline
APT1 & -- & T1059.003 & -- & -- & T1087.001 & -- & T1560.001, T1119, T1005 & -- & --
\tabularnewline\hline
APT3 & -- & T1059.001 & T1547.001 & -- & T1087.001 & -- & T1560.001 & -- & --
\tabularnewline\hline
APT5 & -- & T1059.001, T1059.003 & -- & -- & T1083 & -- & T1560.001, T1074.001 & -- & --
\tabularnewline\hline
APT18 & -- & T1059.003 & T1547.001 & -- & T1083, T1082 & -- & -- & T1071.001, T1105 & --
\tabularnewline\hline
APT28 & T1566.001 & T1204.002 & -- & -- & T1082 & -- & T1005 & T1071.001 & T1567
\tabularnewline\hline
APT29 & -- & T1059.001 & T1547.001 & T1548.002 & -- & -- & T1560.001 & T1071.001 & --
\tabularnewline\hline
APT32 & -- & -- & T1053.005 & -- & T1087.001, T1082, T1016, T1049, T1033 & -- & -- & T1071.001 & --
\tabularnewline\hline
APT33 & -- & T1059.001 & T1547.001 & -- & -- & -- & T1560.001 & T1071.001 & --
\tabularnewline\hline
APT37 & -- & -- & T1547.001 & T1548.002 & -- & -- & T1123 & T1071.001 & --
\tabularnewline\hline
APT39 & -- & -- & T1547.001, T1547.009 & -- & -- & -- & T1560.001, T1115 & T1071.001 & --
\tabularnewline\hline
APT41 & -- & T1059.003 & -- & -- & T1087.001 & -- & T1560.001 & T1071.001 & --
\tabularnewline\hline
APT42 & -- & T1059.001 & T1547 & -- & T1087.001 & -- & -- & T1071.001 & --
\tabularnewline\hline
LockBit3 & -- & T1047, T1569.002 & T1053.005 & T1070.004 & T1016, T1018, T1082, T1135 & T1021.002, T1570 & -- & -- & --
\tabularnewline\hline
VoltTyphoon & -- & T1047 & T1053.005 & T1070.004 & T1016, T1018, T1033, T1049, T1057, T1087, T1135 & T1021.002, T1570 & -- & -- & --
\tabularnewline\hline
PlayCrypt & -- & T1569.002 & T1053.005 & T1070.004 & T1016, T1018, T1057, T1082, T1087, T1135 & T1021.002, T1570 & -- & -- & --
\tabularnewline\hline
ViceSociety & -- & T1047 & T1053.005 & T1070.004 & T1018, T1082, T1087 & T1021.002 & -- & -- & --
\tabularnewline\hline
BlackMatter & -- & T1569.002 & T1053.005 & T1070.004 & T1018, T1057, T1135 & T1021.002, T1570 & -- & -- & --
\tabularnewline\hline
Wocao & -- & T1047, T1569.002 & T1053.005 & T1070.004 & T1007, T1016, T1018, T1033, T1046, T1049, T1057, T1082, T1087, T1135 & T1021.002, T1570 & T1005, T1119, T1560.001 & -- & --
\tabularnewline\hline

\end{tabular}
\label{tab:kill_chain_capture_apt}
\end{table*}
\clearpage
\begingroup
\scriptsize
\setlength{\tabcolsep}{2pt}
\renewcommand{\arraystretch}{1.02}
\begin{table*}[h!]
\centering
\caption{Per-run statistics of the CAPTure Mix-2 subset.}
\label{tab:capture_mix2_stats}
\begin{tabularx}{\textwidth}{>{\raggedright\arraybackslash}Xrrrrrr}
\toprule
Profile Mix & \#Profiles & Total Events & Size (GB) & Benign Events & Malicious Events & Malicious Ratio (\%) \\
\midrule
Admin338 + APT37 & 2 & 2,306,354 & 0.807 & 2,305,420 & 934 & 0.04 \\
APT1 + APT39 & 2 & 1,355,367 & 0.518 & 1,324,476 & 30,891 & 2.28 \\
APT1 + Higaisa & 2 & 3,227,772 & 1.125 & 3,199,425 & 28,347 & 0.88 \\
APT28 + APT41 & 2 & 2,162,205 & 0.789 & 2,133,500 & 28,705 & 1.33 \\
APT28 + GorgonGroup & 2 & 3,379,498 & 1.220 & 3,347,877 & 31,621 & 0.94 \\
APT29 + APT39 & 2 & 2,331,258 & 0.840 & 2,327,550 & 3,708 & 0.16 \\
APT29 + APT3 & 2 & 2,404,039 & 0.859 & 2,402,799 & 1,240 & 0.05 \\
APT32 + Higaisa & 2 & 2,272,401 & 0.806 & 2,271,772 & 629 & 0.03 \\
APT33 + Admin338 & 2 & 3,196,201 & 1.121 & 3,194,739 & 1,462 & 0.05 \\
APT33 + CobaltGroup & 2 & 3,716,427 & 1.341 & 3,714,823 & 1,604 & 0.04 \\
APT37 + APT33 & 2 & 2,043,992 & 0.734 & 2,043,154 & 838 & 0.04 \\
APT39 + GorgonGroup & 2 & 3,260,124 & 1.178 & 3,253,812 & 6,312 & 0.19 \\
APT3 + APT5 & 2 & 3,932,072 & 1.382 & 3,930,327 & 1,745 & 0.04 \\
APT3 + Patchwork & 2 & 3,309,466 & 1.178 & 3,280,959 & 28,507 & 0.86 \\
CobaltGroup + APT37 & 2 & 2,655,100 & 0.980 & 2,654,013 & 1,087 & 0.04 \\
CobaltGroup + APT41 & 2 & 2,654,367 & 0.989 & 2,652,741 & 1,626 & 0.06 \\
FIN7 + APT37 & 2 & 2,960,863 & 1.015 & 2,960,388 & 475 & 0.02 \\
Gamaredon + APT33 & 2 & 3,187,196 & 1.118 & 3,186,134 & 1,062 & 0.03 \\
GorgonGroup + Higaisa & 2 & 3,474,731 & 1.237 & 3,470,728 & 4,003 & 0.12 \\
Patchwork + Higaisa & 2 & 2,558,055 & 0.919 & 2,529,778 & 28,277 & 1.11 \\
BlackMatter + APT29 & 2 & 2,707,963 & 1.797 & 2,707,343 & 620 & 0.02 \\
BlackMatter + APT33 & 2 & 2,679,448 & 1.760 & 2,678,840 & 608 & 0.02 \\
BlackMatter + APT3 & 2 & 2,675,032 & 1.771 & 2,667,517 & 7,515 & 0.28 \\
BlackMatter + Patchwork & 2 & 2,790,921 & 1.859 & 2,790,328 & 593 & 0.02 \\
LockBit3 + APT18 & 2 & 2,798,183 & 1.861 & 2,797,261 & 922 & 0.03 \\
LockBit3 + APT1 & 2 & 3,525,485 & 2.291 & 3,517,681 & 7,804 & 0.22 \\
LockBit3 + APT29 & 2 & 3,564,039 & 2.311 & 3,562,989 & 1,050 & 0.03 \\
LockBit3 + APT33 & 2 & 2,881,821 & 1.905 & 2,880,902 & 919 & 0.03 \\
LockBit3 + APT39 & 2 & 3,498,847 & 2.293 & 3,497,854 & 993 & 0.03 \\
LockBit3 + FIN7 & 2 & 3,740,177 & 2.447 & 3,739,266 & 911 & 0.02 \\
LockBit3 + Gamaredon & 2 & 3,109,491 & 2.039 & 3,108,581 & 910 & 0.03 \\
PlayCrypt + Admin338 & 2 & 3,593,514 & 2.341 & 3,592,318 & 1,196 & 0.03 \\
PlayCrypt + APT29 & 2 & 3,690,580 & 2.396 & 3,689,361 & 1,219 & 0.03 \\
PlayCrypt + GorgonGroup & 2 & 3,179,240 & 2.130 & 3,178,036 & 1,204 & 0.04 \\
VoltTyphoon + Admin338 & 2 & 3,504,962 & 2.339 & 3,501,134 & 3,828 & 0.11 \\
VoltTyphoon + APT32 & 2 & 2,840,904 & 1.936 & 2,838,184 & 2,720 & 0.10 \\
Wocao + APT18 & 2 & 4,292,792 & 2.839 & 4,290,930 & 1,862 & 0.04 \\
Wocao + APT5 & 2 & 4,825,393 & 3.139 & 4,816,664 & 8,729 & 0.18 \\
Wocao + FIN7 & 2 & 5,159,571 & 3.356 & 5,157,632 & 1,939 & 0.04 \\
Wocao + Gamaredon & 2 & 4,205,987 & 2.806 & 4,204,137 & 1,850 & 0.04 \\
\midrule
 & & \textbf{125,651,838} & \textbf{65.772} & \textbf{125,401,373} & \textbf{250,465} & \\
\bottomrule
\end{tabularx}
\end{table*}
\begin{table*}[h!]
\centering
\caption{Per-run statistics of the CAPTure Mix-3 subset.}
\label{tab:capture_mix3_stats}
\begin{tabularx}{\textwidth}{>{\raggedright\arraybackslash}Xrrrrrr}
\toprule
Profile Mix & \#Profiles & Total Events & Size (GB) & Benign Events & Malicious Events & Malicious Ratio (\%) \\
\midrule
Admin338 + APT33 + APT39 & 3 & 3,303,061 & 1.172 & 3,299,034 & 4,027 & 0.12 \\
Admin338 + FIN7 + APT1 & 3 & 2,678,556 & 0.959 & 2,648,268 & 30,288 & 1.13 \\
Admin338 + FIN7 + APT41 & 3 & 2,653,823 & 0.935 & 2,651,996 & 1,827 & 0.07 \\
Admin338 + Gamaredon + APT37 & 3 & 2,657,632 & 0.923 & 2,656,262 & 1,370 & 0.05 \\
APT1 + CobaltGroup + APT29 & 3 & 3,787,442 & 1.390 & 3,757,448 & 29,994 & 0.79 \\
APT29 + APT41 + APT37 & 3 & 2,229,016 & 0.796 & 2,227,998 & 1,018 & 0.05 \\
APT29 + Gamaredon + APT5 & 3 & 5,323,224 & 1.846 & 5,320,946 & 2,278 & 0.04 \\
APT32 + APT33 + APT39 & 3 & 2,426,438 & 0.870 & 2,421,859 & 4,579 & 0.19 \\
APT33 + Patchwork + APT5 & 3 & 5,458,338 & 1.912 & 5,428,503 & 29,835 & 0.55 \\
APT37 + APT5 + Higaisa & 3 & 4,231,794 & 1.474 & 4,230,177 & 1,617 & 0.04 \\
APT3 + APT41 + Gamaredon & 3 & 4,066,861 & 1.395 & 4,065,218 & 1,643 & 0.04 \\
APT42 + GorgonGroup + Admin338 & 3 & 4,526,849 & 1.585 & 4,522,518 & 4,331 & 0.10 \\
APT5 + APT39 + APT33 & 3 & 4,262,577 & 1.502 & 4,257,628 & 4,949 & 0.12 \\
CobaltGroup + FIN7 + APT28 & 3 & 2,945,836 & 1.106 & 2,916,651 & 29,185 & 0.99 \\
FIN7 + APT28 + APT18 & 3 & 2,563,269 & 0.919 & 2,534,012 & 29,257 & 1.14 \\
Gamaredon + APT29 + CobaltGroup & 3 & 4,039,742 & 1.449 & 4,037,781 & 1,961 & 0.05 \\
GorgonGroup + APT42 + Admin338 & 3 & 4,493,568 & 1.578 & 4,488,835 & 4,733 & 0.11 \\
GorgonGroup + APT5 + CobaltGroup & 3 & 5,780,041 & 2.081 & 5,774,164 & 5,877 & 0.10 \\
Higaisa + FIN7 + Patchwork & 3 & 3,195,873 & 1.130 & 3,167,320 & 28,553 & 0.89 \\
Patchwork + APT37 + APT42 & 3 & 3,508,912 & 1.233 & 3,480,586 & 28,326 & 0.81 \\
BlackMatter + APT18 + APT41 & 3 & 2,656,504 & 1.756 & 2,655,817 & 687 & 0.03 \\
BlackMatter + APT29 + APT3 & 3 & 2,764,387 & 1.814 & 2,756,899 & 7,488 & 0.27 \\
BlackMatter + GorgonGroup + APT18 & 3 & 2,861,536 & 1.893 & 2,860,913 & 623 & 0.02 \\
LockBit3 + APT18 + APT5 & 3 & 2,987,338 & 1.982 & 2,979,523 & 7,815 & 0.26 \\
LockBit3 + APT1 + GorgonGroup & 3 & 3,240,101 & 2.159 & 3,232,311 & 7,790 & 0.24 \\
LockBit3 + GorgonGroup + Gamaredon & 3 & 3,143,219 & 2.095 & 3,142,295 & 924 & 0.03 \\
PlayCrypt + APT39 + APT37 & 3 & 3,050,336 & 2.041 & 3,049,097 & 1,239 & 0.04 \\
PlayCrypt + APT42 + APT29 & 3 & 3,011,936 & 2.004 & 3,010,644 & 1,292 & 0.04 \\
PlayCrypt + APT5 + APT1 & 3 & 3,093,249 & 2.070 & 3,078,388 & 14,861 & 0.48 \\
ViceSociety + APT42 + GorgonGroup & 3 & 3,227,995 & 2.136 & 3,225,861 & 2,134 & 0.07 \\
ViceSociety + CobaltGroup + Admin338 & 3 & 3,352,010 & 2.225 & 3,349,948 & 2,062 & 0.06 \\
ViceSociety + FIN7 + APT33 & 3 & 2,912,768 & 1.964 & 2,910,689 & 2,079 & 0.07 \\
VoltTyphoon + APT28 + APT18 & 3 & 3,976,090 & 2.605 & 3,973,371 & 2,719 & 0.07 \\
VoltTyphoon + APT37 + APT29 & 3 & 3,313,141 & 2.219 & 3,310,426 & 2,715 & 0.08 \\
Wocao + APT1 + APT3 & 3 & 4,843,042 & 3.155 & 4,827,378 & 15,664 & 0.32 \\
Wocao + APT28 + APT32 & 3 & 4,338,289 & 2.874 & 4,336,390 & 1,899 & 0.04 \\
Wocao + APT28 + APT37 & 3 & 4,608,308 & 3.023 & 4,606,420 & 1,888 & 0.04 \\
Wocao + APT29 + APT39 & 3 & 5,147,918 & 3.354 & 5,146,014 & 1,904 & 0.04 \\
Wocao + APT3 + APT18 & 3 & 4,849,147 & 3.149 & 4,840,400 & 8,747 & 0.18 \\
Wocao + CobaltGroup + APT33 & 3 & 5,385,924 & 3.500 & 5,384,046 & 1,878 & 0.03 \\
\midrule
 & & \textbf{146,896,090} & \textbf{74.273} & \textbf{146,564,034} & \textbf{332,056} & \\
\bottomrule
\end{tabularx}
\end{table*}
\begin{table*}[h!]
\centering
\caption{Per-run statistics of the CAPTure Mix-4 subset.}
\label{tab:capture_mix4_stats}
\begin{tabularx}{\textwidth}{>{\raggedright\arraybackslash}Xrrrrrr}
\toprule
Profile Mix & \#Profiles & Total Events & Size (GB) & Benign Events & Malicious Events & Malicious Ratio (\%) \\
\midrule
Admin338 + APT41 + GorgonGroup + APT28 & 4 & 4,816,815 & 1.742 & 4,784,258 & 32,557 & 0.68 \\
Admin338 + Higaisa + APT1 + APT37 & 4 & 3,012,450 & 1.054 & 2,982,743 & 29,707 & 0.99 \\
APT18 + APT3 + Patchwork + Admin338 & 4 & 4,752,760 & 1.706 & 4,722,495 & 30,265 & 0.64 \\
APT18 + APT42 + Patchwork + APT33 & 4 & 6,076,437 & 2.146 & 6,046,676 & 29,761 & 0.49 \\
APT1 + APT32 + APT39 + FIN7 & 4 & 3,039,155 & 1.082 & 3,007,791 & 31,364 & 1.03 \\
APT1 + GorgonGroup + Admin338 + APT3 & 4 & 4,522,085 & 1.611 & 4,489,101 & 32,984 & 0.73 \\
APT29 + Gamaredon + APT33 + Patchwork & 4 & 6,094,284 & 2.152 & 6,064,704 & 29,580 & 0.49 \\
APT37 + APT18 + APT1 + APT28 & 4 & 2,805,418 & 0.998 & 2,748,387 & 57,031 & 2.03 \\
APT37 + APT33 + CobaltGroup + APT1 & 4 & 3,818,831 & 1.402 & 3,788,454 & 30,377 & 0.80 \\
APT37 + APT3 + FIN7 + APT28 & 4 & 3,673,256 & 1.298 & 3,643,701 & 29,555 & 0.80 \\
APT37 + FIN7 + APT18 + APT1 & 4 & 2,942,046 & 1.034 & 2,912,264 & 29,782 & 1.01 \\
APT3 + APT42 + FIN7 + APT28 & 4 & 3,324,085 & 1.194 & 3,295,044 & 29,041 & 0.87 \\
APT42 + APT32 + APT39 + Higaisa & 4 & 4,609,563 & 1.650 & 4,605,741 & 3,822 & 0.08 \\
APT42 + APT32 + Higaisa + Patchwork & 4 & 5,173,324 & 1.837 & 5,144,703 & 28,621 & 0.55 \\
APT42 + APT39 + Admin338 + Gamaredon & 4 & 4,446,380 & 1.521 & 4,441,969 & 4,411 & 0.10 \\
APT42 + Gamaredon + APT33 + APT3 & 4 & 4,694,657 & 1.613 & 4,693,360 & 1,297 & 0.03 \\
APT5 + FIN7 + APT42 + APT1 & 4 & 5,574,927 & 1.944 & 5,544,953 & 29,974 & 0.54 \\
APT5 + Patchwork + CobaltGroup + APT37 & 4 & 5,631,959 & 2.075 & 5,601,024 & 30,935 & 0.55 \\
GorgonGroup + Higaisa + APT32 + APT42 & 4 & 5,688,324 & 2.035 & 5,683,941 & 4,383 & 0.08 \\
Patchwork + APT37 + APT32 + APT42 & 4 & 8,595,232 & 3.267 & 8,567,072 & 28,160 & 0.33 \\
BlackMatter + APT1 + APT28 + APT37 & 4 & 2,925,744 & 1.940 & 2,918,248 & 7,496 & 0.26 \\
BlackMatter + APT33 + GorgonGroup + Higaisa & 4 & 3,052,723 & 2.013 & 3,052,085 & 638 & 0.02 \\
BlackMatter + APT39 + Admin338 + APT28 & 4 & 2,800,469 & 1.862 & 2,799,829 & 640 & 0.02 \\
BlackMatter + APT41 + APT39 + APT29 & 4 & 2,993,340 & 1.981 & 2,992,622 & 718 & 0.02 \\
BlackMatter + APT5 + FIN7 + Admin338 & 4 & 2,896,582 & 1.920 & 2,889,120 & 7,462 & 0.26 \\
BlackMatter + Gamaredon + APT28 + APT29 & 4 & 2,967,851 & 1.952 & 2,967,201 & 650 & 0.02 \\
LockBit3 + APT42 + APT3 + Higaisa & 4 & 2,987,462 & 1.993 & 2,979,699 & 7,763 & 0.26 \\
LockBit3 + APT5 + Higaisa + GorgonGroup & 4 & 4,728,145 & 3.016 & 4,720,344 & 7,801 & 0.16 \\
PlayCrypt + APT42 + Higaisa + APT39 & 4 & 4,262,129 & 2.747 & 4,260,821 & 1,308 & 0.03 \\
ViceSociety + APT18 + APT39 + APT41 & 4 & 3,056,184 & 2.040 & 3,054,029 & 2,155 & 0.07 \\
ViceSociety + APT33 + Admin338 + APT39 & 4 & 3,173,128 & 2.111 & 3,171,023 & 2,105 & 0.07 \\
ViceSociety + APT37 + APT5 + APT28 & 4 & 3,114,334 & 2.092 & 3,105,344 & 8,990 & 0.29 \\
ViceSociety + APT42 + FIN7 + APT3 & 4 & 3,110,217 & 2.086 & 3,101,207 & 9,010 & 0.29 \\
ViceSociety + APT5 + Patchwork + FIN7 & 4 & 3,532,733 & 2.319 & 3,523,773 & 8,960 & 0.25 \\
VoltTyphoon + Admin338 + Gamaredon + APT42 & 4 & 3,652,701 & 2.401 & 3,649,937 & 2,764 & 0.08 \\
VoltTyphoon + APT39 + APT1 + APT33 & 4 & 4,122,780 & 2.703 & 4,113,208 & 9,572 & 0.23 \\
Wocao + APT41 + CobaltGroup + Patchwork & 4 & 5,406,607 & 3.541 & 5,404,666 & 1,941 & 0.04 \\
Wocao + APT5 + Higaisa + APT39 & 4 & 6,638,475 & 4.314 & 6,629,728 & 8,747 & 0.13 \\
Wocao + FIN7 + Gamaredon + APT18 & 4 & 4,155,940 & 2.779 & 4,154,052 & 1,888 & 0.05 \\
Wocao + Higaisa + CobaltGroup + FIN7 & 4 & 5,028,499 & 3.285 & 5,026,621 & 1,878 & 0.04 \\
\midrule
 & & \textbf{167,898,031} & \textbf{82.456} & \textbf{167,281,938} & \textbf{616,093} & \\
\bottomrule
\end{tabularx}
\end{table*}
\begin{table*}[h!]
\centering
\caption{Per-run statistics of the CAPTure Mix-5 subset.}
\label{tab:capture_mix5_stats}
\begin{tabularx}{\textwidth}{>{\raggedright\arraybackslash}Xrrrrrr}
\toprule
Profile Mix & \#Profiles & Total Events & Size (GB) & Benign Events & Malicious Events & Malicious Ratio (\%) \\
\midrule
APT18 + APT29 + Gamaredon + APT3 + APT39 & 5 & 6,804,103 & 2.377 & 6,797,930 & 6,173 & 0.09 \\
APT18 + APT33 + APT42 + APT5 + APT39 & 5 & 4,768,345 & 1.663 & 4,762,296 & 6,049 & 0.13 \\
APT1 + APT18 + Higaisa + APT32 + APT41 & 5 & 3,060,962 & 1.077 & 3,029,335 & 31,627 & 1.03 \\
APT1 + Higaisa + APT18 + Admin338 + APT28 & 5 & 4,312,318 & 1.538 & 4,254,571 & 57,747 & 1.34 \\
APT28 + APT41 + APT18 + APT1 + APT5 & 5 & 4,352,374 & 1.562 & 4,293,700 & 58,674 & 1.35 \\
APT28 + APT42 + FIN7 + APT29 + Gamaredon & 5 & 7,502,961 & 2.559 & 7,473,562 & 29,399 & 0.39 \\
APT29 + Admin338 + APT28 + APT1 + FIN7 & 5 & 4,135,016 & 1.467 & 4,077,416 & 57,600 & 1.39 \\
APT32 + APT33 + APT41 + APT5 + APT29 & 5 & 5,330,546 & 1.952 & 5,325,991 & 4,555 & 0.09 \\
APT32 + FIN7 + Gamaredon + Admin338 + APT18 & 5 & 6,073,935 & 2.113 & 6,070,565 & 3,370 & 0.06 \\
APT33 + APT1 + APT29 + Admin338 + FIN7 & 5 & 6,052,455 & 2.149 & 6,022,109 & 30,346 & 0.50 \\
APT37 + APT28 + APT33 + APT3 + APT1 & 5 & 5,335,117 & 1.872 & 5,278,782 & 56,335 & 1.06 \\
APT39 + APT5 + FIN7 + Patchwork + APT33 & 5 & 5,958,465 & 2.098 & 5,925,430 & 33,035 & 0.55 \\
APT3 + GorgonGroup + APT33 + APT1 + APT37 & 5 & 7,361,896 & 2.581 & 7,327,982 & 33,914 & 0.46 \\
APT41 + APT33 + APT18 + APT28 + FIN7 & 5 & 3,856,212 & 1.359 & 3,825,675 & 30,537 & 0.79 \\
APT42 + APT39 + CobaltGroup + APT3 + APT1 & 5 & 4,754,049 & 1.729 & 4,721,951 & 32,098 & 0.68 \\
CobaltGroup + Admin338 + Patchwork + APT29 + APT37 & 5 & 5,464,044 & 1.928 & 5,433,659 & 30,385 & 0.56 \\
CobaltGroup + APT1 + Higaisa + APT5 + APT28 & 5 & 5,444,703 & 1.962 & 5,386,462 & 58,241 & 1.07 \\
CobaltGroup + APT39 + APT18 + Patchwork + Gamaredon & 5 & 6,998,637 & 2.440 & 6,964,466 & 34,171 & 0.49 \\
FIN7 + Patchwork + APT33 + Higaisa + APT37 & 5 & 5,836,774 & 1.997 & 5,807,347 & 29,427 & 0.50 \\
GorgonGroup + Higaisa + APT3 + FIN7 + APT32 & 5 & 5,097,225 & 1.783 & 5,092,432 & 4,793 & 0.09 \\
BlackMatter + APT1 + APT18 + APT29 + Higaisa & 5 & 4,067,472 & 2.598 & 4,059,960 & 7,512 & 0.18 \\
BlackMatter + APT28 + Patchwork + APT29 + APT32 & 5 & 3,453,099 & 2.245 & 3,452,378 & 721 & 0.02 \\
BlackMatter + APT29 + APT42 + Admin338 + Higaisa & 5 & 3,071,105 & 2.018 & 3,070,375 & 730 & 0.02 \\
BlackMatter + APT41 + APT1 + APT28 + Patchwork & 5 & 3,126,973 & 2.075 & 3,119,324 & 7,649 & 0.24 \\
BlackMatter + APT42 + APT18 + Higaisa + APT1 & 5 & 2,088,202 & 1.407 & 2,080,650 & 7,552 & 0.36 \\
BlackMatter + GorgonGroup + APT29 + APT3 + APT18 & 5 & 3,204,510 & 2.094 & 3,196,939 & 7,571 & 0.24 \\
LockBit3 + APT18 + APT28 + CobaltGroup + APT32 & 5 & 5,032,080 & 3.175 & 5,031,050 & 1,030 & 0.02 \\
LockBit3 + APT32 + APT42 + APT18 + Patchwork & 5 & 4,190,196 & 2.697 & 4,189,113 & 1,083 & 0.03 \\
LockBit3 + FIN7 + APT39 + APT42 + APT41 & 5 & 3,116,341 & 2.063 & 3,115,255 & 1,086 & 0.03 \\
LockBit3 + GorgonGroup + Higaisa + APT3 + APT1 & 5 & 3,328,508 & 2.223 & 3,313,741 & 14,767 & 0.44 \\
ViceSociety + APT29 + APT3 + APT18 + APT41 & 5 & 3,227,762 & 2.147 & 3,218,667 & 9,095 & 0.28 \\
ViceSociety + APT29 + CobaltGroup + APT1 + APT37 & 5 & 3,978,966 & 2.615 & 3,969,545 & 9,421 & 0.24 \\
ViceSociety + FIN7 + APT39 + Higaisa + APT5 & 5 & 3,341,123 & 2.244 & 3,332,004 & 9,119 & 0.27 \\
ViceSociety + Patchwork + CobaltGroup + APT32 + APT33 & 5 & 1,946,728 & 1.258 & 1,946,544 & 184 & 0.01 \\
VoltTyphoon + Admin338 + APT5 + Patchwork + APT42 & 5 & 3,655,564 & 2.437 & 3,645,741 & 9,823 & 0.27 \\
VoltTyphoon + APT28 + APT18 + APT33 + CobaltGroup & 5 & 4,076,329 & 2.693 & 4,073,607 & 2,722 & 0.07 \\
Wocao + Admin338 + Gamaredon + APT32 + CobaltGroup & 5 & 5,513,339 & 3.574 & 5,511,353 & 1,986 & 0.04 \\
Wocao + APT33 + FIN7 + APT41 + GorgonGroup & 5 & 4,370,091 & 2.918 & 4,368,123 & 1,968 & 0.05 \\
Wocao + APT39 + CobaltGroup + APT32 + FIN7 & 5 & 5,008,322 & 3.260 & 5,006,368 & 1,954 & 0.04 \\
Wocao + Higaisa + APT29 + APT41 + APT39 & 5 & 6,867,287 & 4.389 & 6,865,298 & 1,989 & 0.03 \\
\midrule
 & & \textbf{185,164,134} & \textbf{88.336} & \textbf{184,437,696} & \textbf{726,438} & \\
\bottomrule
\end{tabularx}
\end{table*}
\begin{table*}[h!]
\centering
\caption{Per-run statistics of the CAPTure Mix-6 subset.}
\label{tab:capture_mix6_stats}
\begin{tabularx}{\textwidth}{>{\raggedright\arraybackslash}Xrrrrrr}
\toprule
Profile Mix & \#Profiles & Total Events & Size (GB) & Benign Events & Malicious Events & Malicious Ratio (\%) \\
\midrule
APT29 + Higaisa + Patchwork + APT42 + APT33 + GorgonGroup & 6 & 8,598,606 & 3.032 & 8,562,726 & 35,880 & 0.42 \\
APT32 + APT5 + APT1 + APT3 + APT18 + FIN7 & 6 & 6,274,837 & 2.213 & 6,243,805 & 31,032 & 0.49 \\
APT33 + APT5 + GorgonGroup + APT42 + APT37 + APT32 & 6 & 7,704,325 & 2.728 & 7,698,034 & 6,291 & 0.08 \\
APT33 + GorgonGroup + APT5 + APT18 + APT29 + APT32 & 6 & 8,591,321 & 3.019 & 8,584,343 & 6,978 & 0.08 \\
APT37 + APT41 + APT39 + Gamaredon + CobaltGroup + APT32 & 6 & 8,233,608 & 2.888 & 8,228,117 & 5,491 & 0.07 \\
APT3 + APT29 + Higaisa + GorgonGroup + APT18 + APT41 & 6 & 8,510,696 & 2.988 & 8,503,506 & 7,190 & 0.08 \\
APT42 + APT32 + APT37 + APT3 + Admin338 + Higaisa & 6 & 6,665,756 & 2.356 & 6,663,304 & 2,452 & 0.04 \\
APT42 + APT5 + FIN7 + CobaltGroup + Admin338 + APT1 & 6 & 7,066,807 & 2.546 & 7,034,826 & 31,981 & 0.45 \\
APT5 + APT3 + FIN7 + APT1 + APT42 + APT29 & 6 & 5,349,124 & 1.971 & 5,317,930 & 31,194 & 0.58 \\
APT5 + FIN7 + APT41 + GorgonGroup + Patchwork + APT28 & 6 & 8,591,580 & 2.997 & 8,530,344 & 61,236 & 0.71 \\
CobaltGroup + Higaisa + APT41 + APT42 + APT1 + APT29 & 6 & 6,583,359 & 2.364 & 6,552,632 & 30,727 & 0.47 \\
FIN7 + APT32 + APT33 + APT18 + APT1 + Higaisa & 6 & 4,433,411 & 1.628 & 4,403,251 & 30,160 & 0.68 \\
Gamaredon + APT3 + Patchwork + APT41 + APT1 + APT32 & 6 & 5,794,465 & 2.068 & 5,706,769 & 87,696 & 1.51 \\
Gamaredon + FIN7 + APT1 + CobaltGroup + APT41 + GorgonGroup & 6 & 5,864,818 & 2.151 & 5,831,047 & 33,771 & 0.58 \\
GorgonGroup + APT33 + CobaltGroup + APT28 + Higaisa + APT37 & 6 & 6,440,205 & 2.357 & 6,405,911 & 34,294 & 0.53 \\
GorgonGroup + APT37 + APT42 + APT18 + APT41 + APT28 & 6 & 5,899,232 & 2.133 & 5,865,609 & 33,623 & 0.57 \\
Higaisa + CobaltGroup + Gamaredon + APT37 + APT5 + Patchwork & 6 & 8,599,101 & 2.970 & 8,568,407 & 30,694 & 0.36 \\
Patchwork + APT1 + APT32 + APT37 + FIN7 + APT33 & 6 & 7,120,848 & 2.579 & 7,063,744 & 57,104 & 0.80 \\
Patchwork + APT33 + Higaisa + APT29 + FIN7 + Gamaredon & 6 & 8,597,963 & 2.955 & 8,567,378 & 30,585 & 0.36 \\
Patchwork + APT41 + CobaltGroup + APT32 + APT5 + APT37 & 6 & 6,220,925 & 2.250 & 6,189,859 & 31,066 & 0.50 \\
BlackMatter + APT37 + APT41 + APT18 + FIN7 + APT33 & 6 & 3,580,253 & 2.303 & 3,579,498 & 755 & 0.02 \\
BlackMatter + APT37 + GorgonGroup + Admin338 + CobaltGroup + APT29 & 6 & 3,943,742 & 2.540 & 3,943,036 & 706 & 0.02 \\
BlackMatter + FIN7 + APT39 + Higaisa + Gamaredon + APT18 & 6 & 3,177,662 & 2.091 & 3,176,973 & 689 & 0.02 \\
LockBit3 + APT42 + APT39 + Admin338 + CobaltGroup + APT5 & 6 & 3,738,272 & 2.458 & 3,730,252 & 8,020 & 0.21 \\
LockBit3 + CobaltGroup + APT5 + APT32 + APT33 + Patchwork & 6 & 4,650,241 & 2.973 & 4,642,228 & 8,013 & 0.17 \\
PlayCrypt + APT32 + Higaisa + APT41 + Admin338 + APT42 & 6 & 4,593,902 & 2.915 & 4,592,456 & 1,446 & 0.03 \\
PlayCrypt + APT37 + APT39 + APT32 + APT33 + APT3 & 6 & 3,550,829 & 2.328 & 3,542,563 & 8,266 & 0.23 \\
PlayCrypt + APT39 + Gamaredon + CobaltGroup + APT18 + APT32 & 6 & 4,265,750 & 2.766 & 4,264,415 & 1,335 & 0.03 \\
PlayCrypt + APT41 + APT33 + Higaisa + APT18 + CobaltGroup & 6 & 5,197,036 & 3.332 & 5,195,709 & 1,327 & 0.03 \\
ViceSociety + APT1 + APT32 + APT41 + Higaisa + APT39 & 6 & 3,319,007 & 2.214 & 3,310,313 & 8,694 & 0.26 \\
ViceSociety + APT41 + APT29 + APT39 + Admin338 + APT1 & 6 & 3,417,285 & 2.279 & 3,408,247 & 9,038 & 0.26 \\
ViceSociety + APT5 + FIN7 + Gamaredon + APT32 + APT28 & 6 & 3,055,230 & 2.051 & 3,046,282 & 8,948 & 0.29 \\
ViceSociety + CobaltGroup + APT5 + APT33 + APT42 + Higaisa & 6 & 3,552,021 & 2.367 & 3,542,987 & 9,034 & 0.25 \\
ViceSociety + CobaltGroup + Higaisa + Patchwork + APT41 + FIN7 & 6 & 3,112,240 & 2.119 & 3,110,069 & 2,171 & 0.07 \\
ViceSociety + Higaisa + APT29 + Patchwork + APT18 + APT1 & 6 & 3,699,245 & 2.430 & 3,690,680 & 8,565 & 0.23 \\
Wocao + APT32 + APT37 + APT29 + APT33 + GorgonGroup & 6 & 5,255,592 & 3.409 & 5,253,577 & 2,015 & 0.04 \\
Wocao + APT33 + Admin338 + APT41 + Higaisa + Gamaredon & 6 & 7,030,536 & 4.543 & 7,028,526 & 2,010 & 0.03 \\
Wocao + APT5 + Patchwork + APT18 + APT3 + Gamaredon & 6 & 5,038,448 & 3.276 & 5,029,571 & 8,877 & 0.18 \\
Wocao + Gamaredon + APT32 + APT39 + Higaisa + APT3 & 6 & 6,750,317 & 4.352 & 6,740,822 & 9,495 & 0.14 \\
Wocao + GorgonGroup + APT33 + Patchwork + APT39 + APT29 & 6 & 7,408,257 & 4.892 & 7,406,321 & 1,936 & 0.03 \\
\midrule
 & & \textbf{229,476,852} & \textbf{107.831} & \textbf{228,756,067} & \textbf{720,785} & \\
\bottomrule
\end{tabularx}
\end{table*}
\endgroup

\clearpage
\section{Detailed \textsc{CAPTure} End-to-End Results}
\label{app:capture_upstream_details}
This appendix provides the full metric breakdown for the \textsc{CAPTure} unknown-$K$ end-to-end evaluation after pooling single-host and multi-host runs. The compact main-paper summary is reported in Table~\ref{tab:capture_upstream}.
\begin{table*}[h!]
\centering
\caption{Detailed end-to-end results on \textsc{CAPTure} after pooling single-host and multi-host evaluations with Up-to-6 decoding ($K_{\max}=6$). For each upstream system, we report clustering diagnostics, occurrence-level assignment quality, and inference time across Mix-2 to Mix-6. Each cell is reported as mean$_{\text{std}}$. Superscript $\dagger$ marks a statistically favored cell when the 95\% confidence interval is separated from the counterpart in the favorable direction.}
\label{tab:capture_upstream_detail}
\resizebox{\textwidth}{!}{
\begin{tabular}{l|c|c|c|c|c|c|c|c|c}
\toprule
\textbf{Upstream} & \textbf{Model} & \textbf{Mix} & \textbf{Acc} $\uparrow$ & \textbf{FMI} $\uparrow$ & \textbf{NMI} $\uparrow$ & \textbf{P} $\uparrow$ & \textbf{R} $\uparrow$ & \textbf{Macro-F1} $\uparrow$ & \textbf{Time} $\downarrow$ \\
\midrule
\multirow{10}{*}{\textbf{SFM}} & \multirow{5}{*}{TGCM} & 2 & $\mathbf{0.617}_{\mathbf{0.028}}^{\dagger}$ & $\mathbf{0.686}_{\mathbf{0.049}}^{\dagger}$ & $0.059_{0.039}$ & $\mathbf{0.415}_{\mathbf{0.070}}$ & $\mathbf{0.484}_{\mathbf{0.037}}^{\dagger}$ & $\mathbf{0.419}_{\mathbf{0.039}}$ & $\mathbf{0.007}_{\mathbf{0.000}}^{\dagger}$ \\
 & & 3 & $\mathbf{0.431}_{\mathbf{0.056}}$ & $\mathbf{0.558}_{\mathbf{0.045}}^{\dagger}$ & $0.105_{0.031}$ & $\mathbf{0.246}_{\mathbf{0.033}}$ & $\mathbf{0.335}_{\mathbf{0.010}}^{\dagger}$ & $\mathbf{0.250}_{\mathbf{0.012}}$ & $\mathbf{0.007}_{\mathbf{0.000}}^{\dagger}$ \\
 & & 4 & $\mathbf{0.394}_{\mathbf{0.029}}$ & $\mathbf{0.490}_{\mathbf{0.024}}^{\dagger}$ & $0.139_{0.023}$ & $\mathbf{0.198}_{\mathbf{0.022}}$ & $\mathbf{0.272}_{\mathbf{0.017}}^{\dagger}$ & $\mathbf{0.202}_{\mathbf{0.010}}$ & $\mathbf{0.008}_{\mathbf{0.000}}^{\dagger}$ \\
 & & 5 & $\mathbf{0.281}_{\mathbf{0.050}}$ & $\mathbf{0.418}_{\mathbf{0.021}}$ & $0.160_{0.015}$ & $\mathbf{0.141}_{\mathbf{0.008}}$ & $\mathbf{0.209}_{\mathbf{0.005}}$ & $\mathbf{0.144}_{\mathbf{0.009}}$ & $\mathbf{0.008}_{\mathbf{0.001}}^{\dagger}$ \\
 & & 6 & $\mathbf{0.244}_{\mathbf{0.057}}$ & $\mathbf{0.380}_{\mathbf{0.013}}$ & $0.181_{0.030}$ & $\mathbf{0.112}_{\mathbf{0.007}}$ & $\mathbf{0.175}_{\mathbf{0.004}}^{\dagger}$ & $\mathbf{0.114}_{\mathbf{0.008}}$ & $\mathbf{0.009}_{\mathbf{0.001}}^{\dagger}$ \\
\cline{2-10}
 & \multirow{5}{*}{DANet} & 2 & $0.484_{0.036}$ & $0.526_{0.018}$ & $\mathbf{0.118}_{\mathbf{0.011}}$ & $0.403_{0.035}$ & $0.368_{0.043}$ & $0.360_{0.038}$ & $0.118_{0.033}$ \\
 & & 3 & $0.357_{0.025}$ & $0.452_{0.016}$ & $\mathbf{0.153}_{\mathbf{0.020}}$ & $0.240_{0.008}$ & $0.266_{0.007}$ & $0.225_{0.009}$ & $0.077_{0.010}$ \\
 & & 4 & $0.323_{0.038}$ & $0.433_{0.018}$ & $\mathbf{0.199}_{\mathbf{0.015}}^{\dagger}$ & $0.188_{0.022}$ & $0.227_{0.015}$ & $0.181_{0.017}$ & $0.088_{0.019}$ \\
 & & 5 & $0.262_{0.015}$ & $0.388_{0.008}$ & $\mathbf{0.195}_{\mathbf{0.028}}$ & $0.135_{0.016}$ & $0.194_{0.013}$ & $0.140_{0.011}$ & $0.118_{0.041}$ \\
 & & 6 & $0.214_{0.021}$ & $0.360_{0.005}$ & $\mathbf{0.226}_{\mathbf{0.042}}$ & $0.100_{0.011}$ & $0.162_{0.005}$ & $0.106_{0.007}$ & $0.131_{0.020}$ \\
\midrule
\multirow{10}{*}{\textbf{Zoomer}} & \multirow{5}{*}{TGCM} & 2 & $\mathbf{0.572}_{\mathbf{0.049}}$ & $\mathbf{0.617}_{\mathbf{0.034}}$ & $0.089_{0.026}$ & $0.464_{0.045}$ & $0.476_{0.027}$ & $0.435_{0.039}$ & $\mathbf{0.008}_{\mathbf{0.001}}^{\dagger}$ \\
 & & 3 & $0.392_{0.047}$ & $0.485_{0.031}$ & $0.113_{0.019}$ & $0.243_{0.013}$ & $0.288_{0.022}$ & $0.246_{0.021}$ & $\mathbf{0.007}_{\mathbf{0.001}}^{\dagger}$ \\
 & & 4 & $\mathbf{0.322}_{\mathbf{0.043}}$ & $0.415_{0.026}$ & $0.140_{0.019}$ & $\mathbf{0.194}_{\mathbf{0.011}}^{\dagger}$ & $\mathbf{0.250}_{\mathbf{0.008}}$ & $\mathbf{0.190}_{\mathbf{0.019}}$ & $\mathbf{0.007}_{\mathbf{0.000}}^{\dagger}$ \\
 & & 5 & $0.234_{0.017}$ & $0.369_{0.019}$ & $0.155_{0.021}$ & $\mathbf{0.145}_{\mathbf{0.013}}$ & $\mathbf{0.200}_{\mathbf{0.010}}$ & $\mathbf{0.144}_{\mathbf{0.010}}$ & $\mathbf{0.007}_{\mathbf{0.001}}^{\dagger}$ \\
 & & 6 & $0.226_{0.022}$ & $0.321_{0.022}$ & $0.205_{0.021}$ & $\mathbf{0.120}_{\mathbf{0.010}}$ & $0.162_{0.014}$ & $\mathbf{0.117}_{\mathbf{0.008}}$ & $\mathbf{0.007}_{\mathbf{0.000}}^{\dagger}$ \\
\cline{2-10}
 & \multirow{5}{*}{DANet} & 2 & $0.550_{0.064}$ & $0.604_{0.015}$ & $\mathbf{0.235}_{\mathbf{0.025}}^{\dagger}$ & $\mathbf{0.495}_{\mathbf{0.085}}$ & $\mathbf{0.478}_{\mathbf{0.087}}$ & $\mathbf{0.460}_{\mathbf{0.079}}$ & $0.066_{0.018}$ \\
 & & 3 & $\mathbf{0.449}_{\mathbf{0.035}}$ & $\mathbf{0.533}_{\mathbf{0.022}}$ & $\mathbf{0.227}_{\mathbf{0.051}}^{\dagger}$ & $\mathbf{0.294}_{\mathbf{0.021}}^{\dagger}$ & $\mathbf{0.325}_{\mathbf{0.027}}$ & $\mathbf{0.287}_{\mathbf{0.022}}$ & $0.093_{0.017}$ \\
 & & 4 & $0.306_{0.038}$ & $\mathbf{0.458}_{\mathbf{0.012}}$ & $\mathbf{0.218}_{\mathbf{0.013}}^{\dagger}$ & $0.167_{0.011}$ & $0.234_{0.010}$ & $0.171_{0.013}$ & $0.084_{0.010}$ \\
 & & 5 & $\mathbf{0.237}_{\mathbf{0.021}}$ & $\mathbf{0.389}_{\mathbf{0.011}}$ & $\mathbf{0.240}_{\mathbf{0.014}}^{\dagger}$ & $0.133_{0.015}$ & $0.189_{0.008}$ & $0.141_{0.013}$ & $0.131_{0.091}$ \\
 & & 6 & $\mathbf{0.253}_{\mathbf{0.061}}$ & $\mathbf{0.394}_{\mathbf{0.019}}^{\dagger}$ & $\mathbf{0.264}_{\mathbf{0.010}}^{\dagger}$ & $0.108_{0.027}$ & $\mathbf{0.162}_{\mathbf{0.015}}$ & $0.114_{0.021}$ & $0.085_{0.021}$ \\
\midrule
\multirow{10}{*}{\textbf{TREC}} & \multirow{5}{*}{TGCM} & 2 & $\mathbf{0.605}_{\mathbf{0.051}}$ & $\mathbf{0.666}_{\mathbf{0.066}}$ & $0.064_{0.028}$ & $0.402_{0.030}$ & $0.481_{0.036}$ & $0.412_{0.018}$ & $\mathbf{0.008}_{\mathbf{0.000}}^{\dagger}$ \\
 & & 3 & $\mathbf{0.447}_{\mathbf{0.056}}$ & $\mathbf{0.537}_{\mathbf{0.037}}$ & $0.103_{0.026}$ & $0.224_{0.012}$ & $0.306_{0.029}$ & $0.239_{0.021}$ & $\mathbf{0.008}_{\mathbf{0.001}}^{\dagger}$ \\
 & & 4 & $\mathbf{0.340}_{\mathbf{0.031}}$ & $0.437_{0.019}$ & $0.141_{0.021}$ & $0.162_{0.008}$ & $\mathbf{0.238}_{\mathbf{0.021}}$ & $0.173_{0.011}$ & $\mathbf{0.007}_{\mathbf{0.000}}^{\dagger}$ \\
 & & 5 & $\mathbf{0.265}_{\mathbf{0.015}}$ & $\mathbf{0.366}_{\mathbf{0.018}}$ & $0.172_{0.010}$ & $\mathbf{0.141}_{\mathbf{0.012}}$ & $\mathbf{0.206}_{\mathbf{0.009}}$ & $0.141_{0.009}$ & $\mathbf{0.007}_{\mathbf{0.001}}^{\dagger}$ \\
 & & 6 & $\mathbf{0.234}_{\mathbf{0.013}}$ & $0.319_{0.020}$ & $0.193_{0.009}$ & $\mathbf{0.114}_{\mathbf{0.018}}$ & $\mathbf{0.171}_{\mathbf{0.021}}$ & $\mathbf{0.118}_{\mathbf{0.014}}$ & $\mathbf{0.008}_{\mathbf{0.001}}^{\dagger}$ \\
\cline{2-10}
 & \multirow{5}{*}{DANet} & 2 & $0.570_{0.090}$ & $0.630_{0.029}$ & $\mathbf{0.343}_{\mathbf{0.051}}^{\dagger}$ & $\mathbf{0.515}_{\mathbf{0.105}}$ & $\mathbf{0.493}_{\mathbf{0.111}}$ & $\mathbf{0.484}_{\mathbf{0.106}}$ & $0.092_{0.041}$ \\
 & & 3 & $0.375_{0.040}$ & $0.488_{0.018}$ & $\mathbf{0.258}_{\mathbf{0.036}}^{\dagger}$ & $\mathbf{0.280}_{\mathbf{0.043}}$ & $\mathbf{0.310}_{\mathbf{0.039}}$ & $\mathbf{0.266}_{\mathbf{0.036}}$ & $0.135_{0.074}$ \\
 & & 4 & $0.299_{0.069}$ & $\mathbf{0.454}_{\mathbf{0.006}}$ & $\mathbf{0.279}_{\mathbf{0.011}}^{\dagger}$ & $\mathbf{0.181}_{\mathbf{0.039}}$ & $0.235_{0.037}$ & $\mathbf{0.181}_{\mathbf{0.041}}$ & $0.110_{0.046}$ \\
 & & 5 & $0.241_{0.038}$ & $0.350_{0.017}$ & $\mathbf{0.261}_{\mathbf{0.009}}^{\dagger}$ & $0.133_{0.021}$ & $0.197_{0.016}$ & $\mathbf{0.141}_{\mathbf{0.019}}$ & $0.070_{0.008}$ \\
 & & 6 & $0.217_{0.040}$ & $\mathbf{0.348}_{\mathbf{0.007}}$ & $\mathbf{0.279}_{\mathbf{0.008}}^{\dagger}$ & $0.104_{0.018}$ & $0.167_{0.018}$ & $0.114_{0.017}$ & $0.097_{0.016}$ \\
\bottomrule
\end{tabular}}
\vspace{0.25em}
\begin{minipage}{0.98\textwidth}
\footnotesize Bold indicates the better mean between TGCM and DANet for the same upstream extractor, mixture size, and metric. Time is measured as wall-clock inference time; lower is better.
\end{minipage}
\end{table*}

\clearpage
\section{Unknown-$K$ Budget Sensitivity on \textsc{CAPTure}}
\label{app:kmax_budget_sensitivity}
To quantify the cost of unknown-$K$ inference, we perform an inference-only sensitivity study using the same trained checkpoint without retraining. For each pooled \textsc{CAPTure} mixture size, we compare Known-$K$ decoding, where the true number of latent episodes is provided during decoding, with Up-to-6 decoding, where the model may use at most six episode slots. Table~\ref{tab:capture_kmax_sensitivity} reports the resulting accuracy and Macro-F1 differences. This sensitivity table reports aggregate inference-only differences and does not attach confidence intervals; statistical comparisons for the end-to-end \textsc{CAPTure} evaluation are reported in Tables~\ref{tab:capture_upstream} and~\ref{tab:capture_upstream_detail}.
\begin{table}[h!]
\centering
\caption{\textbf{Cost of unknown-$K$ inference on \textsc{CAPTure} after pooling single-host and multi-host evaluations.} Known-$K$ decoding gives the model the true number of latent episodes during decoding. Up-to-6 decoding is the default unknown-$K$ setting, where the model may use at most six episode slots. $\Delta$ reports Up-to-6 minus Known-$K$ performance.}
\label{tab:capture_kmax_sensitivity}
\footnotesize
\setlength{\tabcolsep}{3.2pt}
\begin{tabular*}{\textwidth}{@{\extracolsep{\fill}}llrrrrrr@{}}
\toprule
Extractor & True $K$ & Known-$K$ Acc & Up-to-6 Acc & $\Delta$Acc & Known-$K$ Macro-F1 & Up-to-6 Macro-F1 & $\Delta$Macro-F1 \\
\midrule
SFM & 2 & $\mathbf{0.630}$ & $0.617$ & $-0.014$ (-2.2\%) & $\mathbf{0.443}$ & $0.419$ & $-0.024$ (-5.4\%) \\
SFM & 3 & $\mathbf{0.434}$ & $0.431$ & $-0.003$ (-0.7\%) & $\mathbf{0.255}$ & $0.250$ & $-0.005$ (-1.9\%) \\
SFM & 4 & $\mathbf{0.394}$ & $0.394$ & $-0.001$ (-0.1\%) & $\mathbf{0.203}$ & $0.202$ & $-0.002$ (-0.8\%) \\
SFM & 5 & $\mathbf{0.282}$ & $0.281$ & $-0.001$ (-0.2\%) & $\mathbf{0.145}$ & $0.144$ & $-0.001$ (-0.7\%) \\
SFM & 6 & $\mathbf{0.244}$ & $\mathbf{0.244}$ & \textsc{No drop} & $\mathbf{0.114}$ & $\mathbf{0.114}$ & \textsc{No drop} \\
\midrule
Zoomer & 2 & $\mathbf{0.596}$ & $0.572$ & $-0.025$ (-4.1\%) & $\mathbf{0.464}$ & $0.435$ & $-0.029$ (-6.2\%) \\
Zoomer & 3 & $\mathbf{0.392}$ & $0.392$ & \textsc{No drop} & $\mathbf{0.247}$ & $0.246$ & $-0.001$ (-0.4\%) \\
Zoomer & 4 & $\mathbf{0.322}$ & $\mathbf{0.322}$ & \textsc{No drop} & $\mathbf{0.191}$ & $0.190$ & $-0.001$ (-0.4\%) \\
Zoomer & 5 & $\mathbf{0.234}$ & $\mathbf{0.234}$ & \textsc{No drop} & $\mathbf{0.145}$ & $0.144$ & \textsc{No drop} \\
Zoomer & 6 & $\mathbf{0.226}$ & $\mathbf{0.226}$ & \textsc{No drop} & $\mathbf{0.117}$ & $\mathbf{0.117}$ & \textsc{No drop} \\
\midrule
TREC & 2 & $\mathbf{0.615}$ & $0.605$ & $-0.010$ (-1.7\%) & $\mathbf{0.425}$ & $0.412$ & $-0.013$ (-3.0\%) \\
TREC & 3 & $\mathbf{0.460}$ & $0.447$ & $-0.013$ (-2.8\%) & $\mathbf{0.248}$ & $0.239$ & $-0.009$ (-3.8\%) \\
TREC & 4 & $\mathbf{0.349}$ & $0.340$ & $-0.009$ (-2.6\%) & $\mathbf{0.177}$ & $0.173$ & $-0.004$ (-2.1\%) \\
TREC & 5 & $\mathbf{0.266}$ & $0.265$ & $-0.001$ (-0.3\%) & $\mathbf{0.141}$ & $0.141$ & \textsc{No drop} \\
TREC & 6 & $\mathbf{0.234}$ & $\mathbf{0.234}$ & \textsc{No drop} & $\mathbf{0.118}$ & $\mathbf{0.118}$ & \textsc{No drop} \\
\bottomrule
\end{tabular*}
\end{table}

\clearpage

\section{Complete DARPA TC-E5 Noise Robustness Results}
\label{app:noise_robustness_full}
\begingroup
\scriptsize
\setlength{\tabcolsep}{1.0pt}
\renewcommand{\arraystretch}{1.03}
\setlength{\LTleft}{0pt}
\setlength{\LTright}{0pt}
\setlength{\LTcapwidth}{\textwidth}
\newcommand{\pcell}[2]{${#1}_{#2}$}
\newcommand{\bcell}[2]{${\mathbf{#1}}_{\mathbf{#2}}$}
\newcommand{\dcell}[2]{${\mathbf{#1}}_{\mathbf{#2}}^{\dagger}$}

\begin{longtable}{@{}>{\raggedright\arraybackslash}p{0.105\textwidth}>{\centering\arraybackslash}p{0.030\textwidth}>{\centering\arraybackslash}p{0.040\textwidth}>{\raggedright\arraybackslash}p{0.060\textwidth}*{7}{>{\centering\arraybackslash}p{0.103\textwidth}}@{}}
\caption{Complete DARPA TC-E5 noise robustness results. Each cell reports mean$_{\mathrm{std}}$ over repeated perturbation samples and seeds. Higher is better for Acc, FMI, NMI, P, R, and Macro-F1; lower is better for Time. Bold marks the better mean between TGCM and DANet for the same noise type, $K$, and $\rho$. Superscript $\dagger$ marks a confidence-interval separation: for Acc, FMI, NMI, P, R, and Macro-F1, the favored model's 95\% CI is entirely higher than the counterpart; for Time, it is entirely lower. This appendix table provides the full values summarized by the compact delta table in Table~\ref{tab:noise_robustness}.}
\label{tab:noise_robustness_full}\\
\toprule
\textbf{Noise} & \textbf{$K$} & $\boldsymbol{\rho}$ & \textbf{Model} & \textbf{Acc} & \textbf{FMI} & \textbf{NMI} & \textbf{P} & \textbf{R} & \textbf{Macro-F1} & \textbf{Time} \\
\midrule
\endfirsthead

\toprule
\textbf{Noise} & \textbf{$K$} & $\boldsymbol{\rho}$ & \textbf{Model} & \textbf{Acc} & \textbf{FMI} & \textbf{NMI} & \textbf{P} & \textbf{R} & \textbf{Macro-F1} & \textbf{Time} \\
\midrule
\endhead

\midrule
\multicolumn{11}{r}{\emph{Continued on next page}}\\
\endfoot

\bottomrule
\multicolumn{11}{p{0.98\textwidth}}{{\footnotesize\emph{Note.} Superscript $\dagger$ is assigned from the reported model-level 95\% confidence intervals and is used only when the two model CIs do not overlap in the favorable direction.}}\\
\endlastfoot
Missing & 2 & 0.1 & TGCM & \bcell{0.538}{0.075} & \pcell{0.593}{0.050} & \pcell{0.157}{0.101} & \bcell{0.549}{0.071} & \bcell{0.513}{0.135} & \bcell{0.469}{0.086} & \dcell{0.016}{0.003} \\
Missing & 2 & 0.1 & DANet & \pcell{0.460}{0.070} & \bcell{0.616}{0.061} & \bcell{0.258}{0.126} & \pcell{0.413}{0.063} & \pcell{0.390}{0.057} & \pcell{0.386}{0.053} & \pcell{1.310}{0.024} \\
Missing & 2 & 0.2 & TGCM & \bcell{0.529}{0.092} & \pcell{0.590}{0.061} & \pcell{0.099}{0.052} & \bcell{0.566}{0.084} & \bcell{0.507}{0.093} & \bcell{0.464}{0.066} & \dcell{0.012}{0.001} \\
Missing & 2 & 0.2 & DANet & \pcell{0.441}{0.094} & \bcell{0.608}{0.034} & \dcell{0.334}{0.030} & \pcell{0.383}{0.091} & \pcell{0.373}{0.093} & \pcell{0.362}{0.096} & \pcell{1.271}{0.088} \\
Missing & 2 & 0.3 & TGCM & \bcell{0.540}{0.099} & \pcell{0.604}{0.070} & \pcell{0.143}{0.108} & \bcell{0.547}{0.051} & \bcell{0.528}{0.075} & \bcell{0.482}{0.055} & \dcell{0.017}{0.003} \\
Missing & 2 & 0.3 & DANet & \pcell{0.479}{0.109} & \bcell{0.634}{0.046} & \bcell{0.305}{0.092} & \pcell{0.434}{0.103} & \pcell{0.430}{0.091} & \pcell{0.407}{0.106} & \pcell{1.245}{0.089} \\
Missing & 2 & 0.4 & TGCM & \bcell{0.493}{0.106} & \pcell{0.580}{0.063} & \pcell{0.096}{0.087} & \bcell{0.495}{0.095} & \bcell{0.487}{0.074} & \bcell{0.426}{0.091} & \dcell{0.018}{0.003} \\
Missing & 2 & 0.4 & DANet & \pcell{0.477}{0.075} & \bcell{0.640}{0.061} & \bcell{0.310}{0.114} & \pcell{0.397}{0.133} & \pcell{0.403}{0.108} & \pcell{0.390}{0.114} & \pcell{1.249}{0.123} \\
Missing & 2 & 0.5 & TGCM & \bcell{0.522}{0.082} & \pcell{0.573}{0.054} & \pcell{0.072}{0.045} & \bcell{0.457}{0.096} & \bcell{0.475}{0.118} & \bcell{0.426}{0.100} & \dcell{0.012}{0.001} \\
Missing & 2 & 0.5 & DANet & \pcell{0.424}{0.118} & \bcell{0.615}{0.072} & \dcell{0.285}{0.080} & \pcell{0.358}{0.155} & \pcell{0.348}{0.119} & \pcell{0.334}{0.128} & \pcell{1.318}{0.030} \\
Missing & 2 & 0.6 & TGCM & \bcell{0.531}{0.081} & \pcell{0.575}{0.086} & \pcell{0.113}{0.112} & \bcell{0.458}{0.093} & \bcell{0.458}{0.120} & \bcell{0.426}{0.089} & \dcell{0.014}{0.002} \\
Missing & 2 & 0.6 & DANet & \pcell{0.441}{0.091} & \bcell{0.637}{0.049} & \bcell{0.302}{0.050} & \pcell{0.352}{0.097} & \pcell{0.361}{0.089} & \pcell{0.334}{0.094} & \pcell{1.228}{0.109} \\
Missing & 2 & 0.7 & TGCM & \bcell{0.577}{0.056} & \bcell{0.634}{0.088} & \pcell{0.134}{0.141} & \bcell{0.413}{0.080} & \bcell{0.467}{0.060} & \bcell{0.417}{0.043} & \dcell{0.014}{0.003} \\
Missing & 2 & 0.7 & DANet & \pcell{0.418}{0.117} & \pcell{0.603}{0.069} & \bcell{0.307}{0.114} & \pcell{0.357}{0.086} & \pcell{0.337}{0.115} & \pcell{0.326}{0.092} & \pcell{1.266}{0.072} \\
Missing & 2 & 0.8 & TGCM & \pcell{0.564}{0.108} & \pcell{0.555}{0.207} & \pcell{0.128}{0.173} & \pcell{0.418}{0.133} & \pcell{0.525}{0.123} & \pcell{0.436}{0.119} & \dcell{0.013}{0.002} \\
Missing & 2 & 0.8 & DANet & \bcell{0.586}{0.161} & \bcell{0.620}{0.157} & \bcell{0.426}{0.237} & \bcell{0.555}{0.162} & \bcell{0.549}{0.169} & \bcell{0.534}{0.152} & \pcell{0.959}{0.030} \\
Missing & 2 & 0.9 & TGCM & \bcell{0.583}{0.078} & \bcell{0.644}{0.061} & \pcell{0.369}{0.175} & \pcell{0.478}{0.098} & \bcell{0.560}{0.092} & \bcell{0.489}{0.083} & \dcell{0.014}{0.001} \\
Missing & 2 & 0.9 & DANet & \pcell{0.543}{0.135} & \pcell{0.594}{0.278} & \bcell{0.587}{0.280} & \bcell{0.500}{0.155} & \pcell{0.486}{0.124} & \pcell{0.484}{0.141} & \pcell{0.425}{0.045} \\

Missing & 4 & 0.1 & TGCM & \bcell{0.492}{0.075} & \pcell{0.564}{0.050} & \pcell{0.159}{0.081} & \bcell{0.519}{0.049} & \bcell{0.476}{0.115} & \bcell{0.432}{0.070} & \dcell{0.010}{0.001} \\
Missing & 4 & 0.1 & DANet & \pcell{0.432}{0.073} & \bcell{0.620}{0.054} & \bcell{0.301}{0.101} & \pcell{0.370}{0.078} & \pcell{0.374}{0.059} & \pcell{0.355}{0.064} & \pcell{0.059}{0.020} \\
Missing & 4 & 0.2 & TGCM & \bcell{0.502}{0.093} & \pcell{0.560}{0.051} & \pcell{0.128}{0.055} & \bcell{0.516}{0.094} & \bcell{0.471}{0.060} & \bcell{0.432}{0.045} & \dcell{0.009}{0.000} \\
Missing & 4 & 0.2 & DANet & \pcell{0.418}{0.118} & \bcell{0.600}{0.037} & \dcell{0.360}{0.042} & \pcell{0.355}{0.115} & \pcell{0.362}{0.104} & \pcell{0.340}{0.116} & \pcell{0.023}{0.003} \\
Missing & 4 & 0.3 & TGCM & \bcell{0.526}{0.118} & \pcell{0.573}{0.066} & \pcell{0.170}{0.120} & \bcell{0.520}{0.083} & \bcell{0.504}{0.050} & \bcell{0.457}{0.067} & \dcell{0.009}{0.000} \\
Missing & 4 & 0.3 & DANet & \pcell{0.447}{0.105} & \bcell{0.627}{0.036} & \bcell{0.338}{0.070} & \pcell{0.389}{0.095} & \pcell{0.408}{0.084} & \pcell{0.375}{0.100} & \pcell{0.021}{0.000} \\
Missing & 4 & 0.4 & TGCM & \bcell{0.462}{0.105} & \pcell{0.558}{0.052} & \pcell{0.123}{0.087} & \bcell{0.458}{0.096} & \bcell{0.450}{0.044} & \bcell{0.384}{0.080} & \dcell{0.009}{0.001} \\
Missing & 4 & 0.4 & DANet & \pcell{0.438}{0.095} & \bcell{0.643}{0.071} & \bcell{0.367}{0.127} & \pcell{0.359}{0.128} & \pcell{0.373}{0.114} & \pcell{0.353}{0.112} & \pcell{0.022}{0.001} \\
Missing & 4 & 0.5 & TGCM & \bcell{0.490}{0.071} & \pcell{0.546}{0.066} & \pcell{0.096}{0.046} & \bcell{0.391}{0.104} & \bcell{0.428}{0.087} & \bcell{0.379}{0.075} & \dcell{0.010}{0.001} \\
Missing & 4 & 0.5 & DANet & \pcell{0.405}{0.095} & \bcell{0.602}{0.068} & \dcell{0.319}{0.072} & \pcell{0.338}{0.138} & \pcell{0.327}{0.100} & \pcell{0.314}{0.110} & \pcell{0.021}{0.001} \\
Missing & 4 & 0.6 & TGCM & \bcell{0.509}{0.074} & \pcell{0.545}{0.100} & \pcell{0.131}{0.120} & \bcell{0.425}{0.066} & \bcell{0.429}{0.094} & \bcell{0.394}{0.063} & \dcell{0.009}{0.000} \\
Missing & 4 & 0.6 & DANet & \pcell{0.398}{0.085} & \bcell{0.623}{0.053} & \bcell{0.341}{0.083} & \pcell{0.304}{0.102} & \pcell{0.335}{0.083} & \pcell{0.295}{0.095} & \pcell{0.022}{0.001} \\
Missing & 4 & 0.7 & TGCM & \bcell{0.550}{0.072} & \bcell{0.595}{0.072} & \pcell{0.132}{0.103} & \bcell{0.384}{0.049} & \bcell{0.446}{0.042} & \bcell{0.389}{0.023} & \dcell{0.009}{0.000} \\
Missing & 4 & 0.7 & DANet & \pcell{0.422}{0.126} & \pcell{0.584}{0.054} & \bcell{0.281}{0.083} & \pcell{0.347}{0.088} & \pcell{0.347}{0.108} & \pcell{0.324}{0.096} & \pcell{0.027}{0.009} \\
Missing & 4 & 0.8 & TGCM & \pcell{0.481}{0.067} & \pcell{0.512}{0.174} & \pcell{0.082}{0.086} & \pcell{0.329}{0.072} & \pcell{0.449}{0.065} & \pcell{0.354}{0.059} & \dcell{0.010}{0.001} \\
Missing & 4 & 0.8 & DANet & \bcell{0.523}{0.139} & \bcell{0.619}{0.157} & \dcell{0.462}{0.195} & \bcell{0.483}{0.147} & \bcell{0.501}{0.150} & \bcell{0.470}{0.135} & \pcell{0.053}{0.032} \\
Missing & 4 & 0.9 & TGCM & \bcell{0.536}{0.096} & \bcell{0.636}{0.108} & \pcell{0.343}{0.200} & \pcell{0.410}{0.145} & \bcell{0.503}{0.118} & \pcell{0.430}{0.118} & \bcell{0.009}{0.000} \\
Missing & 4 & 0.9 & DANet & \pcell{0.519}{0.148} & \pcell{0.615}{0.327} & \bcell{0.595}{0.260} & \bcell{0.463}{0.172} & \pcell{0.470}{0.146} & \bcell{0.459}{0.160} & \pcell{0.030}{0.018} \\

Confusion & 2 & 0.1 & TGCM & \bcell{0.538}{0.111} & \pcell{0.578}{0.060} & \pcell{0.148}{0.088} & \bcell{0.599}{0.111} & \bcell{0.524}{0.155} & \bcell{0.485}{0.124} & \dcell{0.015}{0.002} \\
Confusion & 2 & 0.1 & DANet & \pcell{0.477}{0.085} & \bcell{0.631}{0.046} & \bcell{0.277}{0.107} & \pcell{0.415}{0.112} & \pcell{0.426}{0.102} & \pcell{0.405}{0.097} & \pcell{1.150}{0.220} \\
Confusion & 2 & 0.2 & TGCM & \bcell{0.544}{0.117} & \pcell{0.600}{0.048} & \pcell{0.143}{0.103} & \bcell{0.532}{0.123} & \bcell{0.515}{0.117} & \bcell{0.467}{0.118} & \dcell{0.014}{0.002} \\
Confusion & 2 & 0.2 & DANet & \pcell{0.434}{0.080} & \bcell{0.630}{0.028} & \bcell{0.281}{0.036} & \pcell{0.359}{0.100} & \pcell{0.369}{0.086} & \pcell{0.343}{0.085} & \pcell{1.389}{0.139} \\
Confusion & 2 & 0.3 & TGCM & \bcell{0.556}{0.105} & \bcell{0.602}{0.040} & \pcell{0.161}{0.101} & \dcell{0.579}{0.054} & \dcell{0.526}{0.080} & \dcell{0.477}{0.090} & \dcell{0.017}{0.004} \\
Confusion & 2 & 0.3 & DANet & \pcell{0.394}{0.054} & \pcell{0.597}{0.053} & \bcell{0.250}{0.082} & \pcell{0.324}{0.077} & \pcell{0.311}{0.027} & \pcell{0.290}{0.037} & \pcell{1.357}{0.079} \\
Confusion & 2 & 0.4 & TGCM & \bcell{0.573}{0.088} & \pcell{0.586}{0.050} & \pcell{0.172}{0.108} & \bcell{0.576}{0.127} & \bcell{0.551}{0.117} & \bcell{0.514}{0.099} & \dcell{0.014}{0.003} \\
Confusion & 2 & 0.4 & DANet & \pcell{0.449}{0.075} & \bcell{0.621}{0.041} & \bcell{0.277}{0.097} & \pcell{0.352}{0.113} & \pcell{0.344}{0.095} & \pcell{0.329}{0.102} & \pcell{1.314}{0.020} \\
Confusion & 2 & 0.5 & TGCM & \bcell{0.556}{0.121} & \bcell{0.599}{0.032} & \pcell{0.154}{0.109} & \bcell{0.596}{0.086} & \bcell{0.559}{0.125} & \bcell{0.505}{0.108} & \dcell{0.013}{0.001} \\
Confusion & 2 & 0.5 & DANet & \pcell{0.444}{0.136} & \pcell{0.598}{0.034} & \bcell{0.250}{0.068} & \pcell{0.416}{0.141} & \pcell{0.377}{0.150} & \pcell{0.373}{0.133} & \pcell{1.302}{0.181} \\
Confusion & 2 & 0.6 & TGCM & \bcell{0.577}{0.106} & \pcell{0.600}{0.033} & \pcell{0.146}{0.087} & \bcell{0.551}{0.099} & \bcell{0.546}{0.139} & \bcell{0.514}{0.110} & \dcell{0.013}{0.002} \\
Confusion & 2 & 0.6 & DANet & \pcell{0.402}{0.116} & \bcell{0.607}{0.036} & \bcell{0.307}{0.069} & \pcell{0.353}{0.096} & \pcell{0.323}{0.104} & \pcell{0.322}{0.098} & \pcell{1.282}{0.055} \\
Confusion & 2 & 0.7 & TGCM & \bcell{0.601}{0.070} & \pcell{0.611}{0.039} & \pcell{0.161}{0.143} & \dcell{0.605}{0.072} & \dcell{0.590}{0.077} & \dcell{0.534}{0.055} & \dcell{0.016}{0.004} \\
Confusion & 2 & 0.7 & DANet & \pcell{0.445}{0.089} & \bcell{0.622}{0.041} & \bcell{0.267}{0.085} & \pcell{0.364}{0.088} & \pcell{0.352}{0.104} & \pcell{0.336}{0.090} & \pcell{1.317}{0.028} \\
Confusion & 2 & 0.8 & TGCM & \bcell{0.585}{0.040} & \pcell{0.582}{0.036} & \pcell{0.098}{0.104} & \bcell{0.539}{0.066} & \bcell{0.541}{0.047} & \bcell{0.505}{0.033} & \dcell{0.018}{0.005} \\
Confusion & 2 & 0.8 & DANet & \pcell{0.475}{0.164} & \bcell{0.627}{0.041} & \bcell{0.287}{0.107} & \pcell{0.412}{0.152} & \pcell{0.405}{0.152} & \pcell{0.375}{0.162} & \pcell{1.234}{0.117} \\
Confusion & 2 & 0.9 & TGCM & \bcell{0.590}{0.085} & \bcell{0.594}{0.060} & \pcell{0.119}{0.092} & \dcell{0.558}{0.066} & \dcell{0.525}{0.055} & \dcell{0.489}{0.040} & \dcell{0.013}{0.001} \\
Confusion & 2 & 0.9 & DANet & \pcell{0.437}{0.086} & \pcell{0.572}{0.038} & \bcell{0.202}{0.076} & \pcell{0.369}{0.063} & \pcell{0.354}{0.062} & \pcell{0.339}{0.056} & \pcell{1.377}{0.151} \\

Confusion & 4 & 0.1 & TGCM & \bcell{0.491}{0.110} & \pcell{0.558}{0.055} & \pcell{0.165}{0.093} & \bcell{0.544}{0.092} & \bcell{0.480}{0.142} & \bcell{0.437}{0.114} & \dcell{0.015}{0.002} \\
Confusion & 4 & 0.1 & DANet & \pcell{0.453}{0.073} & \bcell{0.627}{0.049} & \bcell{0.299}{0.107} & \pcell{0.391}{0.117} & \pcell{0.422}{0.093} & \pcell{0.389}{0.098} & \pcell{1.282}{0.036} \\
Confusion & 4 & 0.2 & TGCM & \bcell{0.500}{0.106} & \pcell{0.579}{0.041} & \pcell{0.163}{0.097} & \bcell{0.485}{0.119} & \bcell{0.472}{0.100} & \bcell{0.424}{0.111} & \dcell{0.014}{0.001} \\
Confusion & 4 & 0.2 & DANet & \pcell{0.435}{0.084} & \bcell{0.626}{0.034} & \bcell{0.311}{0.049} & \pcell{0.359}{0.096} & \pcell{0.380}{0.078} & \pcell{0.345}{0.081} & \pcell{1.197}{0.069} \\
Confusion & 4 & 0.3 & TGCM & \bcell{0.526}{0.103} & \pcell{0.577}{0.036} & \pcell{0.173}{0.111} & \dcell{0.519}{0.034} & \dcell{0.501}{0.063} & \dcell{0.441}{0.080} & \dcell{0.012}{0.001} \\
Confusion & 4 & 0.3 & DANet & \pcell{0.379}{0.040} & \bcell{0.585}{0.053} & \bcell{0.264}{0.077} & \pcell{0.305}{0.090} & \pcell{0.311}{0.025} & \pcell{0.279}{0.043} & \pcell{1.103}{0.050} \\
Confusion & 4 & 0.4 & TGCM & \bcell{0.538}{0.081} & \pcell{0.560}{0.043} & \pcell{0.179}{0.120} & \bcell{0.521}{0.122} & \bcell{0.517}{0.100} & \bcell{0.474}{0.093} & \dcell{0.013}{0.001} \\
Confusion & 4 & 0.4 & DANet & \pcell{0.447}{0.061} & \bcell{0.617}{0.043} & \bcell{0.303}{0.080} & \pcell{0.358}{0.085} & \pcell{0.361}{0.059} & \pcell{0.333}{0.067} & \pcell{1.121}{0.127} \\
Confusion & 4 & 0.5 & TGCM & \bcell{0.515}{0.102} & \pcell{0.576}{0.025} & \pcell{0.177}{0.127} & \bcell{0.541}{0.066} & \bcell{0.513}{0.101} & \bcell{0.455}{0.088} & \dcell{0.013}{0.002} \\
Confusion & 4 & 0.5 & DANet & \pcell{0.445}{0.108} & \bcell{0.584}{0.036} & \bcell{0.253}{0.084} & \pcell{0.402}{0.128} & \pcell{0.384}{0.123} & \pcell{0.368}{0.109} & \pcell{1.102}{0.053} \\
Confusion & 4 & 0.6 & TGCM & \bcell{0.537}{0.094} & \pcell{0.583}{0.027} & \pcell{0.181}{0.103} & \bcell{0.492}{0.104} & \bcell{0.493}{0.116} & \bcell{0.455}{0.094} & \dcell{0.011}{0.001} \\
Confusion & 4 & 0.6 & DANet & \pcell{0.379}{0.110} & \bcell{0.603}{0.041} & \bcell{0.340}{0.063} & \pcell{0.315}{0.084} & \pcell{0.315}{0.090} & \pcell{0.292}{0.089} & \pcell{1.074}{0.098} \\
Confusion & 4 & 0.7 & TGCM & \bcell{0.565}{0.060} & \pcell{0.587}{0.046} & \pcell{0.184}{0.151} & \dcell{0.540}{0.062} & \bcell{0.545}{0.061} & \bcell{0.486}{0.046} & \dcell{0.013}{0.002} \\
Confusion & 4 & 0.7 & DANet & \pcell{0.428}{0.110} & \bcell{0.607}{0.044} & \bcell{0.277}{0.094} & \pcell{0.338}{0.088} & \pcell{0.346}{0.110} & \pcell{0.320}{0.093} & \pcell{1.016}{0.084} \\
Confusion & 4 & 0.8 & TGCM & \bcell{0.541}{0.057} & \pcell{0.558}{0.046} & \pcell{0.117}{0.120} & \bcell{0.479}{0.098} & \bcell{0.502}{0.065} & \bcell{0.460}{0.063} & \dcell{0.013}{0.001} \\
Confusion & 4 & 0.8 & DANet & \pcell{0.449}{0.173} & \bcell{0.620}{0.038} & \bcell{0.314}{0.094} & \pcell{0.376}{0.136} & \pcell{0.394}{0.136} & \pcell{0.350}{0.153} & \pcell{1.053}{0.105} \\
Confusion & 4 & 0.9 & TGCM & \bcell{0.537}{0.093} & \bcell{0.567}{0.058} & \pcell{0.133}{0.111} & \bcell{0.476}{0.054} & \dcell{0.468}{0.038} & \dcell{0.429}{0.044} & \dcell{0.012}{0.001} \\
Confusion & 4 & 0.9 & DANet & \pcell{0.388}{0.075} & \pcell{0.554}{0.026} & \bcell{0.241}{0.071} & \pcell{0.327}{0.077} & \pcell{0.312}{0.053} & \pcell{0.297}{0.056} & \pcell{1.044}{0.102} \\

Insertion & 2 & 0.1 & TGCM & \bcell{0.535}{0.106} & \pcell{0.583}{0.043} & \pcell{0.183}{0.111} & \bcell{0.550}{0.115} & \bcell{0.486}{0.140} & \bcell{0.452}{0.123} & \dcell{0.018}{0.004} \\
Insertion & 2 & 0.1 & DANet & \pcell{0.486}{0.094} & \bcell{0.605}{0.020} & \bcell{0.253}{0.070} & \pcell{0.424}{0.110} & \pcell{0.395}{0.116} & \pcell{0.381}{0.101} & \pcell{1.356}{0.037} \\
Insertion & 2 & 0.2 & TGCM & \bcell{0.472}{0.122} & \pcell{0.572}{0.047} & \pcell{0.157}{0.080} & \bcell{0.521}{0.118} & \bcell{0.421}{0.163} & \bcell{0.391}{0.133} & \dcell{0.017}{0.001} \\
Insertion & 2 & 0.2 & DANet & \pcell{0.389}{0.091} & \bcell{0.666}{0.095} & \bcell{0.387}{0.120} & \pcell{0.356}{0.105} & \pcell{0.333}{0.087} & \pcell{0.329}{0.092} & \pcell{1.373}{0.164} \\
Insertion & 2 & 0.3 & TGCM & \pcell{0.446}{0.074} & \pcell{0.564}{0.053} & \pcell{0.168}{0.049} & \pcell{0.466}{0.094} & \pcell{0.389}{0.127} & \pcell{0.349}{0.089} & \dcell{0.017}{0.003} \\
Insertion & 2 & 0.3 & DANet & \bcell{0.538}{0.132} & \bcell{0.650}{0.054} & \dcell{0.326}{0.077} & \bcell{0.475}{0.102} & \bcell{0.451}{0.130} & \bcell{0.438}{0.121} & \pcell{1.271}{0.204} \\
Insertion & 2 & 0.4 & TGCM & \bcell{0.447}{0.047} & \pcell{0.563}{0.035} & \pcell{0.166}{0.094} & \bcell{0.471}{0.076} & \bcell{0.377}{0.114} & \bcell{0.343}{0.069} & \dcell{0.017}{0.003} \\
Insertion & 2 & 0.4 & DANet & \pcell{0.428}{0.138} & \bcell{0.617}{0.045} & \bcell{0.284}{0.117} & \pcell{0.352}{0.100} & \pcell{0.329}{0.120} & \pcell{0.323}{0.102} & \pcell{1.308}{0.138} \\
Insertion & 2 & 0.5 & TGCM & \pcell{0.377}{0.059} & \pcell{0.540}{0.026} & \pcell{0.183}{0.075} & \pcell{0.401}{0.050} & \pcell{0.276}{0.075} & \pcell{0.272}{0.044} & \dcell{0.014}{0.002} \\
Insertion & 2 & 0.5 & DANet & \bcell{0.502}{0.094} & \bcell{0.610}{0.037} & \bcell{0.275}{0.053} & \bcell{0.468}{0.119} & \bcell{0.409}{0.115} & \bcell{0.406}{0.113} & \pcell{1.297}{0.117} \\
Insertion & 2 & 0.6 & TGCM & \pcell{0.413}{0.106} & \pcell{0.535}{0.047} & \pcell{0.189}{0.061} & \bcell{0.397}{0.048} & \pcell{0.280}{0.096} & \pcell{0.289}{0.068} & \dcell{0.017}{0.004} \\
Insertion & 2 & 0.6 & DANet & \bcell{0.441}{0.143} & \bcell{0.620}{0.047} & \bcell{0.317}{0.057} & \pcell{0.371}{0.151} & \bcell{0.360}{0.147} & \bcell{0.348}{0.144} & \pcell{1.269}{0.089} \\
Insertion & 2 & 0.7 & TGCM & \pcell{0.432}{0.136} & \pcell{0.544}{0.072} & \pcell{0.210}{0.079} & \bcell{0.403}{0.072} & \pcell{0.304}{0.128} & \pcell{0.299}{0.099} & \dcell{0.016}{0.003} \\
Insertion & 2 & 0.7 & DANet & \bcell{0.456}{0.094} & \bcell{0.611}{0.039} & \bcell{0.314}{0.089} & \pcell{0.399}{0.075} & \bcell{0.379}{0.081} & \bcell{0.370}{0.079} & \pcell{1.324}{0.150} \\
Insertion & 2 & 0.8 & TGCM & \pcell{0.410}{0.136} & \pcell{0.536}{0.060} & \pcell{0.191}{0.067} & \bcell{0.371}{0.073} & \pcell{0.274}{0.124} & \pcell{0.272}{0.100} & \dcell{0.016}{0.003} \\
Insertion & 2 & 0.8 & DANet & \bcell{0.445}{0.132} & \bcell{0.611}{0.042} & \dcell{0.343}{0.045} & \pcell{0.353}{0.110} & \bcell{0.319}{0.082} & \bcell{0.321}{0.097} & \pcell{1.332}{0.022} \\
Insertion & 2 & 0.9 & TGCM & \bcell{0.402}{0.140} & \pcell{0.527}{0.069} & \pcell{0.184}{0.085} & \bcell{0.382}{0.080} & \pcell{0.272}{0.130} & \pcell{0.277}{0.112} & \dcell{0.017}{0.005} \\
Insertion & 2 & 0.9 & DANet & \pcell{0.368}{0.186} & \bcell{0.552}{0.022} & \bcell{0.273}{0.073} & \pcell{0.321}{0.155} & \bcell{0.287}{0.152} & \bcell{0.281}{0.146} & \pcell{1.248}{0.081} \\

Insertion & 4 & 0.1 & TGCM & \pcell{0.471}{0.098} & \pcell{0.558}{0.047} & \pcell{0.184}{0.123} & \bcell{0.489}{0.119} & \bcell{0.435}{0.136} & \bcell{0.399}{0.125} & \dcell{0.026}{0.005} \\
Insertion & 4 & 0.1 & DANet & \bcell{0.478}{0.078} & \bcell{0.605}{0.031} & \bcell{0.303}{0.072} & \pcell{0.413}{0.112} & \pcell{0.400}{0.104} & \pcell{0.377}{0.101} & \pcell{1.077}{0.092} \\
Insertion & 4 & 0.2 & TGCM & \bcell{0.423}{0.108} & \pcell{0.555}{0.040} & \pcell{0.181}{0.083} & \bcell{0.451}{0.085} & \bcell{0.376}{0.129} & \bcell{0.346}{0.112} & \dcell{0.023}{0.003} \\
Insertion & 4 & 0.2 & DANet & \pcell{0.366}{0.089} & \bcell{0.655}{0.083} & \bcell{0.411}{0.114} & \pcell{0.323}{0.110} & \pcell{0.322}{0.080} & \pcell{0.305}{0.093} & \pcell{0.982}{0.087} \\
Insertion & 4 & 0.3 & TGCM & \pcell{0.384}{0.068} & \pcell{0.539}{0.051} & \pcell{0.200}{0.045} & \pcell{0.409}{0.069} & \pcell{0.336}{0.101} & \pcell{0.304}{0.078} & \dcell{0.025}{0.005} \\
Insertion & 4 & 0.3 & DANet & \bcell{0.502}{0.121} & \bcell{0.636}{0.049} & \dcell{0.357}{0.077} & \bcell{0.434}{0.101} & \bcell{0.427}{0.115} & \bcell{0.408}{0.108} & \pcell{0.993}{0.057} \\
Insertion & 4 & 0.4 & TGCM & \pcell{0.387}{0.031} & \pcell{0.538}{0.023} & \pcell{0.195}{0.077} & \bcell{0.403}{0.052} & \bcell{0.329}{0.091} & \pcell{0.297}{0.053} & \dcell{0.022}{0.003} \\
Insertion & 4 & 0.4 & DANet & \bcell{0.402}{0.138} & \bcell{0.605}{0.033} & \bcell{0.316}{0.079} & \pcell{0.326}{0.091} & \pcell{0.318}{0.118} & \bcell{0.304}{0.098} & \pcell{0.988}{0.057} \\
Insertion & 4 & 0.5 & TGCM & \pcell{0.341}{0.056} & \pcell{0.529}{0.031} & \pcell{0.200}{0.064} & \pcell{0.364}{0.035} & \pcell{0.264}{0.069} & \pcell{0.250}{0.038} & \dcell{0.023}{0.001} \\
Insertion & 4 & 0.5 & DANet & \bcell{0.471}{0.097} & \dcell{0.613}{0.029} & \bcell{0.326}{0.057} & \bcell{0.415}{0.130} & \bcell{0.389}{0.112} & \bcell{0.376}{0.111} & \pcell{0.930}{0.089} \\
Insertion & 4 & 0.6 & TGCM & \pcell{0.361}{0.112} & \pcell{0.522}{0.043} & \pcell{0.202}{0.048} & \bcell{0.360}{0.052} & \pcell{0.253}{0.093} & \pcell{0.259}{0.072} & \dcell{0.025}{0.002} \\
Insertion & 4 & 0.6 & DANet & \bcell{0.417}{0.128} & \bcell{0.629}{0.050} & \dcell{0.356}{0.060} & \pcell{0.340}{0.131} & \bcell{0.355}{0.136} & \bcell{0.328}{0.130} & \pcell{0.963}{0.075} \\
Insertion & 4 & 0.7 & TGCM & \pcell{0.379}{0.153} & \pcell{0.529}{0.074} & \pcell{0.216}{0.075} & \bcell{0.373}{0.084} & \pcell{0.274}{0.137} & \pcell{0.267}{0.111} & \dcell{0.024}{0.003} \\
Insertion & 4 & 0.7 & DANet & \bcell{0.428}{0.111} & \bcell{0.602}{0.032} & \bcell{0.342}{0.041} & \pcell{0.364}{0.102} & \bcell{0.361}{0.107} & \bcell{0.345}{0.106} & \pcell{0.927}{0.050} \\
Insertion & 4 & 0.8 & TGCM & \pcell{0.356}{0.138} & \pcell{0.514}{0.067} & \pcell{0.193}{0.064} & \bcell{0.334}{0.080} & \pcell{0.242}{0.120} & \pcell{0.237}{0.100} & \bcell{0.019}{0.005} \\
Insertion & 4 & 0.8 & DANet & \bcell{0.401}{0.123} & \bcell{0.597}{0.031} & \dcell{0.383}{0.056} & \pcell{0.318}{0.094} & \bcell{0.291}{0.072} & \bcell{0.290}{0.084} & \pcell{0.377}{0.449} \\
Insertion & 4 & 0.9 & TGCM & \bcell{0.347}{0.134} & \pcell{0.511}{0.068} & \pcell{0.199}{0.064} & \bcell{0.339}{0.077} & \pcell{0.235}{0.111} & \pcell{0.236}{0.096} & \dcell{0.012}{0.002} \\
Insertion & 4 & 0.9 & DANet & \pcell{0.339}{0.165} & \bcell{0.546}{0.019} & \bcell{0.308}{0.046} & \pcell{0.286}{0.136} & \bcell{0.275}{0.135} & \bcell{0.255}{0.130} & \pcell{0.054}{0.014} \\

\end{longtable}
\endgroup

Table~\ref{tab:noise_robustness_full} reports the complete mean and standard deviation values for all metrics used in the robustness evaluation summarized in Table~\ref{tab:noise_robustness}.

\clearpage
\section{Feature-Importance Diagnostics and Embedding-Type Sensitivity}
\label{app:embedding_type_effect}
This appendix evaluates whether the topic-guided variant depends on a specific topic encoder. To avoid confounding topic-cardinality and encoder effects, Table~\ref{tab:embedding_type_sensitivity} fixes $K_{\mathrm{topic}}=3$ and compares embedding types under the Topic+Consistency setting. The reported Acc ranges vary by less than one percentage point across encoders for every mixture size, and Macro-F1 varies within a similarly narrow band. This suggests that the embedding type does not materially change the main occurrence-level assignment conclusion, although individual encoders can slightly shift macro-balance behavior. Table~\ref{tab:lime_feature_importance} reports a LIME-style local feature-importance diagnostic over validation configurations.
\begin{table}[H]
\centering
\caption{\textbf{Diagnostic LIME feature-importance analysis.}
The contribution weight reflects local importance in the validation configuration space. This diagnostic is reported only as an interpretability check.}
\label{tab:lime_feature_importance}
\footnotesize
\setlength{\tabcolsep}{4pt}
\begin{tabularx}{\columnwidth}{@{}>{\raggedright\arraybackslash}Xr@{}}
\toprule
\textbf{Feature} & \textbf{Weight} \\
\midrule
Mixture cardinality $K$ $\downarrow$ & 0.1323  \\
Sequence count $N$ $\uparrow$ & 0.0322  \\
Mixing step $T$ $\downarrow$ & 0.0249  \\
Embedding method & 0.00104 \\
\bottomrule
\end{tabularx}
\end{table}

\begin{table*}[h]
\centering
\caption{\textbf{Sensitivity to the topic-encoder embedding type at fixed topic cardinality.}}
\label{tab:embedding_type_sensitivity}
\scriptsize
\setlength{\tabcolsep}{2pt}
\renewcommand{\arraystretch}{1.16}
\begin{tabular*}{\textwidth}{@{\extracolsep{\fill}}c c c c c c c c c@{}}
\toprule
\textbf{Mix $K$} & \textbf{ATT\&CK-BERT} & \textbf{CTI-BERT} & \textbf{CYBERT} & \textbf{CySecBERT} & \textbf{SecBERT} & \textbf{SecureBERT} & \textbf{MiniLM} \\
\midrule
2 & $\mathbf{0.580}_{\mathbf{0.007}}$ & $0.573_{0.010}$ & $0.571_{0.009}$ & $0.577_{0.010}$ & $0.574_{0.004}$ & $0.575_{0.012}$ & $0.574_{0.006}$ \\
3 & $\mathbf{0.564}_{\mathbf{0.005}}$ & $0.561_{0.006}$ & $0.560_{0.009}$ & $0.563_{0.008}$ & $0.561_{0.002}$ & $0.561_{0.009}$ & $0.563_{0.005}$ \\
4 & $0.545_{0.003}$ & $0.544_{0.005}$ & $0.543_{0.008}$ & $\mathbf{0.546}_{\mathbf{0.007}}$ & $0.543_{0.003}$ & $0.544_{0.006}$ & $0.546_{0.006}$ \\
5 & $0.527_{0.003}$ & $0.527_{0.003}$ & $0.526_{0.007}$ & $0.528_{0.006}$ & $0.526_{0.003}$ & $0.527_{0.006}$ & $\mathbf{0.528}_{\mathbf{0.005}}$ \\
6 & $0.517_{0.003}$ & $0.516_{0.003}$ & $0.516_{0.006}$ & $0.517_{0.005}$ & $0.516_{0.003}$ & $0.517_{0.005}$ & $\mathbf{0.518}_{\mathbf{0.005}}$ \\
\bottomrule
\end{tabular*}
\vspace{0.25em}
\begin{minipage}{0.98\textwidth}
\footnotesize Each embedding column reports Macro-F1 as mean$_{\mathrm{std}}$ under the same TGCM setting. Bold marks the highest Macro-F1 for each mixture size.
\end{minipage}
\end{table*}

\clearpage
\section{Technique Descriptions Used for FASTopic}
\label{app:tech_desc_mapping}
\begin{longtable}{p{0.16\textwidth}p{0.78\textwidth}}
\caption{Technique descriptions~(from MITRE ATT\&CK) used to construct campaign-level FASTopic documents. Each single-APT campaign is represented by concatenating the descriptions of all techniques appearing in its technique sequence.}
\label{tab:tech_desc_mapping}\\
\toprule
\textbf{Technique} & \textbf{Description} \\
\midrule
\endfirsthead

\toprule
\textbf{Technique} & \textbf{Description} \\
\midrule
\endhead

\midrule
\multicolumn{2}{r}{Continued on next page} \\
\endfoot

\bottomrule
\endlastfoot

T1003 & Adversaries may attempt to dump credentials to obtain account login and credential material, normally in the form of a hash or a clear text password. \\
T1003.001 & Adversaries may attempt to access credential material stored in the process memory of the Local Security Authority Subsystem Service (LSASS). \\
T1003.002 & Adversaries may attempt to extract credential material from the Security Account Manager (SAM) database either through in-memory techniques or through the Windows Registry where the SAM database is stored. \\
T1003.003 & Adversaries may attempt to access or create a copy of the Active Directory domain database in order to steal credential information, as well as obtain other information about domain members such as devices, users, and access rights. \\
T1005 & Adversaries may search local system sources, such as file systems and configuration files or local databases, to find files of interest and sensitive data prior to Exfiltration. \\
T1007 & Adversaries may try to gather information about registered local system services. \\
T1016 & Adversaries may look for details about the network configuration and settings, such as IP and/or MAC addresses, of systems they access or through information discovery of remote systems. \\
T1018 & Adversaries may attempt to get a listing of other systems by IP address, hostname, or other logical identifier on a network that may be used for Lateral Movement from the current system. \\
T1021.001 & Adversaries may use Valid Accounts to log into a computer using the Remote Desktop Protocol (RDP). \\
T1033 & Adversaries may attempt to identify the primary user, currently logged-in user, set of users that commonly use a system, or whether a user is actively using the system. \\
T1036.003 & Adversaries may rename legitimate system utilities to try to evade security mechanisms concerning the usage of those utilities. \\
T1036.004 & Adversaries may attempt to manipulate the name of a task or service to make it appear legitimate or benign. \\
T1037.001 & Adversaries may use Windows logon scripts automatically executed at logon initialization to establish persistence. \\
T1040 & Adversaries may passively sniff network traffic to capture information about an environment, including authentication material passed over the network. \\
T1046 & Adversaries may attempt to get a listing of services running on remote hosts and local network infrastructure devices, including those that may be vulnerable to remote software exploitation. \\
T1047 & Adversaries may abuse Windows Management Instrumentation (WMI) to execute malicious commands and payloads. \\
T1049 & Adversaries may attempt to get a listing of network connections to or from the compromised system they are currently accessing or from remote systems by querying for information over the network. \\
T1053.005 & Renamed from ATT\&CK to be consistent with at, launchd, cron siblings; name as is looks like parent. \\
T1055.001 & Adversaries may inject dynamic-link libraries (DLLs) into processes in order to evade process-based defenses as well as possibly elevate privileges. \\
T1055.002 & Adversaries may inject portable executables (PE) into processes in order to evade process-based defenses as well as possibly elevate privileges. \\
T1057 & Adversaries may attempt to get information about running processes on a system. \\
T1059.001 & Adversaries may abuse PowerShell commands and scripts for execution. \\
T1059.003 & Adversaries may abuse the Windows command shell for execution. \\
T1069.001 & Adversaries may attempt to find local system groups and permission settings. \\
T1069.002 & Adversaries may attempt to find domain-level groups and permission settings. \\
T1070.005 & Adversaries may remove share connections that are no longer useful in order to clean up traces of their operation. \\
T1071.001 & Adversaries may communicate using application layer protocols associated with web traffic to avoid detection or network filtering by blending in with existing traffic. \\
T1074.001 & Adversaries may stage collected data in a central location or directory on the local system prior to Exfiltration. \\
T1078.001 & Adversaries may obtain and abuse credentials of a default account as a means of gaining Initial Access, Persistence, Privilege Escalation, or Defense Evasion. \\
T1082 & An adversary may attempt to get detailed information about the operating system and hardware, including version, patches, hotfixes, service packs, and architecture. \\
T1083 & Adversaries may enumerate files and directories or may search in specific locations of a host or network share for certain information within a file system. \\
T1087.001 & Adversaries may attempt to get a listing of local system accounts. \\
T1090.001 & Adversaries may use an internal proxy to direct command and control traffic between two or more systems in a compromised environment. \\
T1105 & Session is initiated by the client, and may be a custom protocol which is why it is related to generic network traffic instead of file transfer network traffic. \\
T1112 & Adversaries may interact with the Windows Registry to hide configuration information within Registry keys, remove information as part of cleaning up, or aid persistence and execution. \\
T1113 & Adversaries may attempt to take screen captures of the desktop to gather information over the course of an operation. \\
T1115 & Adversaries may collect data stored in the clipboard from users copying information within or between applications. \\
T1119 & Once established within a system or network, an adversary may use automated techniques for collecting internal data. \\
T1120 & Adversaries may attempt to gather information about attached peripheral devices and components connected to a computer system. \\
T1123 & An adversary can leverage a computer's peripheral devices. \\
T1124 & An adversary may gather the system time and/or time zone settings from a local or remote system. \\
T1125 & An adversary can leverage a computer's peripheral devices. \\
T1135 & Adversaries may look for folders and drives shared on remote systems as a means of identifying sources of information to gather as a precursor for Collection and to identify potential systems of interest for Lateral Movement. \\
T1137 & Adversaries may leverage Microsoft Office-based applications for persistence between startups. \\
T1137.002 & Adversaries may abuse the Microsoft Office ``Office Test'' Registry key to obtain persistence on a compromised system. \\
T1201 & Adversaries may attempt to access detailed information about the password policy used within an enterprise network or cloud environment. \\
T1204.002 & An adversary may rely upon a user opening a malicious file in order to gain execution. \\
T1217 & Adversaries may enumerate information about browsers to learn more about compromised environments. \\
T1219 & An adversary may use legitimate desktop support and remote access software to establish an interactive command and control channel to target systems within networks. \\
T1482 & Adversaries may attempt to gather information on domain trust relationships that may be used to identify lateral movement opportunities in Windows multi-domain or forest environments. \\
T1486 & Adversaries may encrypt data on target systems or on large numbers of systems in a network to interrupt availability to system and network resources. \\
T1490 & Adversaries may delete or remove built-in data and turn off services designed to aid in the recovery of a corrupted system to prevent recovery. \\
T1491 & Adversaries may modify visual content available internally or externally to an enterprise network, thus affecting the integrity of the original content. \\
T1496 & Adversaries may leverage the resources of co-opted systems to complete resource-intensive tasks, which may impact system and/or hosted service availability. \\
T1497.001 & Adversaries may employ various system checks to detect and avoid virtualization and analysis environments. \\
T1499 & Adversaries may perform Endpoint Denial of Service (DoS) attacks to degrade or block the availability of services to users. \\
T1518 & Adversaries may attempt to get a listing of software and software versions that are installed on a system or in a cloud environment. \\
T1518.001 & Adversaries may attempt to get a listing of security software, configurations, defensive tools, and sensors that are installed on a system or in a cloud environment. \\
T1531 & Adversaries may interrupt availability of system and network resources by inhibiting access to accounts utilized by legitimate users. \\
T1546.013 & Adversaries may gain persistence and elevate privileges by executing malicious content triggered by PowerShell profiles. \\
T1547 & Adversaries may configure system settings to automatically execute a program during system boot or logon to maintain persistence or gain higher-level privileges on compromised systems. \\
T1547.001 & Adversaries may achieve persistence by adding a program to a startup folder or referencing it with a Registry run key. \\
T1547.004 & Adversaries may abuse features of Winlogon to execute DLLs and/or executables when a user logs in. \\
T1547.009 & Adversaries may create or modify shortcuts that can execute a program during system boot or user login. \\
T1547.010 & Adversaries may use port monitors to run an adversary-supplied DLL during system boot for persistence or privilege escalation. \\
T1548.002 & Adversaries may bypass UAC mechanisms to elevate process privileges on a system. \\
T1552.002 & Adversaries may search the Registry on compromised systems for insecurely stored credentials. \\
T1560 & An adversary may compress and/or encrypt data that is collected prior to exfiltration. \\
T1562.001 & Adversaries may modify and/or disable security tools to avoid possible detection of their malware, tools, and activities. \\
T1562.002 & Adversaries may disable Windows event logging to limit data that can be leveraged for detections and audits. \\
T1562.004 & Adversaries may disable or modify system firewalls in order to bypass controls limiting network usage. \\
T1564 & Adversaries may attempt to hide artifacts associated with their behaviors to evade detection. \\
T1564.001 & Adversaries may set files and directories to be hidden to evade detection mechanisms. \\
T1564.003 & Adversaries may use hidden windows to conceal malicious activity from the plain sight of users. \\
T1564.004 & Adversaries may use NTFS file attributes to hide their malicious data in order to evade detection. \\
T1566.001 & Adversaries may send spearphishing emails with a malicious attachment in an attempt to gain access to victim systems. \\
T1574.001 & Adversaries may execute their own malicious payloads by hijacking the search order used to load DLLs. \\

\end{longtable}

\clearpage
\section{LLM Baseline Prompt Template}
\label{app:llm_prompt_template}

The following template is the exact prompt used for the few-shot prompt-only LLM baseline. It receives a mixed MITRE ATT\&CK technique sequence as input and asks the model to output a strict JSON object containing the inferred number of campaigns, reconstructed segments, and occurrence-level assignment vector. The field name \texttt{probable\_attribution} is part of this fixed baseline prompt only; TGCM itself reports episode assignments rather than final actor-attribution labels.

For reproducibility, the ChatGPT-5.5 baseline was evaluated between June 30 and July 4, and the Gemini-3 baseline was evaluated between June 18 and July 4. Each baseline was executed once over the evaluation set using the provider-default temperature and maximum-token settings. Because proprietary LLM behavior and default decoding settings may change over time, these single-run prompt-only results should be interpreted as time-stamped reference baselines rather than fixed model checkpoints or repeated-run estimates. Malformed, non-JSON, or length-mismatched outputs are counted as invalid.

\begin{Verbatim}[breaklines=true,breakanywhere=true]
prompt_template = """
### Role & Objective
You are a world-class Senior Threat Intelligence Analyst and Digital
Forensics Expert specializing in APT (Advanced Persistent Threat)
attribution and campaign reconstruction.
Your task is to perform Blind Source Separation (BSS) on a raw, mixed
sequence of MITRE ATT&CK Technique IDs.

### Scenario Context
The input sequence represents a chronological timeline of malicious
events observed on a compromised network.
CRITICAL: This timeline contains multiple, concurrent APT campaigns
running simultaneously.
The logs are highly interleaved, irregular, and asymmetric.

Key Characteristics of this Data:
1. Identical Starts: Different attackers often use the same entry
   vector. You might see T1566 (Phishing) appear multiple times at the
   start, initiating distinct chains.
2. Uneven Lengths: One campaign might be a quick "Smash and Grab"
   (2-3 steps), while another is a "Low and Slow" espionage operation
   (10+ steps). Do NOT assume equal length.
3. Bursty Interleaving: Techniques are not perfectly alternated.

### Input Data
Technique Sequence:
{question}

### Analysis Guidelines (Step-by-Step Logic)
1. Disambiguate Start Points: If you see multiple T1566 or T1078
   events, assign them to different IDs (e.g., Campaign 1 and Campaign 2)
   based on the subsequent techniques that follow them.
2. Follow the Kill Chain: Trace the narrative. A short Ransomware chain
   (Access -> Impact) is distinct from a long Espionage chain
   (Access -> Discovery -> LatMov -> Collection -> Exfil).
3. Mandatory Assignment: Every single technique MUST be assigned. No
   zeros allowed.

### Response Templates (Learn from these complex scenarios)

Example 1: K=2 (Identical Start, Uneven Lengths)
Input: ["T1566", "T1566", "T1486", "T1059", "T1057", "T1003", "T1041"]
Scenario: Two attackers both Phish (T1566). Attacker A immediately
          deploys Ransomware (T1486) and leaves. Attacker B stays for a
          long Espionage campaign.
Output:
{
  "analysis_summary": "Separated two phishing-initiated campaigns: Cluster 1 is a short, immediate Ransomware strike, Cluster 2 is a longer persistent Espionage operation.",
  "total_campaigns": 2,
  "campaigns": [
    {
      "id": 1,
      "probable_attribution": "Ransomware (Short)",
      "segment": ["T1566", "T1486"]
    },
    {
      "id": 2,
      "probable_attribution": "Espionage (Long)",
      "segment": ["T1566", "T1059", "T1057", "T1003", "T1041"]
    }
  ],
  "locations": [1, 2, 1, 2, 2, 2, 2]
}

Example 2: K=3 (Common Entry, Divergent Paths)
Input: ["T1078", "T1078", "T1078", "T1098", "T1021", "T1048", "T1562"]
Scenario: Three different actors use Valid Accounts (T1078) to login.
- Actor 1 creates a new account (T1098).
- Actor 2 moves laterally via RDP (T1021) then Exfils (T1048).
- Actor 3 disables defenses (T1562).
Output:
{
  "analysis_summary": "Disambiguated three separate login events leading to Account Manipulation, Lateral Movement/Exfil, and Defense Evasion respectively.",
  "total_campaigns": 3,
  "campaigns": [
    {
      "id": 1,
      "probable_attribution": "Persistence Actor",
      "segment": ["T1078", "T1098"]
    },
    {
      "id": 2,
      "probable_attribution": "Lateral Mov Actor",
      "segment": ["T1078", "T1021", "T1048"]
    },
    {
      "id": 3,
      "probable_attribution": "Defense Evasion Actor",
      "segment": ["T1078", "T1562"]
    }
  ],
  "locations": [1, 2, 3, 1, 2, 2, 3]
}

Example 3: K=3 (Chaos: 1 Long, 2 Short/Isolated)
Input: ["T1190", "T1059", "T1083", "T1190", "T1057", "T1003", "T1190", "T1071"]
Scenario:
- Campaign 1 (Long): Web Exploit -> Cmd -> File Discovery -> Process
  Discovery -> Cred Dump -> Web Traffic.
- Campaign 2 (Short): Web Exploit (Scan only).
- Campaign 3 (Short): Web Exploit (Scan only).
Output:
{
  "analysis_summary": "Identified one fully successful intrusion chain amidst two background scanning/failed attempt campaigns.",
  "total_campaigns": 3,
  "campaigns": [
    {
      "id": 1,
      "probable_attribution": "Successful Intrusion",
      "segment": ["T1190", "T1059", "T1083", "T1057", "T1003", "T1071"]
    },
    {
      "id": 2,
      "probable_attribution": "Scanner A",
      "segment": ["T1190"]
    },
    {
      "id": 3,
      "probable_attribution": "Scanner B",
      "segment": ["T1190"]
    }
  ],
  "locations": [1, 1, 1, 2, 1, 1, 3, 1]
}

Example 4: K=4 (Heavily Interleaved Same-Technique Bursts)
Input: ["T1059", "T1059", "T1059", "T1059", "T1486", "T1041", "T1098", "T1562"]
Scenario: Four PowerShell (T1059) executions happen in a row, triggering
          four different impacts.
Output:
{
  "analysis_summary": "Parsed a burst of four PowerShell executions, each triggering a distinct downstream effect (Ransom, Exfil, Persistence, Evasion).",
  "total_campaigns": 4,
  "campaigns": [
    { "id": 1, "segment": ["T1059", "T1486"] },
    { "id": 2, "segment": ["T1059", "T1041"] },
    { "id": 3, "segment": ["T1059", "T1098"] },
    { "id": 4, "segment": ["T1059", "T1562"] }
  ],
  "locations": [1, 2, 3, 4, 1, 2, 3, 4]
}

Example 5: K=5 (Complex Mix of Long and Short)
Input: ["T1566", "T1003", "T1078", "T1021", "T1566", "T1486", "T1003", "T1048", "T1059", "T1566"]
Reasoning:
- 1: Phishing -> Ransomware (Short).
- 2: Cred Dump -> RDP (Lateral).
- 3: Valid Account -> Cred Dump -> Exfil (Long).
- 4: Phishing -> PowerShell (Medium).
- 5: Phishing (Failed/Isolated).
Output:
{
  "analysis_summary": "Separated 5 clusters: A completed Ransomware chain, a Lateral Movement chain, a full Data Exfiltration chain, a Scripting chain, and an isolated Phishing attempt.",
  "total_campaigns": 5,
  "campaigns": [
    { "id": 1, "segment": ["T1566", "T1486"] },
    { "id": 2, "segment": ["T1003", "T1021"] },
    { "id": 3, "segment": ["T1078", "T1003", "T1048"] },
    { "id": 4, "segment": ["T1566", "T1059"] },
    { "id": 5, "segment": ["T1566"] }
  ],
  "locations": [1, 2, 3, 2, 4, 1, 3, 3, 4, 5]
}

### Output Requirements
You must output the result in Strict JSON format only.
- locations length must EQUAL input length.
- locations must contain ONLY positive integers (1, 2, 3...). NO zeros
  allowed.
- Verify that identical techniques (e.g., multiple T1566s) are assigned
  to different IDs if they belong to different logical chains.
"""
\end{Verbatim}

\clearpage
\makeatletter
\if@twocolumn\else\twocolumn\fi
\makeatother
\bibliographystyleapp{IEEEtran}
\bibliographyapp{ref,appendix_ref}

\end{document}